%% file: main.tex
\documentclass{article}

\usepackage[accepted]{mlsys2025}

\makeatletter
\renewcommand{\mlsys@appearing}{}
\makeatother

\usepackage[utf8]{inputenc}
\usepackage[T1]{fontenc}
\usepackage[hyphens]{url}
\usepackage{booktabs}
\usepackage{tabularx}
\usepackage{arydshln}
\usepackage{graphicx}
\usepackage{subcaption}
\usepackage{amsmath}
\usepackage{amsfonts}
\usepackage{nicefrac}
\usepackage{microtype}
\usepackage{xcolor}
\usepackage{multirow}
\usepackage{threeparttable}
\usepackage{tikz}
\usepackage{pgfplots}
\usepackage{enumitem}

\usepackage{hyperref}
\definecolor{arxivlinkblue}{rgb}{0,0.08,0.45}

\pgfplotsset{compat=1.18}

\newcommand{\todo}[1]{\textcolor{cyan}{[todo: #1]}}

\newcommand{\ours}{\textbf{\textsc{Smart-AIMC}}}

\mlsystitlerunning{\ours{}: Sensitivity-Aware Mixed ADC Resolution with Fine-Grained Tunability for Analog IMC}

\hypersetup{
  pdftitle={SMART-AIMC: Sensitivity-Aware Mixed ADC Resolution with Fine-Grained Tunability for Analog In-Memory Computing},
  pdfauthor={Yung-Chin Chen, Ourania Stouraiti, Naveen Verma},
  pdfsubject={arXiv preprint},
  pdfborder={0 0 0},
  colorlinks=true,
  linkcolor=arxivlinkblue,
  citecolor=arxivlinkblue,
  filecolor=arxivlinkblue,
  urlcolor=arxivlinkblue
}

\makeatletter
\newcommand{\printArxivAffiliations}{%
  \stepcounter{@affiliationcounter}%
  {\let\thefootnote\relax\footnotetext{\hspace*{-\footnotesep}%
  \forloop{@affilnum}{1}{\value{@affilnum} < \value{@affiliationcounter}}{%
    \textsuperscript{\arabic{@affilnum}}%
    \csname @affilname\the@affilnum\endcsname.\space
  }%
  Correspondence to: \mlsyscorrespondingauthor@text.}}
}
\makeatother

\begin{document}

\twocolumn[
\mlsystitle{\ours{}: Sensitivity-Aware Mixed ADC Resolution with Fine-Grained Tunability for Analog In-Memory Computing}

\begin{mlsysauthorlist}
\mlsysauthor{Yung-Chin Chen}{princeton,encharge}
\mlsysauthor{Ourania Stouraiti}{princeton}
\mlsysauthor{Naveen Verma}{princeton,encharge}
\end{mlsysauthorlist}

\mlsysaffiliation{princeton}{Princeton University, NJ, USA}
\mlsysaffiliation{encharge}{EnCharge AI, CA, USA}
\mlsyscorrespondingauthor{Yung-Chin Chen}{yc9182@princeton.edu}
\mlsyskeywords{in-memory computing, quantization, mixed precision, hessian-based sensitivity, transformer inference}

\vskip 0.3in

\begin{abstract}
\input{sections/00_abstract}
\end{abstract}
]

\printArxivAffiliations

\addtocontents{toc}{\protect\setcounter{tocdepth}{-1}}

\input{sections/01_introduction}
\input{sections/02_background_and_related_works}
\input{sections/03_methods}
\input{sections/04_experimental_setup}
\input{sections/05_results_and_analysis}
\input{sections/06_conclusions}

\clearpage
\bibliographystyle{mlsys2025}
\bibliography{references}

\section*{AI Usage Disclosure}
Generative AI tools, including OpenAI ChatGPT, OpenAI Codex, Google Gemini, and GitHub Copilot, were used in the following aspects of this work:

\begin{itemize}
    \item \textbf{Writing assistance.} Generative AI tools were used to aid and polish manuscript writing, including improving clarity, concision, and presentation.

    \item \textbf{Retrieval and discovery.} Generative AI tools were used to help identify and explain relevant prior work and references.

    \item \textbf{Research ideation and execution.} Generative AI tools were used to support research discussions, code implementation, writing scripts to run sweep and ablation experiments, and collecting experimental results.

    \item \textbf{Drafting.} Generative AI tools were used to draft portions of the manuscript, which were subsequently reviewed and revised by the authors.
\end{itemize}

\clearpage
\onecolumn
\appendix
\addtocontents{toc}{\protect\setcounter{tocdepth}{1}}
\renewcommand{\contentsname}{Appendix Contents}
\setcounter{tocdepth}{1}
\tableofcontents
\clearpage
\twocolumn
\input{sections/appendix}

\end{document}

%% file: sections/00_abstract.tex
Analog In-Memory Computing (IMC) is a promising paradigm for accelerating GEMM-based AI workloads by executing computations directly within the memory array in the analog domain. While IMC delivers substantial energy and throughput improvements, it can suffer from significant task accuracy degradation due to analog-to-digital converter (ADC) quantization noise, which occurs at the GEMM output and is thus different from weight/activation quantization, which occurs at the GEMM input. Introducing mixed ADC resolution, where IMC modules can select between a high-resolution, high-energy ADC mode and a low-resolution, low-energy ADC mode, offers a compelling accuracy-energy trade-off. However, systematically selecting the ADC resolution across IMC units to which GEMM computations are mapped, while aligning with the granularity constraints of the hardware to preserve efficiency, remains a key challenge.
In this paper, we propose \ours{}, a \textbf{S}ensitivity-aware \textbf{M}ixed \textbf{A}DC \textbf{R}esolution framework with fine-grained \textbf{T}unability for \textbf{A}nalog \textbf{I}n-\textbf{M}emory \textbf{C}omputing. \ours{} correlates the algorithmic task loss to hardware ADC quantization noise in the form of an ADC Channel Sensitivity (ACS) metric. Using this metric, \ours{} systematically, and in a disciplined manner, allocates ADC resources to critical computations. Furthermore, we introduce a sensitivity-based channel reordering scheme that overcomes the hardware granularity constraints with no runtime overhead and approaches the optimal ADC quantization possible for the model.
Evaluated across diverse workloads, \ours{} unlocks a superior accuracy--energy Pareto frontier, providing (1) higher task accuracy at matched energy budgets and (2) fine-grained tuning across efficiency targets, from aggressive energy savings to near-digital inference accuracy. At the latter operating point, \ours{} achieves $6.99\times$ higher modeled system-level energy efficiency for linear projections than the INT8 digital baseline with negligible accuracy degradation.

%% file: sections/01_introduction.tex
\section{Introduction}
\label{sec:introduction}
Large Language Models (LLMs) have demonstrated unprecedented capabilities across a wide spectrum of natural language and reasoning tasks, but their massive number of compute operations imposes prohibitive computation and memory costs for inference.
To mitigate this bottleneck, the field has widely adopted quantization to trade compute fidelity for substantial hardware efficiency. By representing data with fewer bits and executing computations with low-precision kernels for both weights and activations, this strategy has been highly effective over the past decade. Today, mainstream digital accelerators have successfully pushed this paradigm to the regime of ultra-low-precision compute (e.g., INT4, MXFP4 \citep{microscaling}, NVFP4 \citep{NVFP4}). Yet, the further gains achievable from this trajectory are rapidly diminishing with digital architectures already operating at aggressive quantization points and with accuracy degradation becoming an increasing concern.

\begin{figure*}[t]
  \centering
  \begin{subfigure}[t]{0.66\textwidth}
    \centering
    \includegraphics[width=\linewidth]{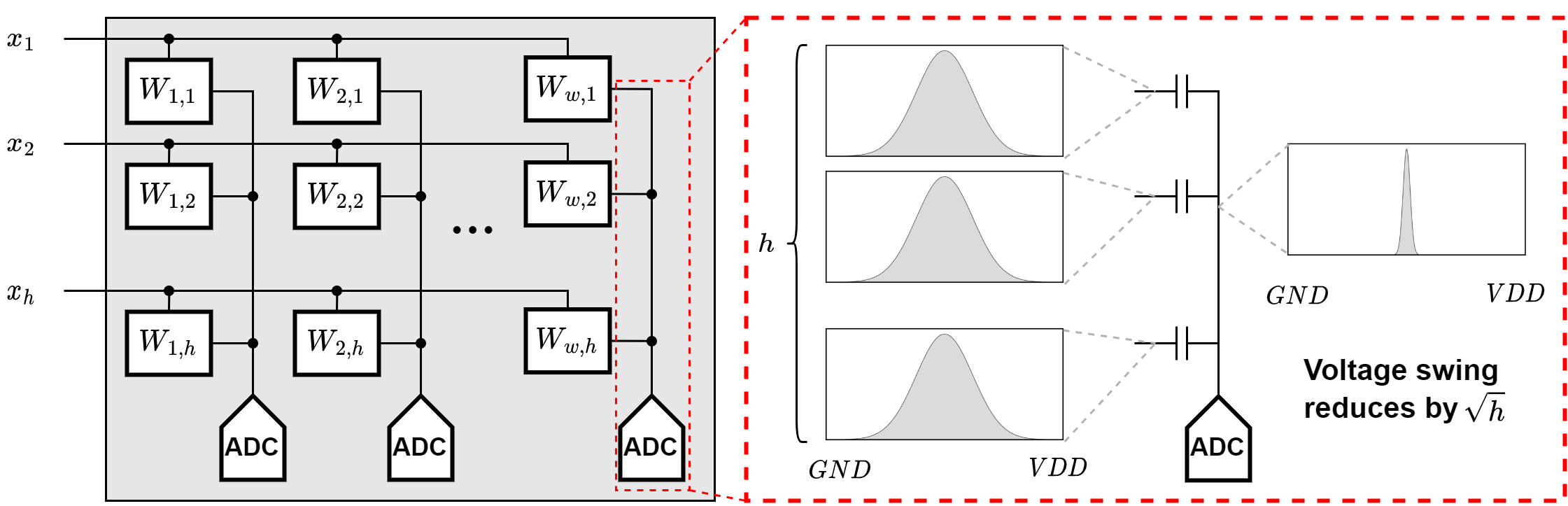}
    \caption{\textbf{IMC architecture and reduced signal swing.} IMC enables MVMs by streaming input vector $x$ to memory, storing weight matrix $W$, and performing multiplication and accumulation to perform reduction across each column in the analog domain. This analog summation achieves \(\sim 10\times\) higher energy efficiency than a digital approach, but reduces the output signal swing and thus exacerbates ADC quantization difficulty.}
    \label{fig:background-imc-intro}
  \end{subfigure}\hfill
  \begin{subfigure}[t]{0.30\textwidth}
    \centering
    \includegraphics[width=\linewidth]{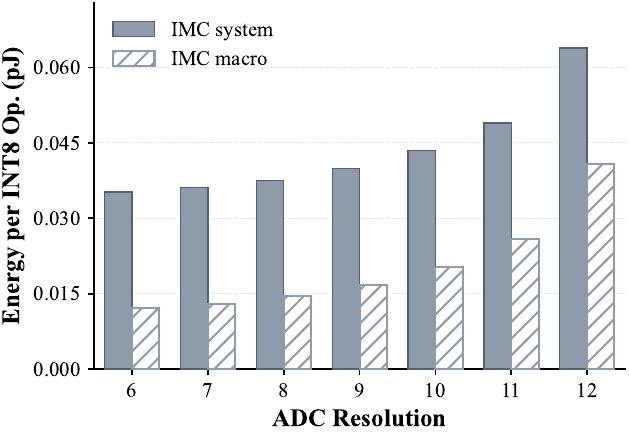}
    \caption{\textbf{IMC macro- and system-level energy vs. ADC resolution.} Energy increases exponentially with ADC resolution (details are provided in Appendix \ref{appx:figure_experiment_details}).}
    \label{fig:background-adc-precision-energy-cost}
  \end{subfigure}
  \caption{IMC architecture, signal swing reduction, and ADC resolution-energy trade-off.}
  \label{fig:background-imc-swing-adc}
\end{figure*}

Analog In-Memory Computing (IMC) has emerged as a compelling approach for enhancing compute efficiency. By executing matrix-vector multiplications (MVMs) directly within the memory array using the physical properties of the circuits, analog IMC: (1) delivers significant improvements in energy efficiency and throughput of compute; and (2) minimizes data transfers between the memory and processing units. Analog IMC thus represents a highly attractive architecture for sustaining the continued hardware efficiency gains for LLM inference.

However, in terms of compute accuracy, analog IMC introduces its own set of non-idealities: \textit{analog noise} and \textit{analog-to-digital (A/D) conversion}. While the challenge of analog noise has been largely overcome by modern switched-capacitor architectures, this advancement has left the A/D conversion as the critical bottleneck. The Analog-to-Digital Converter (ADC) dominates the total macro energy, making it the primary knob in the accuracy-efficiency trade-off space. While utilizing a higher ADC resolution improves compute fidelity by preserving the partial sum accumulation, it can substantially degrade the energy benefits that make analog IMC attractive in the first place.

To navigate this challenge, utilizing mixed ADC resolution presents a highly effective strategy. By optimally setting the ADC resolution across different layers, operations, or channels, it is possible to balance this trade-off, allocating expensive high resolution only where it is necessary. Such resolution configurability can be readily supported within the ADC circuitry typically used for IMC.
Such hardware configurability, however, raises an algorithmic question: how can we systematically allocate ADC resolution across a large model to maximize inference accuracy under different energy-efficiency targets?

\begin{figure*}[t]
  \centering
  \begin{subfigure}[t]{0.45\textwidth}
    \centering
    \includegraphics[width=\linewidth]{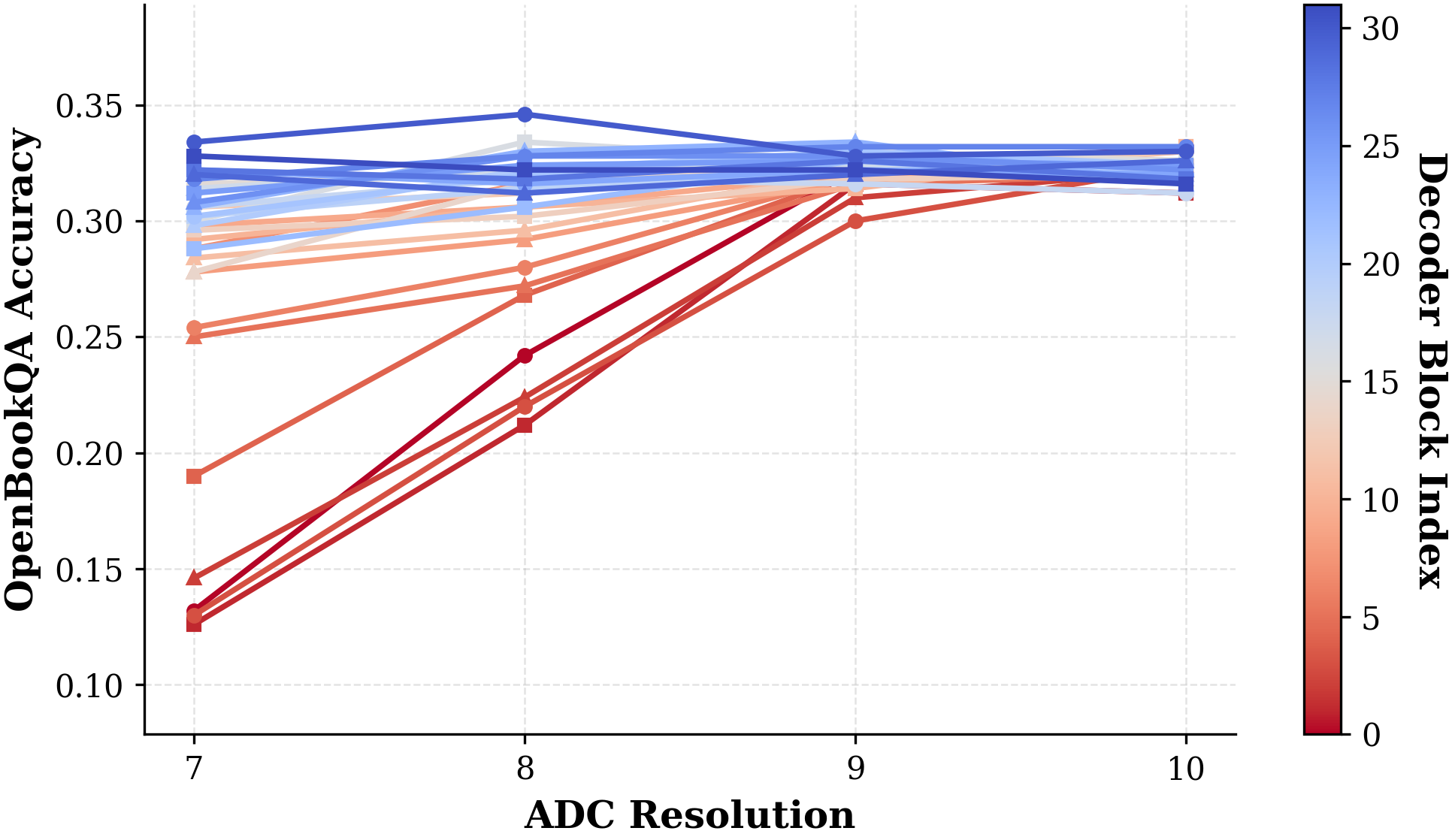}
    \caption{ADC sensitivity of decoder blocks varies}
    \label{fig:layer-sensitivity-curve}
  \end{subfigure}\hfill
  \begin{subfigure}[t]{0.51\textwidth}
    \centering
    \includegraphics[width=\linewidth]{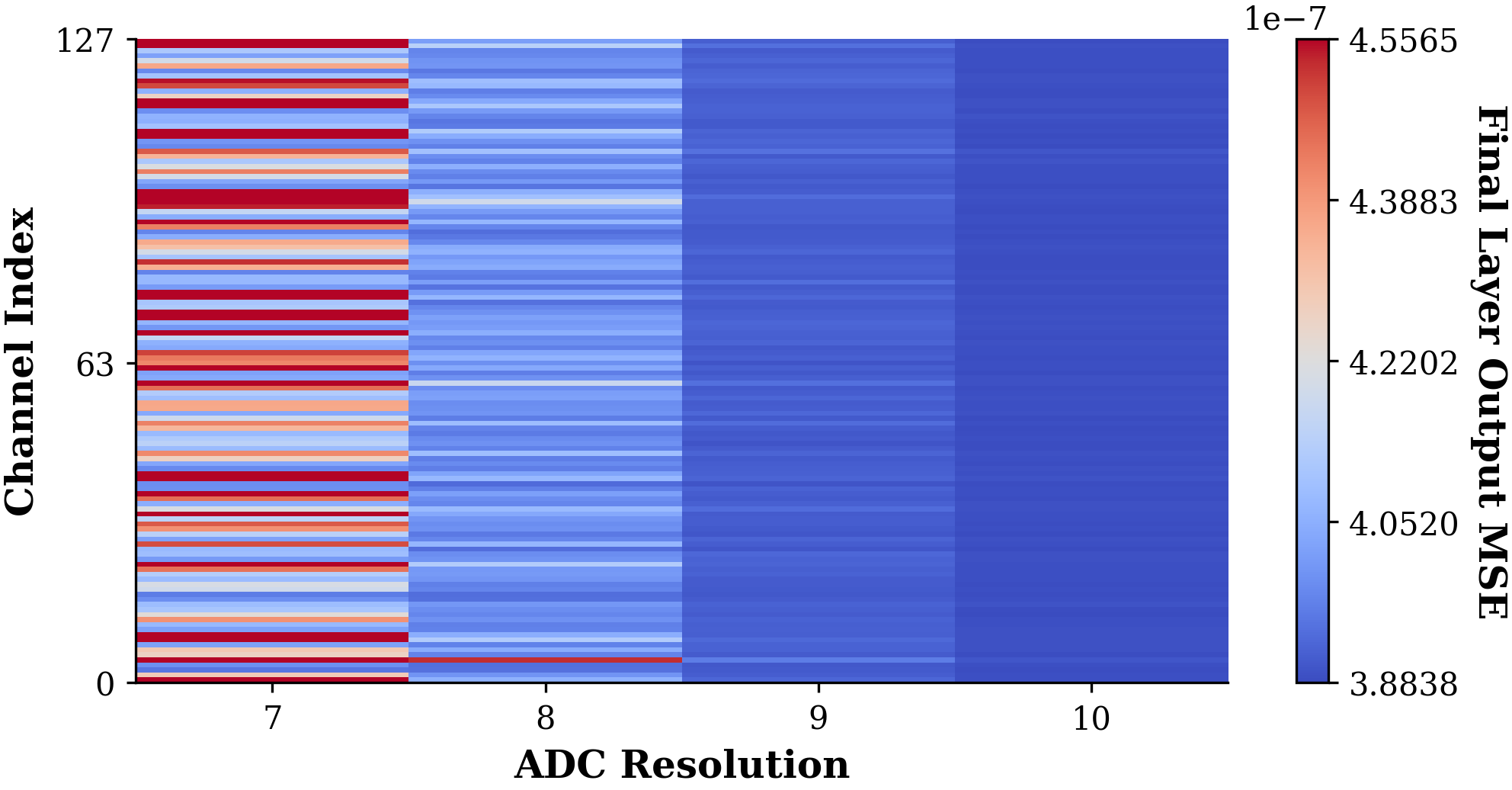}
    \caption{ADC sensitivity of channels varies}
    \label{fig:channel-sensitivity-heatmap}
  \end{subfigure}
  \caption{\textbf{Variations in ADC sensitivity.} Sensitivity to ADC quantization varies significantly not only across decoder blocks, but also among individual channels, necessitating a systematic approach for optimal compute-to-ADC resolution mapping. The experiment setting is detailed in Appendix \ref{appx:figure_experiment_details}.}
  \label{fig:adc-sensitivity}
\end{figure*}

In this paper, we present \ours{}, a \textbf{S}ensitivity-aware \textbf{M}ixed \textbf{A}DC \textbf{R}esolution framework with fine-grained \textbf{T}unability for \textbf{A}nalog \textbf{I}n-\textbf{M}emory \textbf{C}omputing. \ours{} leverages the second-order information of the loss landscape to identify the model channels that are most sensitive to ADC quantization, finely guiding the configuration to high-resolution ADCs where they are most needed. However, IMC hardware constraints require grouping output channels together to amortize hardware overheads, making it challenging to leverage fine-grained optimal resolution configuration. Thus, we further introduce a channel reordering scheme to group channels with similar sensitivity together in the same IMC unit, bridging the performance gap imposed by hardware granularity requirements without incurring any runtime overhead.
We demonstrate that \ours{} establishes a superior accuracy--energy Pareto frontier over conventional allocation strategies, enabling fine-grained tuning across diverse operating points. This flexibility allows \ours{} to achieve higher accuracy under constrained energy budgets, while also approaching digital-level accuracy when higher fidelity is prioritized, without sacrificing the substantial energy-efficiency advantages of analog IMC.
Our contributions are summarized as follows:

\begin{itemize}
    \item We propose \ours{}, a systematic, sensitivity-guided framework for mixed ADC resolution assignment in analog IMC-based Transformer inference, jointly capturing task-loss curvature and ADC quantization-noise scaling to optimize the accuracy--energy trade-off.
    \item We introduce sensitivity-based channel reordering to mitigate the ADC resolution granularity constraint, improving ADC quantization quality without additional runtime overhead or interference with rotation-based activation quantization method.
    \item We demonstrate that \ours{} achieves a superior accuracy--energy Pareto frontier across diverse architectures and tasks, enabling analog IMC frameworks to flexibly support both energy-constrained and accuracy-critical scenarios, retaining an estimated $7$--$10\times$ higher system-level energy efficiency for linear projections than the INT8 digital baseline.
\end{itemize}

%% file: sections/02_background_and_related_works.tex
\section{Background and Related Works}
\label{sec:background-related-work}
\subsection{Analog IMC circuits}
\label{subsec:analog-imc-circuit}
The motivation behind analog IMC is to circumvent the energy and latency penalties associated with data movement and processing in MVMs by executing computations directly within the memory array. In a typical analog IMC architecture shown in Figure \ref{fig:background-imc-intro}, weight matrices are loaded into a 2D memory array. Input activations are then streamed in parallel across the memory wordlines. The MVM operation is realized by performing element-wise multiplication locally at the site of each memory cell, followed by an analog accumulation along the bitlines. This in-situ computing paradigm offers two primary advantages: (1) it eliminates the transfer of weight from local memory to compute units, and (2) it improves compute efficiency by performing massive reduction operations in the analog domain. Consequently, IMC can function as a specialized MVM engine, delivering high throughput and energy efficiency required for GEMM-dominated AI workloads.

While various memory technologies (e.g., RRAM \citep{RRAM-IMC}, MRAM \citep{MRAM-IMC}, PCM \citep{PCM-IMC}, eDRAM \citep{edram-IMC}, etc.) and analog accumulation schemes (e.g., current-domain \citep{current-domain-IMC}, time-domain \citep{time-domain-IMC}, etc.) have been explored for IMC, their practicality is limited by accuracy degradation caused by intrinsic analog noise. To overcome this limitation, switched-capacitor based IMC designs \citep{valavi201964,lee2021fully,lee2024switched} have been widely adopted. Leveraging the high precision of lithographically-defined metal capacitors freely available in current chip technologies, these designs dramatically mitigate the impact of analog noise, enabling precise high-dimensionality reduction operations in charge domain. This, however, shifts the compute accuracy bottleneck to the subsequent ADC operation following analog accumulation, which is required to provide digital outputs to the rest of the architecture.

Information loss during ADC quantization primarily stems from a diminishing effective signal swing \cite{Murmann-mixed-signal,Benchmarking-IMC}. When accumulating independent and identically distributed (i.i.d.) products across $h$ rows, the signal's standard deviation scales by $\sqrt{h}$, while the maximum theoretical range scales by $h$. Because charge-domain analog outputs are bounded by fixed supply rails, the effective signal swing narrows proportionally to $1/\sqrt{h}$, illustrated in Figure \ref{fig:background-imc-intro}. Thus, scaling up row parallelism to maximize throughput forces the signal into an increasingly narrow signal band. Compounding this challenge is the lack of dynamic range management: unlike digital quantization, where data distributions can be easily scaled prior to low-precision casting, implementing equivalent analog preprocessing before the ADC increases analog noise significantly \citep{grimm2022neural}. As a result, this inherently fixed-range analog quantization severely underutilizes the dynamic range and degrades compute fidelity \citep{RAOQ}. While increasing the ADC resolution is a straightforward solution, it imposes an exponential penalty in energy (Figure \ref{fig:background-adc-precision-energy-cost}), rapidly negating the efficiency advantages of IMC \citep{Murmann-mixed-signal}.

\subsection{Mixed ADC Resolution IMC}
\label{subsec:mixed-adc-precision-imc}
To better navigate this accuracy-energy tradeoff, employing a mixed ADC resolution scheme emerges as a highly effective strategy. The core principle relies on sensitivity-based resource allocation: provisioning high ADC resolution for the most sensitive computations, while assigning lower resolutions to the remaining operations. Figure \ref{fig:adc-sensitivity} shows why a mixed ADC resolution scheme is highly beneficial: the sensitivities of different decoder blocks, as well as those of individual channels within a specific layer, differ substantially and are distributed irregularly. This irregularity necessitates a fine-grained, mixed-resolution approach to balance model accuracy and hardware efficiency.

At the hardware level, mixed ADC resolution can be implemented with substantially less complexity than heterogeneous fixed-precision computing architectures. In such digital mixed-precision architectures, different operations must be mapped and routed to engines with different precisions. This imposes complexity both in how model layers/operations with different sensitivity requirements must be physically mapped and routed to the hardware, and in how different heterogeneous-precision engines should be provisioned within the architecture.
In contrast, IMC engines can readily be made resolution-configurable by leveraging the Successive Approximation Register (SAR) ADC approach commonly used in IMC hardware. SAR ADCs can be made resolution-configurable to yield substantial energy savings in low-resolution modes by halting the successive-approximation process at a desired bit resolution \citep{verma200625}, physically decoupling the dominant most-significant-bit (MSB) capacitors within the capacitive DAC array \citep{yip2013resolution}, and scaling the comparator power to match the tolerable electronic noise floor \citep{nuzzo2008noise}. Leveraging these mechanisms, the integration of resolution-configurable ADCs with analog IMC has been widely demonstrated as a viable path for optimizing system efficiency \citep{lee2021charge, ando20254541, yoshioka202434}.

Given the feasibility of such configurable hardware, the fundamental challenge shifts to the algorithmic domain: how do we systematically identify and map sensitive computations to higher ADC resolutions?
Existing literature predominantly relies on search-based methods, including greedy search \citep{lee2021charge,edram-IMC} and iterative evolutionary optimization \citep{cim2pq}. These methods require repeated evaluations of candidate resolution assignments, incurring substantial search overhead that can grow with model size and allocation granularity. This makes it challenging to simultaneously achieve low optimization overhead, fine-grained sensitivity-aware resolution allocation, and generalizability across diverse workloads. Consequently, scaling mixed-resolution IMC for large-scale LLM workloads requires a systematic, low-overhead framework capable of automatically determining fine-grained, Pareto-optimal computation-to-resolution mappings.

\subsection{Mapping MVMs to IMC via Tiling}
\label{subsec:imc-tiling-and-extensions}
Deploying a linear layer $\mathbf{y} = \mathbf{W}\mathbf{x}$ on an IMC processor, where the weight matrix $\mathbf{W} \in \mathbb{R}^{d_{\text{out}} \times d_{\text{in}}}$ and the input vector $\mathbf{x} \in \mathbb{R}^{d_{\text{in}}}$, requires partitioning $\mathbf{W}$ to fit within physically constrained IMC units. While the partition can be done temporally or spatially without affecting output, we assume a fully spatial unrolling for notational simplicity. Assuming each unit has a physical capacity of $h$ rows (inputs) and $w$ columns (outputs), $\mathbf{W}$ is thus unrolled into a $K \times L$ grid of sub-matrices $\mathbf{W}^{(k,l)}$ to be stored in the IMC units, where $K = \lceil d_{\text{in}} / h \rceil$ and $L = \lceil d_{\text{out}} / w \rceil$.

This tiling strategy introduces two distinct axes of decomposition:
\begin{itemize}[leftmargin=*]
\item \textbf{Output Extension (Column-wise):} The weight matrix is partitioned across $L$ units to compute distinct subsets of output channels ($\mathbf{y}^{(l)}$). The input vector $\mathbf{x}$ is broadcasted to these units.
\item \textbf{Input Extension (Row-wise):} The weight matrix is partitioned across $K$ units to process segments of the input vector ($\mathbf{x}^{(k)}$). Along this axis, each unit computes a local analog partial sum:
\[
    \mathbf{p}_{\text{analog}}^{(k,l)}
    = \mathbf{W}_{\text{int}}^{(k,l)} \mathbf{x}_{\text{int}}^{(k)}.
\]
Taking symmetric quantization as an example, these integer operands are quantized digitally via
\[
    \mathbf{W}_{\text{int}}^{(k,l)}
    = \left\lfloor \frac{\mathbf{W}^{(k,l)}}{s_w^{(l)}} \right\rceil,
    \qquad
    \mathbf{x}_{\text{int}}^{(k)}
    = \left\lfloor \frac{\mathbf{x}^{(k)}}{s_x} \right\rceil,
\]
where $s_w^{(l)}$ and $s_x$ are the floating-point scaling factors, and $\lfloor \cdot \rceil$ represents rounding to the nearest integer.
\end{itemize}

To produce the final results, these analog partial sums must be digitized by the ADCs and accumulated. Let $Q_{\text{ADC}}^{(k,l)}(\cdot)$ denote the ADC quantization for the $(k, l)$-th unit. Incorporating the floating-point scaling factors for the inputs ($s_x$) and weights ($s_w$), the final dequantized output vector for the $l$-th column group is mathematically reconstructed as:
$$\mathbf{y}^{(l)} = s_x s_w^{(l)} \sum_{k=0}^{K-1} Q_{\text{ADC}}^{(k,l)} \left( \mathbf{W}_{\text{int}}^{(k,l)} \mathbf{x}_{\text{int}}^{(k)} \right)$$

To reflect practical hardware design constraints, mixed ADC resolution is applied at the granularity of individual IMC units. This block-level assignment avoids the prohibitive area overhead of per-column control wiring and ensures that all channels in a unit complete their conversions in the same number of SAR ADC clock cycles, simplifying output synchronization.

While the hardware can support independent configurations for every IMC unit, in this work, we focus on applying mixed ADC resolution exclusively along the output extension dimension. Consequently, all $K$ units along the input extension for a given column group $l$ are constrained to use an identical ADC resolution (i.e., $Q_{\text{ADC}}^{(k,l)} = Q_{\text{ADC}}^{(l)}$ for all $k$).


%% file: sections/03_methods.tex
\section{Methods}
\label{sec:methods}

\subsection{Hessian-based Sensitivity for ADC Quantization}
\label{subsec:hessian-sensitivity-adc-quantization}
Systematic ADC resolution assignment requires quantifying how quantization noise impacts the overall task loss. Unlike standard Hessian-based sensitivity metrics (detailed in Appendix \ref{appx:mixed-precision-computing}), which model perturbations directly at the input or weight tensors, analog IMC architectures introduce ADC quantization noise directly at the GEMM output ($\mathbf{Y} = \mathbf{WX}$) during A/D conversion. Consequently, we must reformulate the sensitivity analysis with respect to the output activations.
Let $\boldsymbol{\epsilon}$ denote the effective ADC quantization noise in the unquantized domain, injected at the GEMM output $\mathbf{Y}$. The resulting perturbation to the total task loss $\mathcal{L}$ can be approximated using a second-order Taylor expansion:$$\delta \mathcal{L} \approx \mathbf{g}^\top \boldsymbol{\epsilon} + \frac{1}{2} \boldsymbol{\epsilon}^\top \mathbf{H} \boldsymbol{\epsilon}$$where $\mathbf{g} = \nabla_{\mathbf{Y}} \mathcal{L}$ represents the first-order gradient and $\mathbf{H} = \frac{\partial^2 \mathcal{L}}{\partial \mathbf{Y}^2}$ is the Hessian matrix of the loss with respect to the output activations.
As will be established in Sec. \ref{subsec:modeling-adc-noise-propagation}, the underlying ADC quantization introduces a zero-mean error that is statistically independent of the input-dependent gradients $\mathbf{g}$ when the signal has variance much larger than the quantization step size. Consequently, the expected value of the effective unquantized perturbation is also zero ($\mathbb{E}[\boldsymbol{\epsilon}] = 0$), causing the first-order term to vanish in expectation. Furthermore, because this quantization noise can be modeled as statistically independent across different output channels (as detailed in Appendix \ref{appx:out-channel-independence}), the off-diagonal elements of $\mathbf{H}$ cancel out. The expected loss degradation thus simplifies to a function of the diagonal Hessian elements $H_{ii}$ and the variance of the effective unquantized noise $\epsilon_i$:$$\mathbb{E}[\delta \mathcal{L}] \approx \frac{1}{2} \sum_i H_{ii} \mathbb{E}[\epsilon_i^2]$$

\subsection{Modeling ADC Noise Propagation}
\label{subsec:modeling-adc-noise-propagation}
To evaluate the expected noise variance $\mathbb{E}[\epsilon_i^2]$, we must project the noise from the physical ADC domain back to the unquantized floating-point domain. First, consider the quantization occurring at a single IMC unit. The ADC digitizes an analog partial sum using a uniform step size $\Delta_i$ assigned to the computations producing the $i$-th output channel. According to the statistical theory of quantization \citep{widrow-quantization}, assuming the analog signal variation is sufficiently large relative to the step size, the raw quantization error for the $k$-th IMC unit can be modeled as an independent uniform distribution $n_{\text{adc}}^{(k)} \sim \mathcal{U}(-\frac{\Delta_i}{2}, \frac{\Delta_i}{2})$. This hardware noise has a zero mean and a known variance of $\text{Var}(n_{\text{adc}}^{(k)}) = \frac{\Delta_i^2}{12}$. Next, we aggregate this noise across the input extension. Because the computation of a single output channel requires accumulating partial sums from $K$ different IMC units, and the physical noise sources across these units can be modeled as statistically independent (as detailed in Appendix \ref{appx:input-imcu-independence}), their variances add linearly. The total aggregated noise in the quantized domain, $n_{\text{total}} = \sum_{k=1}^K n_{\text{adc}}^{(k)}$, has a combined variance of:$$\text{Var}(n_{\text{total}}) = \sum_{k=1}^K \text{Var}(n_{\text{adc}}^{(k)}) = K \frac{\Delta_i^2}{12}$$Finally, we map this aggregated hardware noise back to the unquantized domain. To recover the unquantized activation, the quantized output is multiplied by the corresponding scaling factors for the inputs ($s_x$) and weights ($s_w$). Therefore, the effective unquantized noise is $\epsilon_i = s_x s_w n_{\text{total}}$, yielding a final variance of:$$\mathbb{E}[\epsilon_i^2] = (s_x s_w)^2 \text{Var}(n_{\text{total}}) = K \frac{\Delta_i^2}{12} (s_x s_w)^2$$

\begin{figure*}[t]
    \centering
    \includegraphics[width=\textwidth]{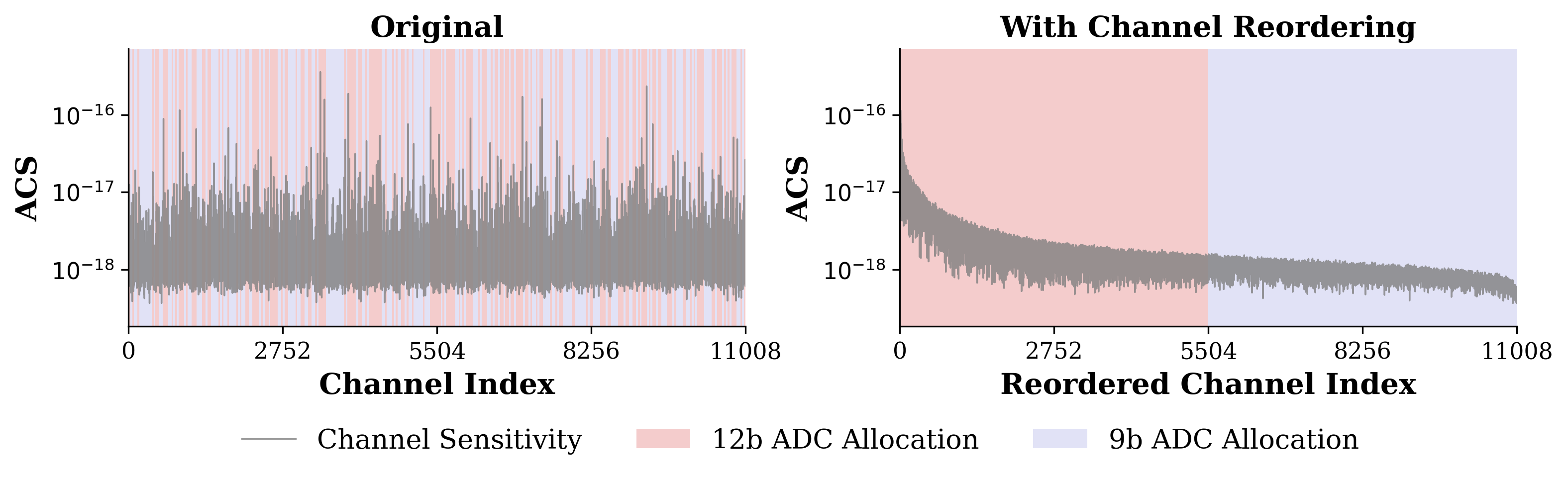}
    \caption{\textbf{Channel reordering for ADC allocation.} Without reordering, high-sensitivity channels can be inadvertently mapped to low-resolution IMC units. Channel reordering strictly ensures that the high-sensitivity channels are mapped to high-resolution ADCs. The setting is detailed in Appendix \ref{appx:figure_experiment_details}.}
    \label{fig:sensitivity-with-reordering}
\end{figure*}

\subsection{ACS and IMC Unit Sensitivity}
\label{subsec:channel-imc-unit-sensitivity}

Substituting the effective noise variance back into our expected loss equation yields:$$\mathbb{E}[\delta \mathcal{L}] \approx \frac{1}{2} \sum_i H_{ii} \left( K \frac{\Delta_i^2}{12} (s_x s_w)^2 \right)$$Because the required number of partial sums, $K$, scales proportionally with the layer's input dimension $d_{\text{in}}$ ($K = \lceil d_{\text{in}}/h \rceil$), we can abstract away constant hardware scaling factors to express the total expected loss perturbation proportionally:$$\mathbb{E}[\delta \mathcal{L}] \propto \sum_i H_{ii} \cdot d_{\text{in}} \cdot (s_x s_w)^2 \cdot \Delta_i^2 = \sum_i \Omega_i \cdot \Delta_i^2$$Here, we extract the \textbf{ADC Channel Sensitivity (ACS)} as $\Omega_i = H_{ii} \cdot d_{\text{in}} \cdot (s_x s_w)^2$. This formulation deliberately isolates the intrinsic algorithmic and architectural sensitivity ($\Omega_i$) from the controllable hardware parameter ($\Delta_i$). By decoupling these terms, we establish a rigorous foundation for mixed-resolution allocation: the expected degradation in task loss depends strictly on the channel's inherent sensitivity and the chosen ADC step size.

If we take a closer look at the composition of ACS, we can see that it captures both (1) Hessian information and (2) the scale of ADC-induced quantization noise, providing a more complete measure of quantization sensitivity than Hessian information alone. We further evaluate the benefit of incorporating ADC-specific noise information in Appendix~\ref{appx:sensitivity-estimation-quality}.

While $\Omega_i$ quantifies the sensitivity of individual output channels, a single ADC resolution must be applied across all output channels processed by the same IMC unit, as detailed in Sec. \ref{subsec:imc-tiling-and-extensions}. This enforces a uniform step size ($\Delta^{(l)}$) across all $w$ output channels processed by Unit $l$. To align with this constraint, we average the constituent ACS values (denoted by the index set $\mathcal{C}_l$) to define the \textbf{IMC unit sensitivity}:$$\bar{\Omega}^{(l)} = \frac{1}{w} \sum_{i \in \mathcal{C}_l} \Omega_i$$This aggregated metric ($\bar{\Omega}^{(l)}$) serves as a practical proxy to evaluate the global importance of the physical macro.

To allocate ADC resolution under a limited energy budget, we adopt a cost-aware greedy strategy that prioritizes the expected reduction in task loss per unit hardware cost. Specifically, increasing the ADC resolution of an IMC output group requires additional ADC conversions proportional to the number of input tiles, $K=\lceil d_{\mathrm{in}}/h\rceil$. We therefore rank candidate groups by their predicted loss reduction normalized by the corresponding incremental ADC conversion cost, allocating high-resolution ADCs in descending order until the budget is exhausted.


\subsection{Sensitivity-aware Channel Reordering}
\label{subsec:channel-reordering}
The architectural requirement of enforcing a uniform ADC resolution across an entire IMC unit inherently degrades the model's accuracy-efficiency trade-off. As depicted in Figure \ref{fig:sensitivity-with-reordering} (left), this degradation occurs because channels with highly divergent sensitivities are forced to share the same ADC resolution, leading to either over-provisioning of insensitive channels (wasting energy) or under-provisioning of sensitive ones (harming accuracy).

To minimize this degradation, we utilize the channel reordering method introduced by \citet{RPTQ}, but guided by the ACS metric ($\Omega_i$). As shown in Figure \ref{fig:sensitivity-with-reordering} (right), sorting output channels by sensitivity ensures sensitive channels receive high-resolution ADCs, minimizing the accuracy degradation caused by sharing a single ADC resolution across the IMC unit. Mathematically, this is achieved by applying a permutation matrix $\mathbf{P}$ to reorder the output channels of the current layer's weight matrix ($\tilde{\mathbf{W}}_1 = \mathbf{P}\mathbf{W}_1$). To preserve mathematical equivalence, the inverse permutation $\mathbf{P}^{-1}$ is identically applied to reorder the input channels of the subsequent layer's weight matrix ($\tilde{\mathbf{W}}_2 = \mathbf{W}_2\mathbf{P}^{-1}$). This structural reorganization is fused directly into the static weights offline, incurring zero runtime overhead. Furthermore, while aggressive channel-mixing orthogonal transformations could be employed, we deliberately restrict our intervention to strict channel permutation. This ensures our output-side reordering preserves compatibility with the rotation-based methods commonly used to optimize input activation quantization \citep{Quarot, Spinquant}.

\begin{figure*}[t]
	\centering
	\includegraphics[width=\textwidth]{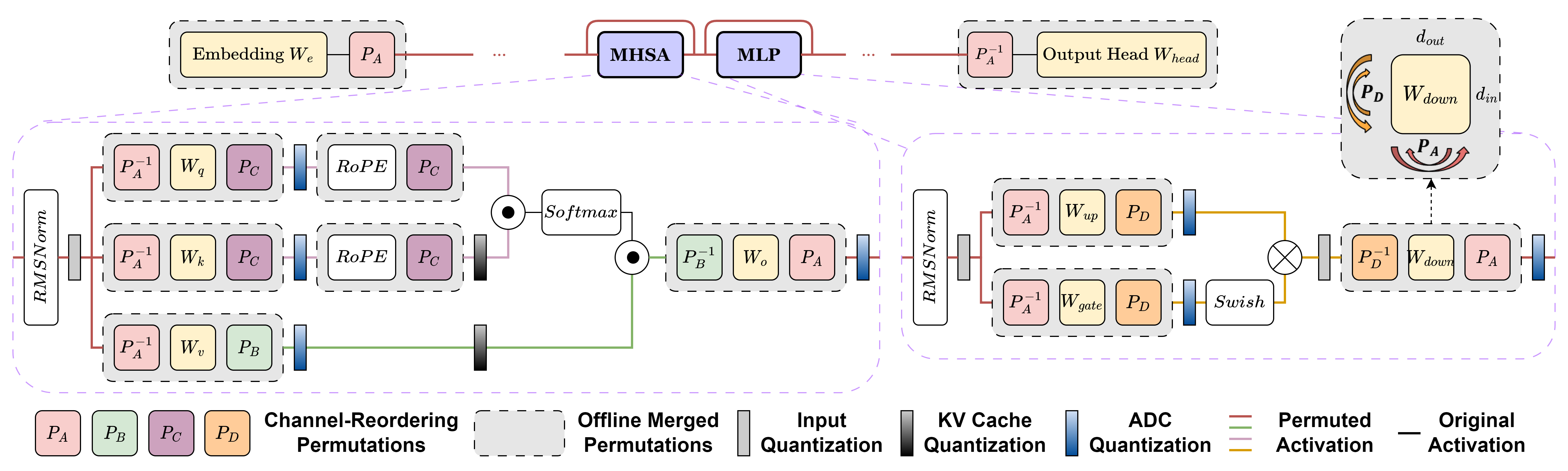}
  \caption{\textbf{Sensitivity-aware channel reordering across a decoder layer.} Output channels with similar sensitivity are grouped to reduce IMC unit-level mixed-resolution loss during ADC quantization.}
	\label{fig:decoder-layer-reordering}
\end{figure*}

As illustrated in Figure~\ref{fig:decoder-layer-reordering}, this channel reordering strategy can be systematically applied prior to every ADC quantization step across the transformer architecture to minimize the degradation imposed by the IMC unit granularity constraint. A comprehensive breakdown of how these distinct permutation matrices (e.g., $\mathbf{P}_A, \mathbf{P}_B, \mathbf{P}_C, \mathbf{P}_D$) are formulated, integrated across various architectural components, and co-designed with rotation-based methods, is detailed in Appendix \ref{appx:permutation-construction}.

%% file: sections/04_experimental_setup.tex
\section{Experimental Setup}
\label{sec:experiments}

\subsection{Mixed ADC Resolution IMC Framework}
\label{subsec:mixed-adc-imc-framework}
\paragraph{IMC Configuration.}
Our evaluation framework simulates a practical IMC architecture where both input activations and weights are quantized to 8-bit precision. Following common IMC architectures \citep{lee2024switched,cambricon-cim}, we emulate an execution scheme that divides these 8-bit operands into 4-bit slices---a sweet spot for balancing analog circuit accuracy and efficiency \citep{cambricon-cim}. These architectures demonstrate superior compute efficiency, but suffer substantial accuracy degradation due to ADC quantization. To evaluate \ours{}, we model each physical IMC unit with a capacity of $h=512$ rows and $w=32$ columns, where the ADC resolution per unit is configurable to either 9 or 12 bits. These represent bit resolutions where the commonly used SAR ADC falls in the different operating regimes of moderate versus high energy-resolution scaling (Fig. \ref{fig:background-adc-precision-energy-cost}). Further details regarding the underlying IMC modeling and computation pipeline are provided in Appendix \ref{appx:imc-encoding-mapping}.

\paragraph{Sensitivity Estimation.}
To efficiently compute the ACS ($\Omega_i$) without incurring prohibitive computational overhead, we approximate the Hessian diagonals ($H_{ii}$) with the Empirical Fisher Information Matrix (EFIM) \citep{woodfisher, martens2020new}---a low-cost proxy widely adopted in LLM quantization literature \citep{Squeezellm, MixLLM, FGMP}. Given a small calibration dataset of $N$ samples, the diagonal elements with respect to the output activations $\mathbf{Y}$ are approximated using the squared first-order gradients:
$$H_{ii} \approx \frac{1}{N} \sum_{n=1}^N \left( \frac{\partial \mathcal{L}_n}{\partial Y_i} \right)^2$$

\paragraph{Optimization Pipeline.}
Our evaluation pipeline first applies rotation \citep{Quarot} to optimize input activation quantization. We then compute the ACS ($\Omega_i$) for each output channel using a small calibration dataset. This metric drives a two-stage optimization process: mitigating IMC unit granularity constraints via offline channel reordering, followed by cost-aware mixed ADC resolution allocation across physical IMC units. Full settings for accuracy evaluation and energy simulation are detailed in Appendices \ref{appx:accuracy-evaluation} and \ref{appx:energy-simulation}. We evaluate our framework through a Pareto analysis of the resulting accuracy-efficiency tradeoffs (Sec. \ref{subsec:pareto-analysis}) and a comprehensive task-level evaluation across diverse architectures (Sec. \ref{subsec:results}).

\paragraph{Heuristic Baseline.}
Prior studies have identified the value, attention output, and down projections as particularly sensitive under various quantization and analog-computing settings~\citep{MixLLM, ashkboos2024quik}. Based on these findings, we construct a heuristic ADC allocation baseline that assigns high-resolution ADCs to different combinations of these three projection types, while keeping the remaining projections at low resolution. We exhaustively evaluate all seven non-empty combinations, applying each configuration across all blocks and output channels. This provides a strong projection-type-aware baseline without using fine-grained sensitivity estimation.

\subsection{Models and Datasets}
\label{subsec:models-and-datasets}

We evaluate \ours{} across a diverse suite of architectures, including LLaMA-2-7B \citep{LLAMA2-7B}, LLaMA-3.2-3B \citep{LLaMA-3.2}, Mistral-7B-v0.3 \citep{MISTRAL-7B-v0.3}, and Qwen2-1.5B-Instruct \citep{QWEN2-1.5B-Instruct} for natural language tasks, alongside ViT-Small, ViT-Base, and ViT-Large (all with a patch size of $16\times16$) for image classification. For the LLMs, we report perplexity on the WikiText-2 test set \citep{WikiText-2} and zero-shot accuracy on seven widely used reasoning and question-answering benchmarks: ARC-Challenge and ARC-Easy \citep{ARC}, BoolQ \citep{BoolQ}, HellaSwag \citep{HellaSwag}, PIQA \citep{PIQA}, OpenBookQA \citep{openbookQA}, and Winogrande \citep{winogrande}. The main text reports their unweighted average accuracy, while the complete per-task results are provided in Table \ref{tab:allocation_schemes_full_accuracy} in Appendix \ref{appx:extended-results}. For the Pareto analysis, we compute $D_{\mathrm{KL}}(p_{\mathrm{FP16}}\|p_{\mathrm{IMC}})$ between the complete next-token distributions of the FP16 and IMC models at each of 8,192 token positions from fixed WikiText-2 samples and average uniformly over positions; log probabilities are evaluated in FP32. For ViTs, we report top-1 accuracy and output-distribution KL divergence on the same 1000-image subset of the ImageNet-1K validation set \citep{ImageNet-1K}, averaging KL over images and all 1000 classes.

\subsection{Accelerator Architecture and System-Level Energy Modeling}
\label{subsec:imc-accelerator}

We evaluate the accuracy--energy trade-off of \ours{} using a system-level IMC accelerator energy model. Our goal is to assess how ADC resolution allocation affects energy consumption along the entire processing path, rather than considering only IMC macro-level energy. To this end, our model accounts for all relevant energy sources, including IMC computation, peripheral operations, memory accesses, and interconnect communication.

\begin{figure}[t]
    \centering
    \includegraphics[width=\columnwidth]{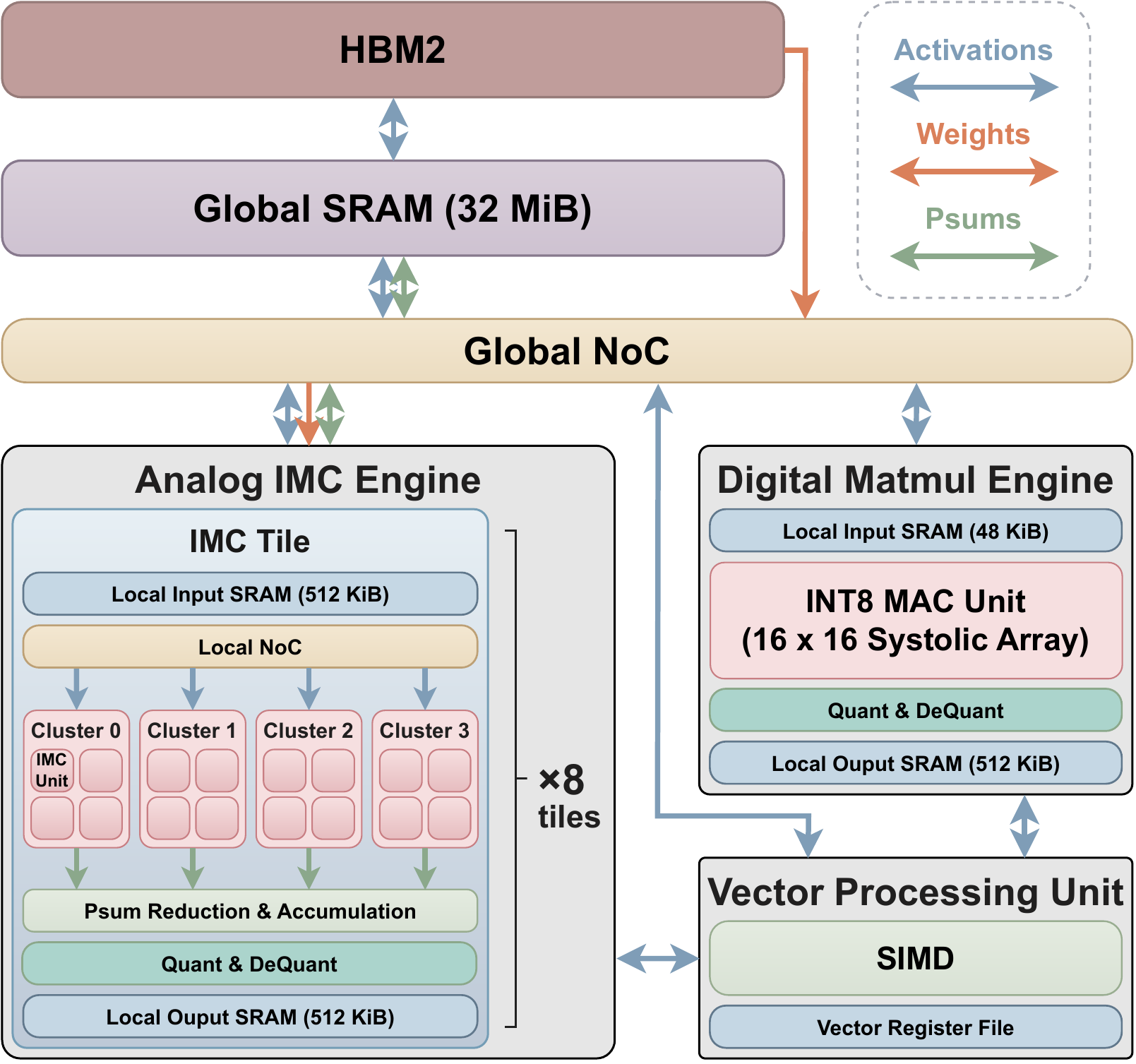}
    \caption{\textbf{IMC accelerator architecture.} Static-weight linear projections are mapped to analog IMC, while attention matrix multiplications and vector operations are executed on digital engines.}
    \label{fig:imc-accelerator-architecture}
\end{figure}

\paragraph{Accelerator Architecture.} We consider a heterogeneous accelerator comprising an analog IMC engine, a digital matrix multiplication engine, a vector processing unit (VPU), and a hierarchical memory and interconnect system, as illustrated in Figure~\ref{fig:imc-accelerator-architecture}. The key design principle is to exploit the high compute efficiency and weight reuse of IMC for static-weight linear projections, while retaining digital engines for operations less suited to IMC. Following a representative IMC mapping scheme \citep{analog-foundation-models}, we map the query, key, value, attention output, gate, up, and down projections to IMC, while executing attention BMMs and vector operations digitally. In particular, attention BMMs operate on dynamically generated operands and have relatively short reduction dimensions under blocked execution, making them less suitable for conventional weight-stationary IMC.


\paragraph{Energy Simulation.} We build upon CiMLoop \citep{cimloop}, an established framework for system-level energy modeling of IMC accelerators, through its integration into AccelForge \citep{accelforge}. To support mixed-ADC-resolution IMC, we extend the framework with custom component energy models. Detailed component models, provenance, and assumptions are provided in Appendix \ref{appx:energy-component-modeling}.
We use batch size 1 and a sequence length of 2048 during prefill for energy simulation; detailed architecture and mapping configurations are provided in Appendices \ref{appx:energy-accelerator-architecture} and \ref{appx:energy-workload-mapping-scheduling}.

\paragraph{Energy Metric.}

We report $E_{\mathrm{proj}}$, the system-level energy attributable to the seven linear projections mapped to IMC, including their computation and associated system overheads. This excludes operators mapped to digital units, such as attention BMMs and RMSNorm, as they fall outside the scope of \ours{}'s optimization. The metric therefore directly captures the accuracy--energy trade-off for these linear projections. We count one multiply-accumulate as two operations and report $2N_{\mathrm{MAC}}/E_{\mathrm{proj}}$ in TOPS/W. The Pareto plots report the normalized gain $G_\eta=\eta_{\mathrm{proj}}/\eta_{\mathrm{INT8}}$, while task-level tables retain absolute TOPS/W. For mixed 9/12-bit configurations, energy is interpolated between the fixed-resolution endpoints according to the allocated fraction of high-resolution ADC operations. Detailed energy attribution and breakdowns are provided in Appendix \ref{appx:energy-accounting-evaluation-methodology}.

%% file: sections/05_results_and_analysis.tex
\section{Results and Analysis}
\label{sec:results-and-analysis}

\begin{figure}[t]
    \centering
    \includegraphics[width=\columnwidth]{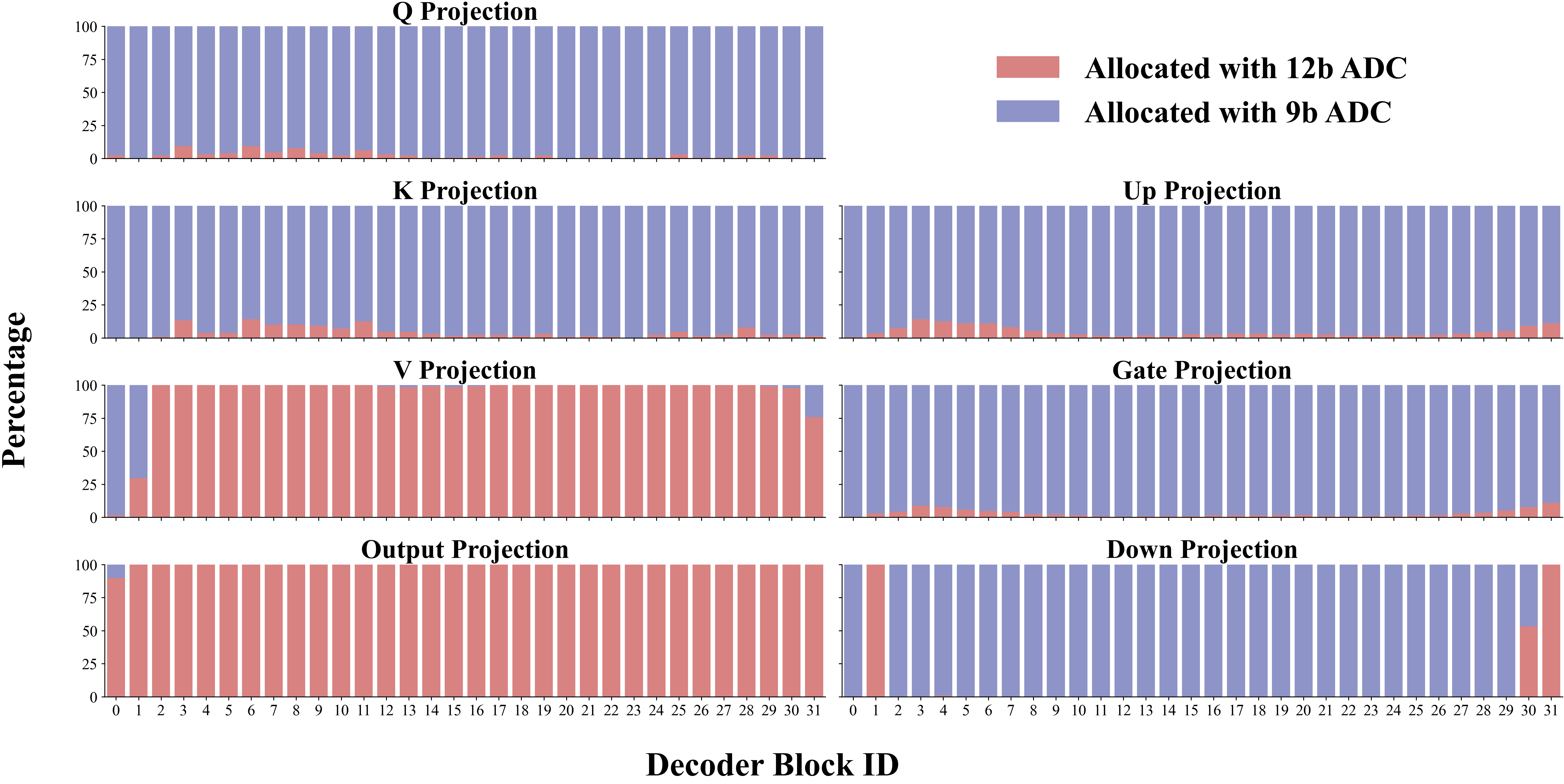}
    \caption{\textbf{\ours{} ADC-resolution allocation.} \ours{} identifies distinct sensitivities across projections and blocks to optimize fine-grained ADC assignments under a budget of 20\% 12-bit and 80\% 9-bit ADCs.}
    \label{fig:allocation-breakdown}
\end{figure}

\subsection{ADC Resolution Allocation}
\label{subsec:allocation-results}

To demonstrate the granularity of our approach, we apply the Hessian-guided framework to LLaMA-2-7B \citep{LLAMA2-7B} under a budget constraint of 20\% 12-bit and 80\% 9-bit ADCs. Figure \ref{fig:allocation-breakdown} illustrates that \ours{} successfully identifies distinct sensitivities across projection types, assigning most of the high-resolution budget to value, attention output, and down projections. The query, key, up, and gate projections receive sparse, block-dependent 12-bit assignments, demonstrating that \ours{} also captures depth-wise sensitivities in a fine-grained manner.


\subsection{Fidelity--Energy Efficiency Pareto Frontier}
\label{subsec:pareto-analysis}

\begin{figure*}[t]
    \centering
    \begin{subfigure}[t]{0.49\textwidth}
        \centering
        \includegraphics[width=\linewidth]{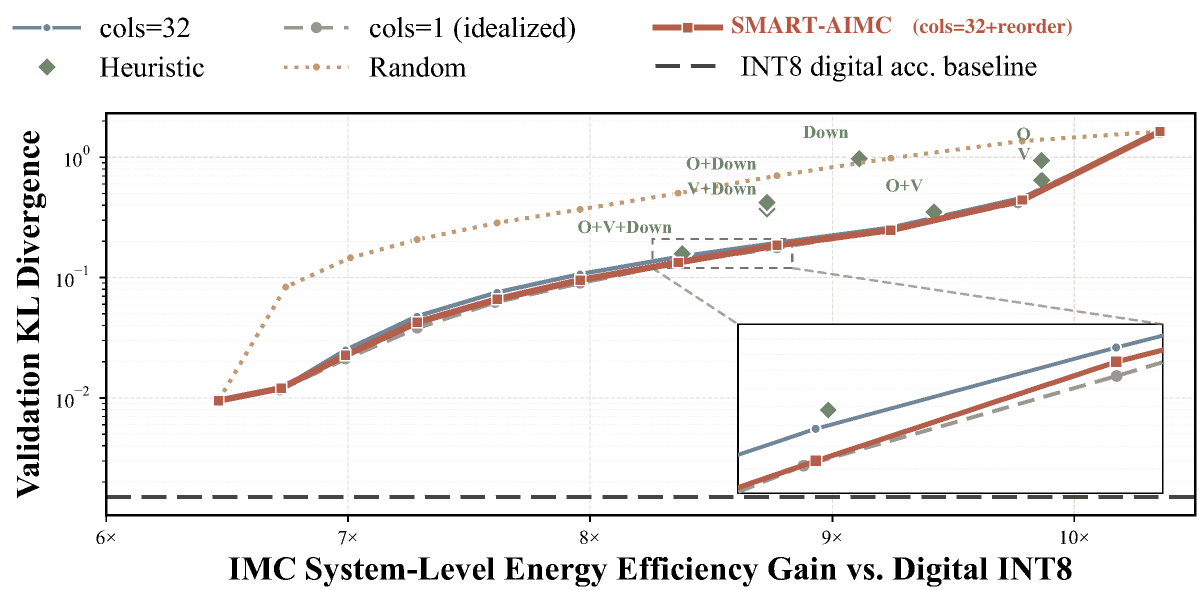}
        \caption{Validation KL divergence.}
        \label{fig:pareto-curve-kl}
    \end{subfigure}\hfill
    \begin{subfigure}[t]{0.49\textwidth}
        \centering
        \includegraphics[width=\linewidth]{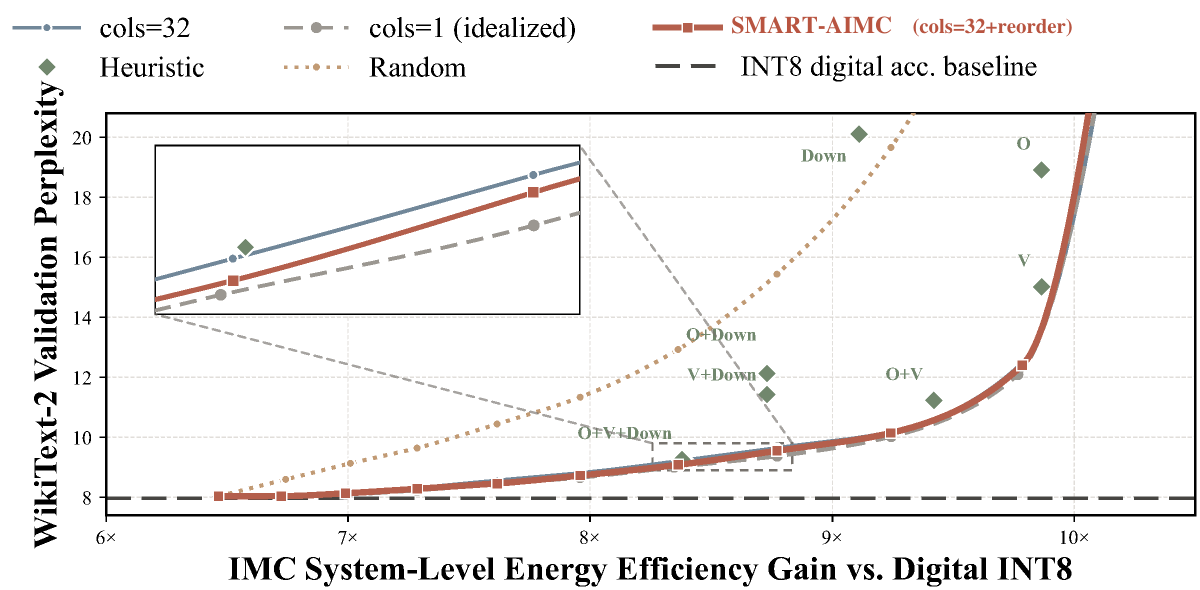}
        \caption{WikiText-2 validation perplexity.}
        \label{fig:pareto-curve-ppl}
    \end{subfigure}
    \caption{\textbf{\ours{} fidelity--energy efficiency Pareto analysis.} KL divergence and perplexity versus system-level energy-efficiency gain for linear projections over digital INT8 for LLaMA-2-7B on WikiText-2. Dashed lines show digital INT8 fidelity; lower fidelity metrics and higher gain are better. \ours{} establishes a superior frontier over the heuristic and random baselines.}
    \label{fig:pareto-curve}
\end{figure*}

To evaluate the fidelity--energy trade-off, Figure \ref{fig:pareto-curve} plots token-level KL divergence and WikiText-2 validation perplexity against $G_\eta=\eta_{\mathrm{proj}}/\eta_{\mathrm{INT8}}$ for LLaMA-2-7B. We compare five configurations: (1) \texttt{cols=32}, our standard sensitivity-guided allocation for an IMC unit with $w = 32$ (i.e., \ours{} without reordering); (2) \texttt{cols=32+reorder}, enhanced with sensitivity-aware channel reordering (i.e., \ours{}); (3) \texttt{cols=1}, an idealized, hardware-impractical per-channel allocation for an IMC unit with $w = 1$, serving as the algorithmic upper bound; (4) the \texttt{Heuristic} defined in Section~\ref{subsec:mixed-adc-imc-framework}; and (5) \texttt{Random} allocation. Digital INT8 quality is the horizontal dashed reference, and $\eta_{\mathrm{INT8}}$ is the denominator of $G_\eta$. The sensitivity-based and Random curves sweep high-ADC budgets from 0\% to 100\% in 10-percentage-point increments. Each point computes $\eta_{\mathrm{proj}}$ from its physical ADC allocation, including input tiling and output grouping; the seven Heuristic markers cover all nonempty subsets of attention output, value, and down projections. Appendix \ref{appx:figure_experiment_details} provides the complete protocol.

The results first demonstrate the superiority of \ours{}: even under the IMC unit granularity constraints, it achieves lower KL divergence and perplexity than the Heuristic and Random baselines at comparable energy-efficiency gain. Furthermore, unlike the heuristic's discrete projection-type configurations, \ours{} supports fine-grained allocation across arbitrary energy budgets, providing a smooth fidelity--energy trade-off. This flexibility is further demonstrated in Section~\ref{subsec:near-digital-accuracy}. However, comparing this hardware-constrained assignment against the idealized per-channel allocation upper bound reveals a fidelity gap between the curves, confirming that forcing a shared ADC resolution across all output channels within each IMC unit degrades the accuracy. Crucially, incorporating our offline sensitivity-aware channel reordering almost entirely bridges the gap for both fidelity metrics. By grouping channels with similar sensitivities into shared IMC units, we reclaim the Pareto-optimal fidelity--energy trade-off that would otherwise be lost to coarse-grained hardware constraints.

\input{sections/05_results_and_analysis_tables}



\subsection{Task-level Evaluation}
\label{subsec:results}

\paragraph{Results}
Table \ref{tab:allocation_schemes} compares \ours{} against three baselines across four LLMs: (1) an ideal digital INT8 baseline with no ADC quantization, (2) random ADC resolution allocation, and (3) coarse-grained, architecture-based heuristic allocation. We report KL divergence, WikiText-2 perplexity, seven-task average accuracy, and TOPS/W; complete per-task accuracy results are provided in Table~\ref{tab:allocation_schemes_full_accuracy} in Appendix~\ref{appx:extended-results}. To ensure a fair comparison, all mixed-ADC methods use the matched O+V budget, which provides exactly enough energy for the heuristic to assign high-resolution ADCs to all attention output and value projections; they therefore share the same model-specific system-level energy efficiency. The INT8 baseline replaces the analog IMC compute macros with digital MAC and accumulator units while preserving the same attributable system-level overheads, including modeled precision conversion, memory, and communication traffic, as detailed in Appendix \ref{appx:energy-accounting-evaluation-methodology}. Table \ref{tab:vit_allocation} reports ImageNet accuracy and output-distribution KL divergence for three ViT model sizes under the O+V and O+Down budgets.

\ours{} achieves the lowest KL divergence and WikiText-2 perplexity among the mixed-ADC allocation schemes across all four LLMs. It also achieves the highest seven-task average accuracy for LLaMA-2-7B, LLaMA-3.2-3B, and Mistral-7B-v0.3; for Qwen2-1.5B-Instruct, it is 0.20 percentage points below the heuristic while retaining lower KL divergence and perplexity. For ViT-Large, the O+V budget---constructed by assigning high-resolution ADCs to all attention output and value projections in the heuristic configuration---leads to low accuracy for all three mixed-ADC methods. We therefore also evaluate the larger O+Down budget. Across these two budgets, \ours{} achieves the best mixed-ADC KL divergence and ImageNet accuracy for ViT-Small and ViT-Base. For ViT-Large, \ours{} achieves the best accuracy under both budgets; the heuristic has marginally lower KL divergence under O+Down.

\subsection{Calibration Efficiency and Overhead}
\label{subsec:calibration-efficiency}

The offline calibration of \ours{} is fast and does not incur any inference-time overhead. For LLaMA-2-7B, the complete calibration process, including empirical Fisher estimation, ACS computation, and channel reordering, takes approximately 75 seconds, with a peak GPU memory usage of 31.4 GiB on a single NVIDIA A100-80GB GPU. Furthermore, the resulting model quality remains stable across calibration dataset sizes, sequence lengths, and data domains, as detailed in Appendix~\ref{appx:calibration-robustness}.

\subsection{Flexible Accuracy--Energy Trade-offs}
\label{subsec:near-digital-accuracy}

While the preceding results demonstrate the benefits of \ours{} under constrained ADC-energy budgets, the fine-grained tunability of \ours{} also enables accuracy-critical operating points, which the heuristic method does not support. As shown in Table~\ref{tab:near-digital-accuracy}, allocating high-resolution ADCs to 80\% of IMC computation brings LLaMA-2-7B close to the digital INT8 baseline (5.57 vs. 5.47 perplexity; 65.84\% vs. 66.64\% average accuracy), while retaining $6.99\times$ higher modeled system-level energy efficiency for linear projections. This demonstrates the flexibility of \ours{} to support both energy-constrained and accuracy-critical inference scenarios using a unified mixed ADC resolution IMC framework.

\begin{table}[t]
\centering
\small
\setlength{\tabcolsep}{4pt}
\caption{LLaMA-2-7B quality and system-level energy efficiency for linear projections at the higher-ADC-budget operating point.}
\label{tab:near-digital-accuracy}
\resizebox{\columnwidth}{!}{%
\begin{tabular}{lcccc}
\toprule
Method & KL ($\downarrow$) & PPL ($\downarrow$) & Avg. Acc. ($\uparrow$) & TOPS/W ($\uparrow$) \\
\midrule
INT8 Baseline & 0.0015 & 5.47 & 66.64 & 2.42 \\
\ours{} & 0.0224 & 5.57 & 65.84 & 16.93 \\
\bottomrule
\end{tabular}
}
\end{table}

%% file: sections/05_results_and_analysis_tables.tex
\begin{table*}[t]
\centering
\footnotesize
\setlength{\tabcolsep}{1pt}
\caption{LLM quality and system-level energy efficiency for linear projections across four models under the matched O+V budget. Bold values denote the best quality among mixed-ADC allocation schemes.}
\label{tab:allocation_schemes}
\begingroup
\renewcommand{\todo}[1]{\textcolor{cyan}{TODO}}
\renewcommand{\arraystretch}{1.08}
\resizebox{\textwidth}{!}{%
\begin{tabular}{l*{4}{cccc}}
\toprule
\multirow{2}{*}{Method}
& \multicolumn{4}{c}{LLaMA-2-7B}
& \multicolumn{4}{c}{LLaMA-3.2-3B}
& \multicolumn{4}{c}{Mistral-7B-v0.3}
& \multicolumn{4}{c}{Qwen2-1.5B-Instruct} \\
\cmidrule(lr){2-5}\cmidrule(lr){6-9}\cmidrule(lr){10-13}\cmidrule(lr){14-17}
& KL ($\downarrow$) & PPL ($\downarrow$) & Avg. Acc. ($\uparrow$) & TOPS/W ($\uparrow$)
& KL ($\downarrow$) & PPL ($\downarrow$) & Avg. Acc. ($\uparrow$) & TOPS/W ($\uparrow$)
& KL ($\downarrow$) & PPL ($\downarrow$) & Avg. Acc. ($\uparrow$) & TOPS/W ($\uparrow$)
& KL ($\downarrow$) & PPL ($\downarrow$) & Avg. Acc. ($\uparrow$) & TOPS/W ($\uparrow$) \\
\midrule
INT8 Baseline
& 0.0015 & 5.47 & 66.64 & 2.42
& 0.0024 & 7.82 & 64.99 & 2.41
& 0.0008 & 5.32 & 71.58 & 2.42
& 0.0123 & 10.12 & 60.76 & 2.40 \\
\hdashline
Random Alloc.
& 1.0535 & 11.94 & 47.93 & 22.82
& 2.4698 & 73.01 & 40.63 & 22.58
& 0.5423 & 8.63 & 57.45 & 23.32
& 1.3563 & 32.22 & 42.80 & 21.91 \\
Heuristic Alloc.
& 0.3510 & 7.33 & 59.24 & 22.82
& 0.7756 & 16.78 & 50.81 & 22.58
& 0.2946 & 6.91 & 64.22 & 23.32
& 0.9115 & 21.57 & \textbf{47.92} & 21.91 \\
\ours{}
& \textbf{0.2917} & \textbf{7.01} & \textbf{59.91} & 22.82
& \textbf{0.6762} & \textbf{14.92} & \textbf{52.37} & 22.58
& \textbf{0.2642} & \textbf{6.76} & \textbf{64.85} & 23.32
& \textbf{0.7791} & \textbf{19.39} & 47.72 & 21.91 \\
\bottomrule
\end{tabular}
}
\endgroup
\end{table*}

\begin{table}[t]
\centering
\footnotesize
\setlength{\tabcolsep}{3pt}
\caption{ImageNet quality under the O+V and O+Down mixed-ADC budgets. Bold values denote the best mixed-ADC result for each model and budget.}
\label{tab:vit_allocation}
\resizebox{\columnwidth}{!}{%
\begin{tabular}{l*{3}{cc}}
\toprule
\multirow{2}{*}{Method}
& \multicolumn{2}{c}{ViT-Small}
& \multicolumn{2}{c}{ViT-Base}
& \multicolumn{2}{c}{ViT-Large} \\
\cmidrule(lr){2-3}\cmidrule(lr){4-5}\cmidrule(lr){6-7}
& KL ($\downarrow$) & Acc. ($\uparrow$)
& KL ($\downarrow$) & Acc. ($\uparrow$)
& KL ($\downarrow$) & Acc. ($\uparrow$) \\
\midrule
INT8 Baseline
& 0.0011 & 79.80
& 0.0004 & 81.80
& 0.0006 & 83.40 \\
\hdashline
\multicolumn{7}{l}{\textit{O+V budget}} \\
Random Alloc.
& 2.1804 & 40.34
& 0.2754 & 78.18
& 6.5593 & 0.10 \\
Heuristic Alloc.
& 2.2133 & 41.60
& 0.2169 & 79.90
& 6.5754 & 0.10 \\
\ours{}
& \textbf{0.5863} & \textbf{68.30}
& \textbf{0.0964} & \textbf{80.90}
& \textbf{6.1645} & \textbf{2.30} \\
\midrule
\multicolumn{7}{l}{\textit{O+Down budget}} \\
Random Alloc.
& 1.4196 & 52.68
& 0.2453 & 78.26
& 6.5573 & 0.10 \\
Heuristic Alloc.
& 0.3984 & 73.40
& 0.0531 & 80.40
& \textbf{0.0347} & 82.20 \\
\ours{}
& \textbf{0.2384} & \textbf{74.40}
& \textbf{0.0373} & \textbf{81.70}
& 0.0349 & \textbf{83.10} \\
\bottomrule
\end{tabular}
}
\end{table}

%% file: sections/06_conclusions.tex
\section{Conclusions}
\label{sec:conclusions}

In this paper, we proposed \ours{} to optimize the energy-accuracy trade-off for analog IMC inference. By mathematically bridging algorithmic loss curvature with physical ADC quantization noise, we derived an ACS metric to systematically guide ADC resolution allocation. To overcome the hardware constraint of uniform ADC resolution within an IMC unit, a sensitivity-aware channel reordering strategy is applied to group channels of similar importance into the same physical macros without incurring runtime overhead. Comprehensive evaluations across diverse Transformer architectures demonstrate that \ours{} establishes a superior accuracy--energy Pareto frontier over heuristic and random allocation strategies. This enables IMC accelerators to (1) improve inference accuracy under tight ADC-energy budgets and (2) approach digital-level accuracy when fidelity is prioritized, while preserving substantial energy-efficiency advantages over digital computing.

%% file: sections/appendix.tex
\input{appendices/app00_related_works.tex}
\input{appendices/app01_analysis_on_adc_quantization_noise.tex}
\input{appendices/app02_imc_details_for_processing_mvms.tex}
\input{appendices/app03_accuracy_evaluation.tex}
\input{appendices/app04_energy_simulation.tex}
\input{appendices/app05_permutation_construction_for_channel_reordering.tex}
\input{appendices/app06_extended_experiments_and_results.tex}
\input{appendices/app07_experiment_setup_for_the_figures.tex}
\input{appendices/app08_discussions.tex}
\input{appendices/app09_documentation_of_assets_and_licenses.tex}

%% file: appendices/app00_related_works.tex
\section{Related Works}


\subsection{Mixed Precision Digital Computing}
\label{appx:mixed-precision-computing}
Mixed-precision computing has been extensively explored in the digital domain for transformer inference, encompassing mixed-precision assignments for weights \citep{LLM-INT8, SpQR, slim-llm}, input activations \citep{ResQ}, or both \citep{FGMP}. Conventional allocation strategies frequently rely on local information, such as layer-wise reconstruction error \citep{SpQR} or activation magnitude thresholding \citep{ATOM}. These local metrics, however, are unable to capture the model's global sensitivity to quantization noise, thus failing to account for the precision requirements of different architectural constructs (e.g., query vs. down projections) and different layer depths (e.g., intermediate vs. final blocks). Alternative strategies leverage global information, typically utilizing Hessian-based metrics with respect to the final task loss, to achieve a more holistic precision assignment \citep{MixLLM, Squeezellm}. \ours{} further extends the Hessian-based mixed-precision scheme beyond digital inputs/weights, establishing a systematic framework for mixed-resolution outputs via ADC resolution allocation.

\subsection{Saliency-aware Computing for IMC}
Saliency-aware computing has been actively explored within the domain of IMC to optimize the energy-accuracy trade-off. While \ours{} focuses on establishing a mixed ADC resolution scheme for analog charge-domain IMC, prior literature has investigated several alternative directions, such as offloading salient computations to digital components while mapping error-tolerant computations to analog IMC \citep{DIANA,chen2024osa,dash2021robust}, utilizing approximate computing within digital IMC circuits \citep{DIMC,pacim}, and integrating multiple analog IMC technologies that offer varying energy-accuracy profiles \citep{jiang2020device}.

While these methodologies introduce hardware heterogeneity to expand the design space, this added degree of freedom necessitates complex hardware-software co-design. Managing these heterogeneous hardware systems not only complicates accuracy simulation but also introduces complex system-level challenges, including intricate workload scheduling, dynamic load balancing, and more expensive hardware-aware neural architecture search. In contrast, \ours{}---a resolution-configurable mixed ADC framework---operates within a unified analog architecture. This method minimizes these hardware-level overheads and elevates the energy-accuracy trade-off exploration to the software level, providing a highly efficient and scalable solution for saliency-aware IMC execution.

\subsection{Feasibility of Clipping in ADC Quantization}
To mitigate the ADC quantization bottleneck in IMC architectures, circuits that support clipping the output dynamic range—effectively amplifying the distribution—have been proposed \citep{lee2024switched}. Leveraging this hardware capability, recent literature has investigated methods to mathematically optimize these clipping thresholds \citep{OCC,kavishwar2026minimizing}.

However, implementing clipping in the analog domain inherently exacerbates analog noise, especially in high-resolution ADCs where analog noise is significantly more profound relative to the ADC step size \citep{Murmann-mixed-signal,ASIM}. Unlike predictable ADC quantization noise, the distortion from analog clipping is highly unpredictable due to PVT variation. This creates a severe hardware-software mismatch that complicates accurate software-level simulations of LLM inference. Therefore, to ensure simulation fidelity, our mixed ADC framework operates under the constraint that analog clipping is not permitted.

\subsection{Quantization-Aware Training for IMC}
While \ours{} operates under the umbrella of Post-Training Quantization (PTQ), alternative works leverage Quantization-Aware Training (QAT) to optimize models for IMC hardware. These approaches modify the training loop itself, such as injecting simulated random noise to enhance robustness against hardware non-idealities \citep{analog-foundation-models}, or introducing regularization terms to expand partial sum variance prior to ADC quantization \citep{RAOQ}. However, QAT imposes prohibitive computational overhead and memory footprints when scaled to modern large models. By relying exclusively on PTQ, \ours{} avoids these massive training costs, ensuring rapid deployment and scalability for large-scale architectures.

%% file: appendices/app01_analysis_on_adc_quantization_noise.tex
\section{Analysis of ADC Quantization Noise}
\label{appx:noise-assumptions}
The mathematical derivation of our ACS metric relies fundamentally on statistical quantization theory \citep{widrow-quantization}. Specifically, we assume that when the variance of the underlying analog partial sum is sufficiently large compared to the quantization step size ($\Delta$), the resulting ADC quantization error can be modeled as a zero-mean uniform distribution, $\mathcal{U}(-\Delta/2, \Delta/2)$, that is statistically independent of the input signal and other discrete noise sources. This premise enables two critical mathematical simplifications in our framework: output channel independence (as established in Sec. \ref{subsec:hessian-sensitivity-adc-quantization}) and input IMC unit independence (Sec. \ref{subsec:modeling-adc-noise-propagation}). In this section, we empirically validate the physical feasibility and robustness of these theoretical assumptions by analyzing empirical noise distributions and spatial correlation matrices obtained while processing a subset of the WikiText-2 training data with the LLaMA-2-7B model. While the visualizations below use specific configurations (blocks, layers, etc.) for clarity, we observe consistent trends throughout the model.

\subsection{ADC Quantization Noise Distribution}
\label{appx:adc-noise-distribution}
\begin{figure*}[t]
  \centering
  \begin{subfigure}[t]{0.48\textwidth}
    \centering
    \includegraphics[width=\linewidth]{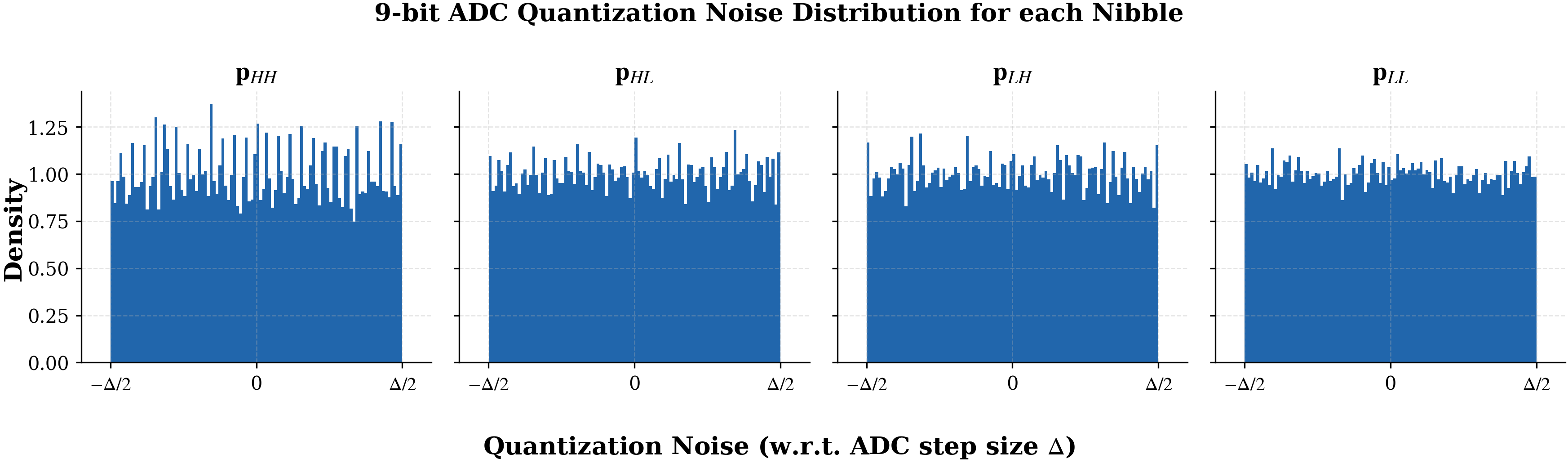}
    \caption{9-bit ADC quantization noise for the four $4\text{-bit} \times 4\text{-bit}$ nibble-product combinations (block 15, down projection)}
    \label{fig:9b_adc_noise_dist}
  \end{subfigure}\hfill
  \begin{subfigure}[t]{0.48\textwidth}
    \centering
    \includegraphics[width=\linewidth]{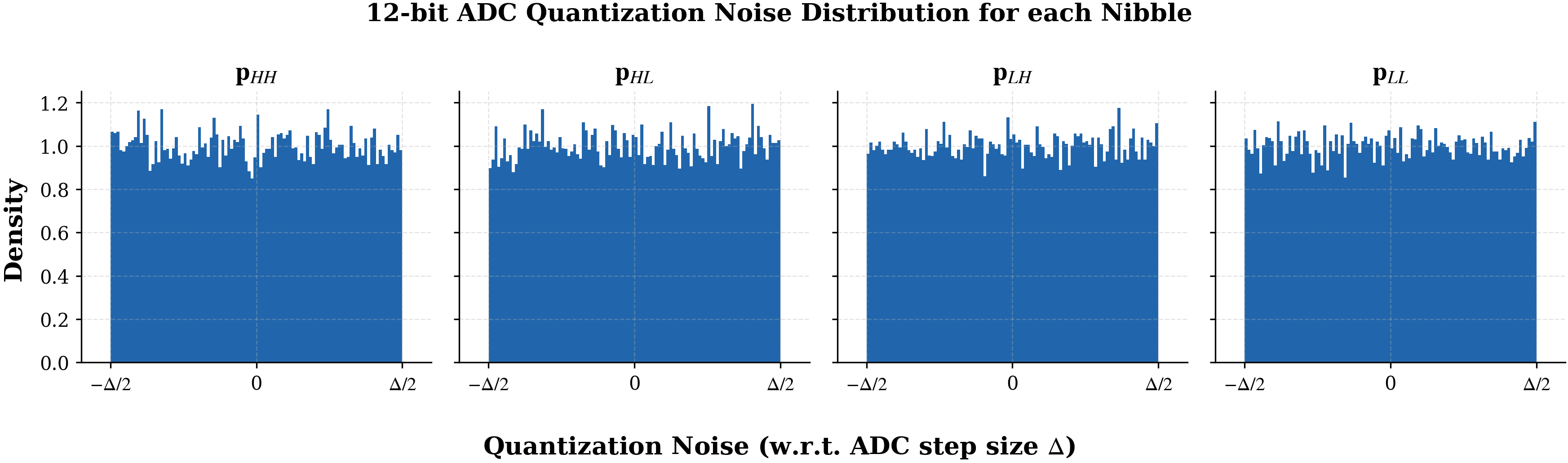}
    \caption{12-bit ADC quantization noise for the four $4\text{-bit} \times 4\text{-bit}$ nibble-product combinations (block 15, down projection)}
    \label{fig:12b_adc_noise_dist}
  \end{subfigure}
  \caption{\textbf{ADC quantization noise distributions.} Both 9-bit and 12-bit ADC configurations exhibit near-uniform quantization noise, aligning with the statistical theory of quantization \cite{widrow-quantization}.}
  \label{fig:adc_noise_dist}
\end{figure*}

Figure \ref{fig:adc_noise_dist} visualizes the ADC quantization noise distributions of the four $4\text{-bit} \times 4\text{-bit}$ nibble-product combinations (as described in Appendix \ref{appx:digital-aggregation}) for a specific IMC unit ($k=0, l=0$) processing the down projection of decoder block 15. The data are collected using a 64-sample calibration set with sequence length 512. For both the 9-bit and 12-bit ADC configurations, the empirical distributions across all nibble-product combinations closely track an ideal uniform density function. This confirms that the analog spatial accumulations are sufficiently wide to pseudo-randomize the quantization error.

\subsection{Output Channel Independence}
\label{appx:out-channel-independence}
\begin{figure}[t]
  \centering
  \begin{subfigure}[t]{0.48\linewidth}
    \centering
    \includegraphics[width=\linewidth]{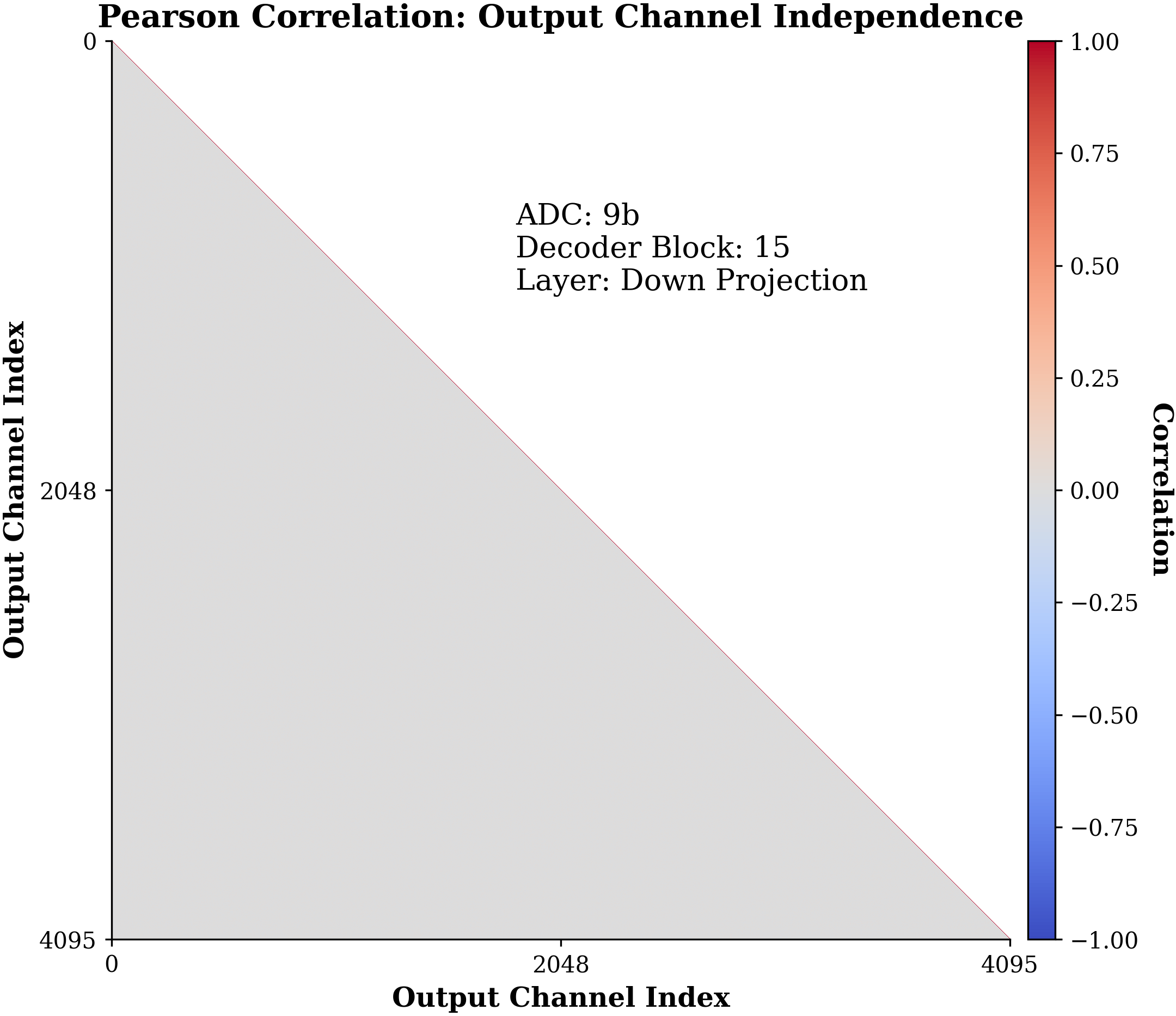}
    \caption{9-bit ADC Pearson correlation matrix for output channels (block 15, down projection)}
    \label{fig:appx-output-correlation-9b}
  \end{subfigure}\hfill
  \begin{subfigure}[t]{0.48\linewidth}
    \centering
    \includegraphics[width=\linewidth]{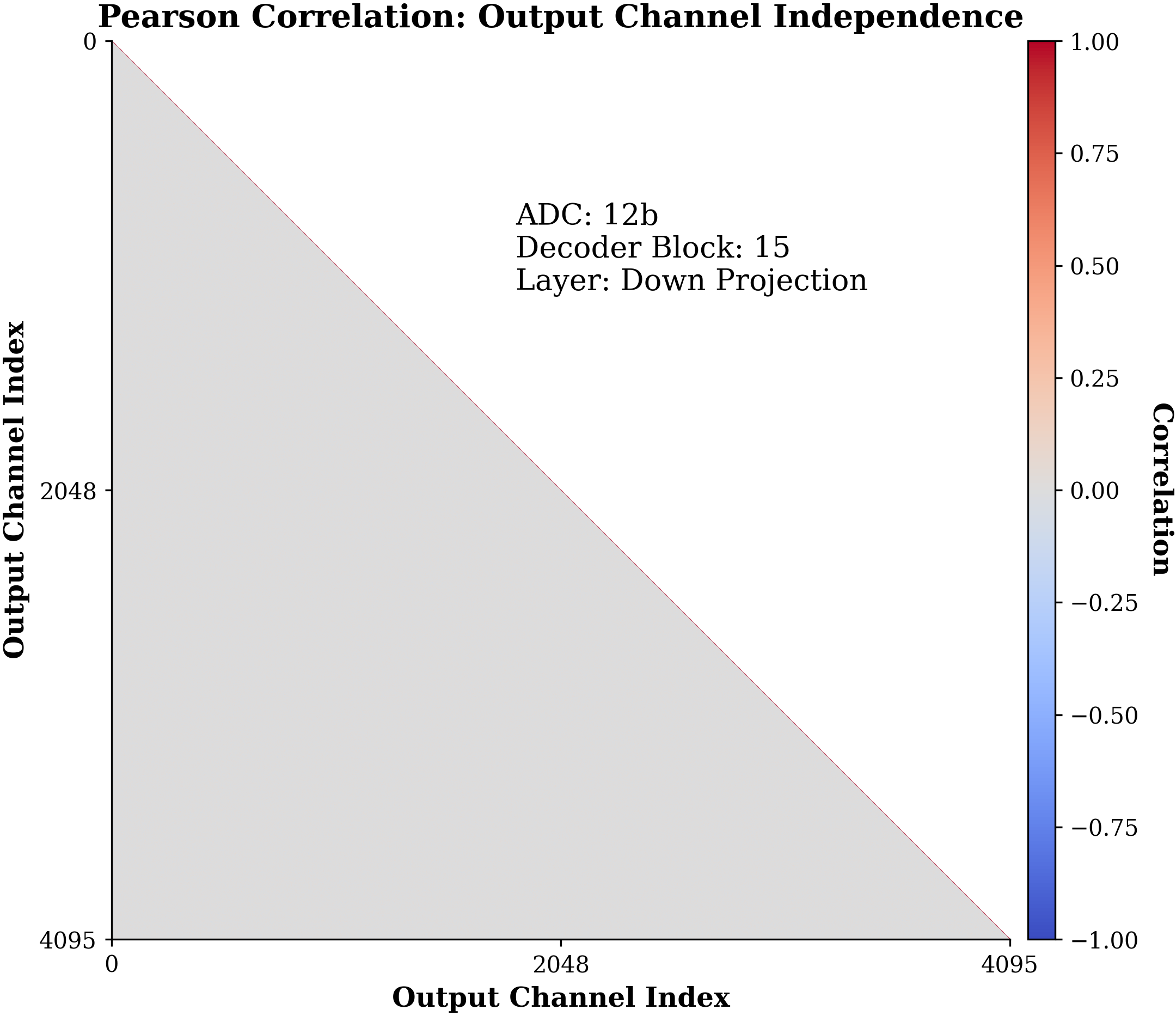}
    \caption{12-bit ADC Pearson correlation matrix for output channels (block 15, down projection)}
    \label{fig:appx-output-correlation-12b}
  \end{subfigure}
  \caption{\textbf{Output channel independence.} Both 9-bit and 12-bit ADC configurations exhibit near-zero correlation of ADC quantization noise between output channels.}
  \label{fig:appx_output_correlation_matrices}
\end{figure}

Our derivation assumes that the quantization noise injected into one output channel does not correlate with the noise injected into adjacent channels, allowing us to ignore the off-diagonal cross-channel terms in the Hessian matrix. To validate this, Figure \ref{fig:appx_output_correlation_matrices} presents the Pearson correlation matrix of the ADC quantization noise across different output channels for the block 15 down projection processing an input sequence of length 512. For both 9-bit and 12-bit ADCs, the off-diagonal terms are negligible, supporting the assumption that output channels are independent.

\subsection{Input IMC Unit Independence}
\label{appx:input-imcu-independence}
\begin{figure}[t]
  \centering
  \begin{subfigure}[t]{0.48\linewidth}
    \centering
    \includegraphics[width=\linewidth]{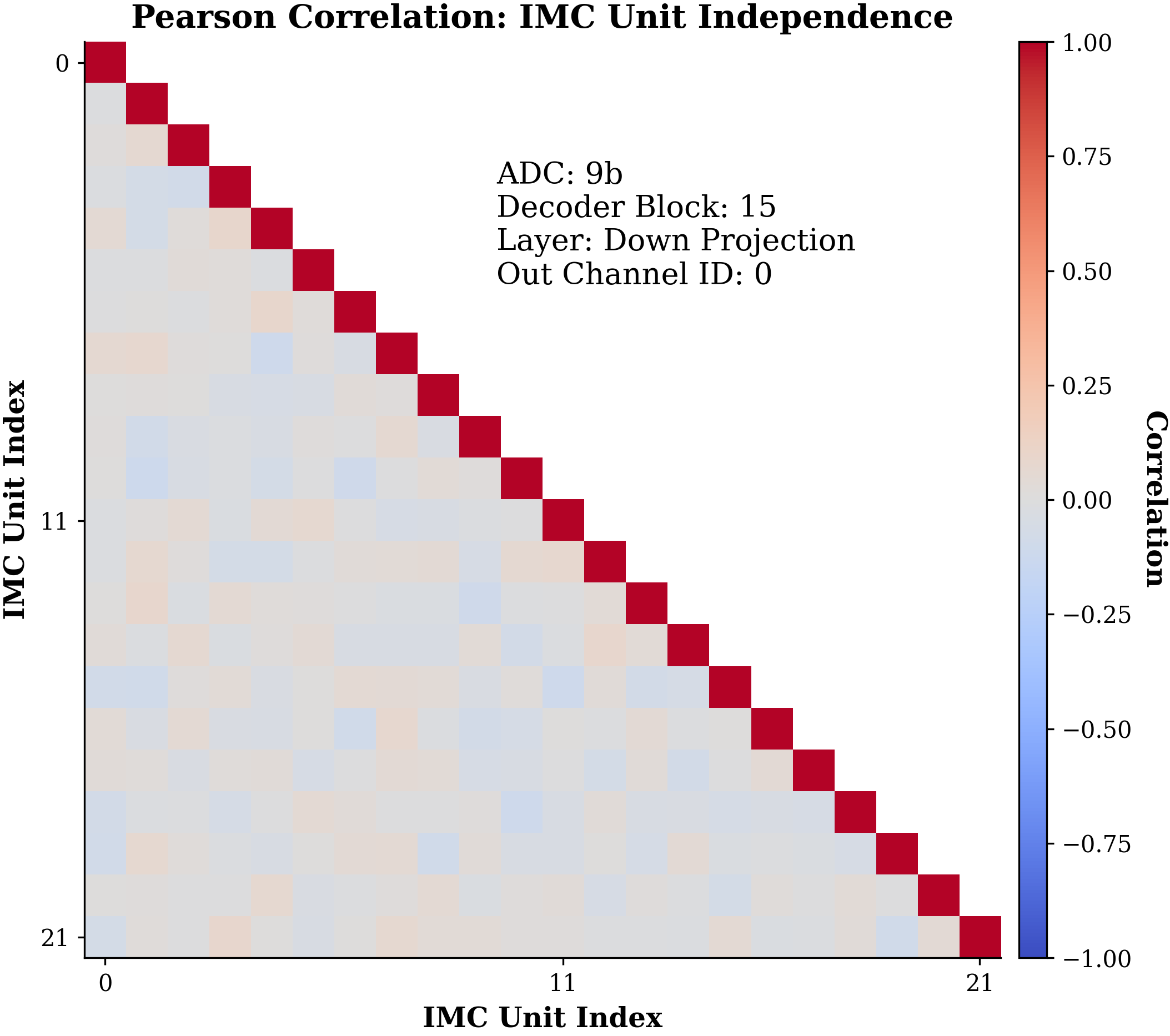}
    \caption{9-bit ADC Pearson correlation matrix for input IMC units (block 15, down projection)}
    \label{fig:appx-input-correlation-9b}
  \end{subfigure}\hfill
  \begin{subfigure}[t]{0.48\linewidth}
    \centering
    \includegraphics[width=\linewidth]{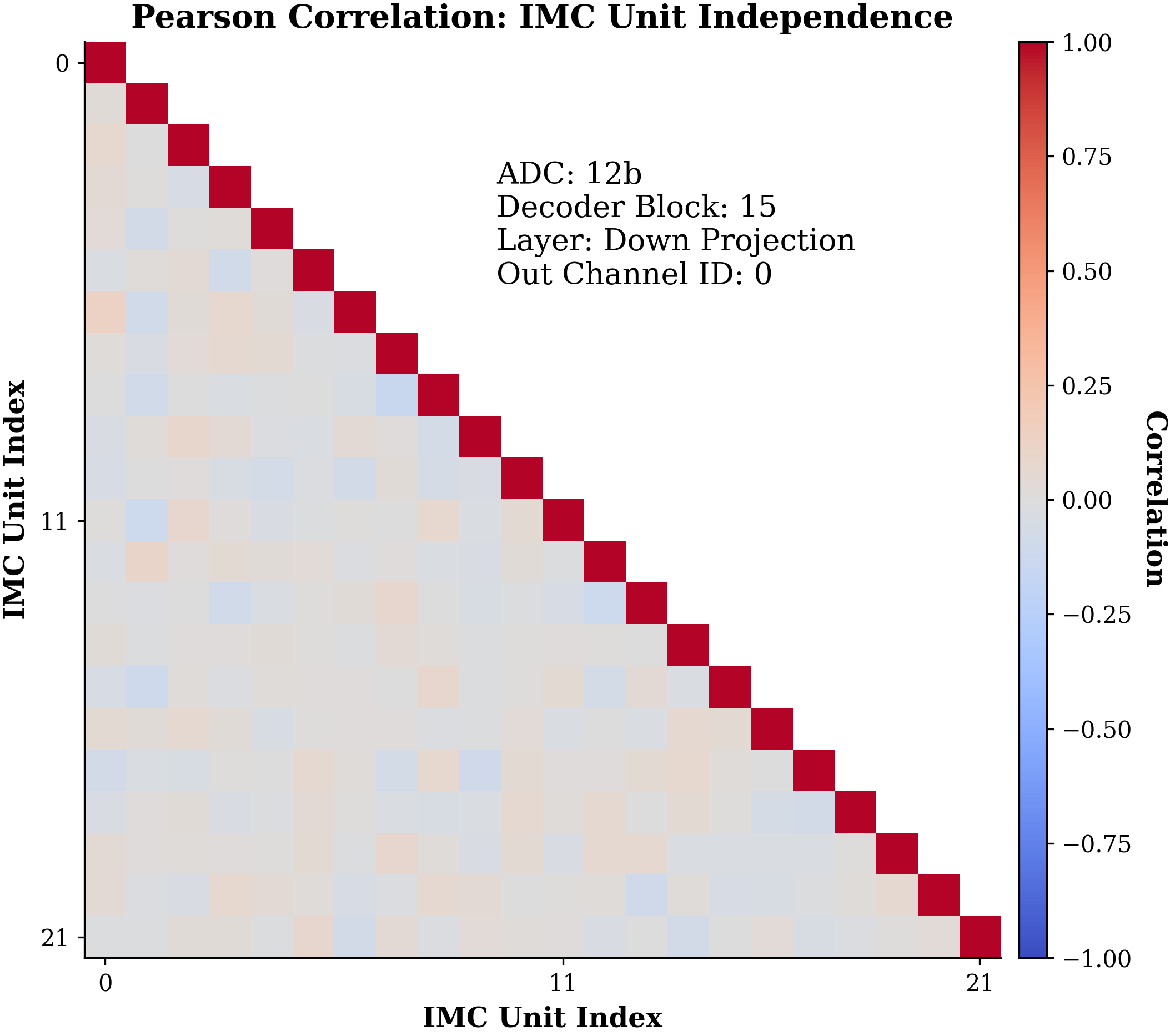}
    \caption{12-bit ADC Pearson correlation matrix for input IMC units (block 15, down projection)}
    \label{fig:appx-input-correlation-12b}
  \end{subfigure}
  \caption{\textbf{Input IMC unit independence.} Both 9-bit and 12-bit ADC configurations exhibit near-zero correlation of ADC quantization noise between input IMC units.}
  \label{fig:appx_input_correlation_matrices}
\end{figure}

Computing the final digital output requires accumulating digitized partial sums across multiple IMC units along the input extension dimension. Our noise aggregation formula assumes these individual ADC errors are uncorrelated. Figure \ref{fig:appx_input_correlation_matrices} displays the Pearson correlation matrix across the $K$ input IMC units contributing to output channel 0 across the calibration set. For both 9-bit and 12-bit ADCs, the off-diagonal terms are all close to zero, supporting the assumption of input IMC unit independence.

%% file: appendices/app02_imc_details_for_processing_mvms.tex
\section{IMC Details for Processing MVMs}
\label{appx:imc-encoding-mapping}

\subsection{Heterogeneous Architecture Mapping}
\label{appx:heterogeneous-architecture-mapping}
Our simulation framework models a heterogeneous processor comprising analog IMC engines for executing high-dimensional GEMMs in low precision, alongside conventional digital units that process non-GEMM operations (e.g., nonlinear activations and RMSNorm) in FP16. Furthermore, we retain the two batch matrix multiplications (BMMs) within scaled dot-product attention in FP16. This mapping scheme is widely adopted in both the IMC \citep{analog-foundation-models} and digital quantization literature \citep{Spinquant, Prefixquant}.

\subsection{8-bit Quantization and 4-bit Slicing Scheme}
\label{appx:8bit-quantization-4bit-slicing}
For IMC-bound GEMM operations, we target 8-bit precision for both input activations and weights. We utilize bit-slicing decomposition, similar to \citet{cambricon-cim}, to achieve 8-bit compute by executing operations at a 4-bit $\times$ 4-bit granularity and subsequently shifting and adding the cross-term partial sums in the digital domain.

To map our workloads onto this scheme, let $x, w \in \mathbb{R}$ denote the floating-point input activation and weight, respectively. We first quantize these values to 8-bit integers using a quantization operator $Q_8(\cdot)$, which maps real values to the integer set $\mathcal{I}_8 = \{-128, \dots, 127\}$. The $Q_8(\cdot)$ operator can implement either standard symmetric quantization or asymmetric quantization with a constant shift. Taking $x_{\text{int}}$ as an example, we then use bit slicing to decompose the 8-bit integer into a 4-bit most significant nibble (MSN, denoted $x_H$) and least significant nibble (LSN, denoted $x_L$). For this signed 8-bit integer $x_{\text{int}}$, our decomposition is mathematically formulated to keep both nibbles within the same range:$$x_{\text{int}} = x_H \cdot 2^4 + (x_L+8)$$where $x_H, x_L \in \{-8, \dots, 7\}$. The analog macro independently processes these slices to generate four distinct partial sums, which are then shifted and accumulated digitally to reconstruct the final equivalent 8-bit output.

\subsection{$-1/1$ Representation and Hardware Synergies}
\label{appx:pm1-representation-synergies}
Instead of utilizing the conventional 2's complement format (often referred to as a $0/1$ representation), we adopt the $-1/1$ representation \citep{programmableIMC,RAOQ}. In this format, individual binary-weighted bits take values of $-1$ or $1$. The 2-bit encoding examples are shown in Tables \ref{tab:encoding_01} and \ref{tab:encoding_pm1}. As illustrated, the $0/1$ representation yields an asymmetric dynamic range of $[-2, 1]$ with a step size of 1. In contrast, the $-1/1$ representation expands this to a symmetric dynamic range of $[-3, 3]$ with a step size of 2.

\begin{table}[h]
\centering
\begin{subtable}[t]{0.48\linewidth}
\centering
\caption{$0/1$ representation}
\label{tab:encoding_01}
\begin{tabular}{@{}cc@{}}
\toprule
Value & Bits ($b_1, b_0$) \\
\midrule
$-2$ & 1, 0 \\
$-1$ & 1, 1 \\
0 & 0, 0 \\
1 & 0, 1 \\
\bottomrule
\end{tabular}
\end{subtable}\hfill
\begin{subtable}[t]{0.48\linewidth}
\centering
\caption{$-1/1$ representation}
\label{tab:encoding_pm1}
\begin{tabular}{@{}cc@{}}
\toprule
Value & Bits ($b_1, b_0$) \\
\midrule
$-3$ & $-1$, $-1$ \\
$-1$ & $-1$, $+1$ \\
1 & $+1$, $-1$ \\
3 & $+1$, $+1$ \\
\bottomrule
\end{tabular}
\end{subtable}
\vspace{0.6em}
\caption{2-bit encoding examples for $0/1$ and $-1/1$ representations.}
\end{table}

At the 4-bit level, this transformation maps the nibble range $[-8, 7]$ to a symmetric range $[-15, 15]$ via the linear relation:$$x_{\text{-1,1}} = 2 x_{\text{nibble}} + 1$$

We utilize this specific encoding scheme due to three primary hardware synergies that optimize analog performance. First, enforcing a symmetric range for both input and weight nibbles minimizes the output numerical dynamic range, which improves the ADC quantization quality. Second, the symmetry of the input encoding enables the use of dynamic range doubling (DRD) techniques \citep{lee2021fully}, effectively doubling the output voltage swing. Finally, the symmetry of the weight encoding supports differential ADC architectures in IMC \citep{lee2024switched}, suppressing common-mode analog noise and variations during conversion.

\subsection{Sub-Block Replication Strategy}
\label{appx:sub-block-replication}
A structural inefficiency arises when a layer's input dimension ($d_{\text{in}}$) is not an exact multiple of the IMC unit's row capacity ($h$). Consequently, the final spatial block ($K = \lceil d_{\text{in}}/h \rceil$) utilizes only a residual number of rows, $d_{\text{rem}} = d_{\text{in}} - (K-1)h$. In the analog domain, activating only a fraction of the rows inherently produces a reduced voltage swing on the accumulation bitlines. This signal attenuation degrades the effective Signal-to-Quantization-Noise Ratio (SQNR) when the partial sum is digitized by the ADC.

To mitigate this degradation, we employ a deterministic replication factor, $\rho$, defined by the available row capacity:
$$\rho = \left\lfloor \frac{h}{d_{\text{rem}}} \right\rfloor$$
For this boundary block $K$, the hardware replicates both the active input activations and the corresponding weight parameters $\rho$ times across the unused rows of the IMC unit. This spatial replication scales the physically accumulated partial sum by $\rho$. By amplifying the analog signal magnitude prior to conversion, this operation enhances the SQNR of the underutilized block against the constant quantization noise floor of a given ADC resolution. To preserve mathematical equivalence, the digitized result is subsequently scaled down by $\rho$ during digital post-processing.

\subsection{Digital Aggregation and Dequantization}
\label{appx:digital-aggregation}

Following the analog execution, the hardware must digitally aggregate the partial sums and map the results back to the floating-point domain. This process involves weighted nibble summation, spatial block accumulation across the IMC units, $-1/1$ format conversion, and final dequantization.

\paragraph{Weighted Nibble Summation}
The first stage of aggregation combines the outputs from the four nibble-product combinations generated by the bit-sliced execution scheme. For a specific IMC unit $k \in \{0, \dots, K-1\}$, the analog macro produces four digitized partial sums corresponding to the input-weight nibble pairs $(u, v) \in \{H, L\}^2$. Let $\mathbf{p}_{u, v}^{(k)}$ denote the digitized output for pair $(u, v)$. To reconstruct the 8-bit integer product for the $k$-th unit, these partial sums are weighted by their corresponding binary significance:
\begin{equation}
\mathbf{p}_{-1, 1}^{(k)} = 2^8 \cdot \mathbf{p}_{HH}^{(k)} + 2^4 \cdot \left( \mathbf{p}_{HL}^{(k)} + \mathbf{p}_{LH}^{(k)} \right) + \mathbf{p}_{LL}^{(k)}
\end{equation}
This weighted sum reconstructs the full integer partial sum within the scope of a single IMC unit in $-1/1$ format. Because the ADC digitization occurs before this summation, the quantization errors from the constituent 4-bit nibble operations natively accumulate into this integer result.

\paragraph{Inter-IMC Unit Aggregation}
The GEMM output requires accumulating these partial results across the $K$ independent spatial partitions along the input dimension $d_{\text{in}}$. Prior to this accumulation, we must correct for the signal replication applied to the final residual block (as described in Appendix \ref{appx:sub-block-replication}). The integer output of the final IMC unit ($k=K-1$) is scaled down by the replication factor $\rho$ to recover the true partial sum magnitude:
\begin{equation}
\mathbf{p'}_{-1,1}^{(K-1)} = \frac{\mathbf{p}_{-1,1}^{(K-1)}}{\rho}
\end{equation}
For all fully utilized units ($k < K-1$), where no replication was applied, the output remains unmodified ($\mathbf{p'}_{-1,1}^{(k)} = \mathbf{p}_{-1,1}^{(k)}$). The accumulated sum across all spatial partitions is then computed as:
\begin{equation}
\mathbf{y}_{-1,1} = \sum_{k=0}^{K-1} \mathbf{p'}_{-1,1}^{(k)}
\end{equation}

\paragraph{Format Conversion and Dequantization}
Because the dot product was physically executed using the $-1/1$ representation, the accumulated result $\mathbf{y}_{-1,1}$ contains cross-term expansions. Expanding the mapped product using the integer operands yields:
\begin{equation}
\begin{aligned}
x_{\text{-1,1}} \cdot w_{\text{-1,1}}
&= (2x_{\text{int}}+1)(2w_{\text{int}}+1) \\
&= 4(x_{\text{int}} \cdot w_{\text{int}})
 + 2x_{\text{int}} + 2w_{\text{int}} + 1
\end{aligned}
\end{equation}
Accumulating this relation over the input dimension $d_{\text{in}}$ produces $\mathbf{y}_{-1,1}$. To recover the standard 2's complement integer sum $\mathbf{y}_{\text{int}} = \sum (x_{\text{int}} \cdot w_{\text{int}})$, we invert the expansion:
\begin{equation}
    \mathbf{y}_{\text{int}} = \frac{\mathbf{y}_{-1,1}}{4} - \frac{\sum x_{\text{int}}}{2} - \frac{\sum w_{\text{int}}}{2} - \frac{d_{\text{in}}}{4}
\end{equation}
where $\sum x_{\text{int}}$ and $\sum w_{\text{int}}$ denote the sums of the elements in the integer input and weight vectors, respectively.

Finally, we map $\mathbf{y}_{\text{int}}$ back to the original floating-point scale. Following our established quantization scheme, the floating-point inputs are quantized asymmetrically ($x \approx s_x (x_{\text{int}} - z_x)$), while the weights are quantized symmetrically ($w \approx s_w w_{\text{int}}$). Substituting these into the floating-point dot product reveals that $\mathbf{y}_{\text{int}}$ contains a bias term proportional to the input zero-point offset $z_x$:
\begin{equation}
    \sum x \cdot w \approx s_x s_w \left( \sum x_{\text{int}} w_{\text{int}} - z_x \sum w_{\text{int}} \right)
\end{equation}
Therefore, the final floating-point output is obtained by subtracting this weight-dependent offset correction ($\mathbf{W}_{\text{sum}} = \sum w_{\text{int}}$) from our recovered 2's complement sum and applying the floating-point scaling factors:
\begin{equation}
    \mathbf{y} = s_x s_w \left( \mathbf{y}_{\text{int}} - z_x \mathbf{W}_{\text{sum}} \right)
\end{equation}

%% file: appendices/app03_accuracy_evaluation.tex
\section{Accuracy Evaluation}
\label{appx:accuracy-evaluation}
\subsection{Calibration Datasets}
\label{appx:calibration-datasets}
We use small calibration sets to compute the EFIM for our ACS metric ($\Omega_i$). For LLMs, we randomly draw 64 fixed-length, 512-token samples from the WikiText-2 training split. For ViT-Small, ViT-Base, and ViT-Large, we randomly sample 64 labeled images from the ImageNet-1K training set. Ground-truth cross-entropy provides the gradients used by the EFIM approximation.

\subsection{Quantization Settings}
\label{appx:quantization-settings}
We employ W8A8 quantization for all evaluated models, using symmetric per-channel weight quantization and asymmetric per-token dynamic activation quantization. LLM runs additionally use dynamic 8-bit KV-cache quantization. Mixed-ADC experiments use 9-bit and 12-bit ADCs with 512 input rows and 32 effective output channels per IMC unit.


For all three standard ViT models, we apply the same rotation pipeline used to mitigate activation outliers. We first absorb LayerNorm affine parameters into adjacent linear layers, center residual-producing projections, replace internal LayerNorms with parameter-free RMSNorms (as introduced in \citet{Slicegpt}), and mean-center and rotate the embedding output. We then compensate the projection and classifier weights, fuse a shared query/key Hadamard rotation offline, and apply the paired MLP/down transformation used by the online Hadamard path. The query/key rotation is shared because standard ViT does not use RoPE.

\subsection{Downstream Task Evaluation}
\label{appx:downstream-evaluation}

For the LLM reasoning benchmarks, we evaluate zero-shot performance across the specified tasks. Following standard evaluation protocols, we report the length-normalized accuracy (acc\_norm) for ARC-Challenge, ARC-Easy, PIQA, HellaSwag, and OpenBookQA. For BoolQ and Winogrande, we report the standard raw accuracy (acc).


For the standard ViT models, the matched comparison uses a deterministic 1000-image subset of the ImageNet-1K validation split, separate from the training images used for calibration. We report top-1 classification accuracy on these images. Output-distribution KL divergence from the FP16 reference is evaluated on the same images over all 1000 classes.

%% file: appendices/app04_energy_simulation.tex
\section{Energy Simulation}
\label{appx:energy-simulation}
In this section, we provide additional details on the accelerator architecture, workload mapping, and energy modeling methodology used in our system-level evaluation.

\subsection{Accelerator Architecture}
\label{appx:energy-accelerator-architecture}
We first introduce the reference accelerator architecture, which comprises an analog IMC engine, a digital matrix multiplication engine, a vector processing unit (VPU), and a hierarchical memory and interconnect system, as illustrated in Figure~\ref{fig:imc-accelerator-architecture}.

\paragraph{Analog IMC Engine.}
The analog IMC engine consists of eight tiles, each containing 16 switched-capacitor SRAM-IMC macros organized into four clusters. Each macro contains 512 rows and 256 physical columns, supporting 32 effective W8A8 output channels. Under four-column ADC sharing \citep{lee2024switched}, each macro performs 64 active conversions per phase. Four macros cooperate in the prescribed spatial-$K$ mapping. Each tile is equipped with 512 KiB weight scratchpad and 512 KiB output/partial-sum SRAM, a local network for operand distribution, and partial-sum reduction and accumulation hardware. A four-lane precision-processing unit supports quantization and dequantization.

\paragraph{Digital Compute Engines.}
The digital matrix multiplication engine contains a $16\times16$ INT8 systolic array with INT32 accumulation, a 48 KiB operand buffer, a 512 KiB local output SRAM, and four-lane precision-conversion and FP16 accumulation units. The VPU contains 16 FP16 SIMD lanes, including eight lanes supporting special functions, and a logical 2 KiB vector register file. It supports FP16 arithmetic, nonlinear functions, reductions, and precision conversions.

\paragraph{Memory and Interconnect.}
The accelerator includes 32 MiB of global SRAM organized as 64 single-read/write-port 512 KiB banks with 256-bit access width, and off-chip HBM2, connected through a shared global network-on-chip (NoC). Static model weights are delivered from HBM2 directly to the IMC tiles through the Global NoC, bypassing Global SRAM. Dedicated 256-bit, 1-GHz half-duplex links connect the VPU to the analog IMC and digital matrix multiplication engines and share one VPU port. Each tile has a 256-Gbit/s ingress; bandwidth stalls are retained. SRAM, NoC, and digital components target 28 nm and 1 GHz subject to their individual timing limits.

\subsection{Workload Mapping and Scheduling}
\label{appx:energy-workload-mapping-scheduling}
We next describe how Transformer operations are mapped and scheduled on the reference accelerator. Figure \ref{fig:background-adc-precision-energy-cost} and the Pareto gain axis target one interior LLaMA-2-7B decoder block under batch-1, 2,048-token dense prefill. The layer has hidden size 4,096, 32 query heads, 32 key/value heads, head dimension 128, and SwiGLU width 11,008. We use a prescribed mapping rather than a mapping search. AccelForge validates its actions, traffic, capacity, bandwidth, latency, and energy at 10-bit ADC resolution, after which the identical mapping and action counts are repriced from 6 through 12 bits. The mapping remains fixed across all ADC allocations.

\paragraph{IMC Linear Projections.}
We map the query, key, value, attention output, gate, up, and down projections onto the analog IMC engine. Each matrix multiplication is partitioned along the input dimension into 512-row tiles and along the output dimension into 32-channel groups, matching the physical IMC macro dimensions. Four macros within each cluster contribute to partial-sum reduction, while additional accumulation is performed across input tiles when necessary. Weights are programmed into IMC macros and reused across activation tokens. Partial sums are retained in local SRAM whenever possible, with additional spills to Global SRAM when required by the prescribed mapping.

\paragraph{FFN Execution.}
The FFN is processed in 1,024-token global tiles, further divided into 128-token fragments to limit intermediate storage and support streamed execution. The gate and up projections execute on IMC, followed by SiLU and elementwise multiplication on the VPU. Their outputs are dynamically quantized to INT8 using per-token, per-group scales (1,024 channels per group) before being transferred to the VPU. Since dynamic quantization requires determining the scale across each group, the VPU temporarily buffers the FP16 SwiGLU outputs in Global SRAM before quantizing and materializing them as INT8 for the down projection. Before the down projection, the intermediate is converted to a common per-token scale to ensure compatible accumulation across input-channel tiles.

\paragraph{Attention and Vector Operations.}
Attention QK$^\top$ and probability--value matrix multiplications execute on the digital INT8 systolic array using $128\times128$ query/key blocks. The VPU performs attention scaling and online softmax, exchanging intermediate tensors with the digital engine. The attention output is accumulated in FP16 and quantized before the IMC attention output projection. Other vector operations, including RMSNorm, RoPE, and residual additions, are also executed on the VPU.

\paragraph{Scheduling Constraints.}
The prescribed execution schedule respects memory capacity, data dependencies, and hardware resource constraints. We keep the workload mapping and scheduling policy fixed across ADC resolution allocations to ensure a fair energy comparison.

\subsection{Component Energy Modeling}
\label{appx:energy-component-modeling}
Building upon AccelForge~\citep{accelforge}, we detail the component-level energy models used to estimate computation, memory, and communication costs.

\paragraph{IMC and ADC.}
We model switched-capacitor IMC computation, macro control, and programming from the measured macro reported by \citet{lee2024switched}. Resolution-dependent ADC conversion, reference, and leakage energy use the absolute, unscaled HWComponents-ADC prediction, informed by the converter survey in \citet{adc_survey}; Lee-anchored ADC curves are used only for sensitivity analysis. The measured macro area is retained as a complete physical-macro reference, so the diagnostic ADC-area estimate is not added to it.

\paragraph{Memory.}
On-chip SRAM dynamic energy, leakage, and area are estimated per physical bank using CACTI~\citep{muralimanohar2007cacti}, accounting for capacity and access width. Off-chip HBM2 dynamic energy is modeled at 3.97 pJ/bit~\citep{oconnor2017fgdram}, including model-weight and activation transfers; its throughput is supplied by the HWComponents HBM primitive.

\paragraph{Interconnect.}
We model on-chip communication with ISAAC-based router primitives~\citep{shafiee2016isaac}. The model accounts for data-transfer volume and traversal through the global NoC, the hierarchical tile-distribution fabric, and the dedicated engine--VPU links.

\paragraph{Digital Computation and Leakage.}
Digital computation energy is estimated using Aladdin-based models~\citep{shao2014aladdin} for INT8 matrix multiplication, registers, reductions, conversions, and FP16 vector operations. We additionally account for component leakage over each simulated execution interval, assuming no power gating.

\subsection{Energy Accounting and Evaluation Methodology}
\label{appx:energy-accounting-evaluation-methodology}
Finally, we explain how component-level energy is aggregated and attributed to obtain the reported system-level energy efficiency for linear projections.

\begin{table*}[t]
    \centering
    \caption{Operator-based attribution of system-level energy for linear projections.}
    \label{tab:imc-energy-attribution}
    \small
    \begin{tabularx}{\textwidth}{@{}lX@{}}
        \toprule
        \textbf{Category} & \textbf{Operations and Associated Costs} \\
        \midrule
        \textbf{IMC-Mapped Operations} &
        Q, K, V, O, Gate, Up, and Down, including input/output
        precision conversion, weight accesses, partial-sum
        accumulation, and associated data movement. \\
        \midrule
        \textbf{Other Operations} &
        Attention BMMs, RMSNorm, RoPE, softmax, SwiGLU,
        residual additions, and their associated costs. \\
        \bottomrule
    \end{tabularx}
\end{table*}

\paragraph{Energy Attribution.}
As summarized in Table~\ref{tab:imc-energy-attribution}, the reported metric covers the seven IMC-mapped linear projections and all system-level energy directly attributable to them, including analog computation, ADC conversion, precision conversion and accumulation, memory and interconnect traffic, and leakage. It excludes attention BMMs, normalization and other vector operations executed on the digital engines. The macro-only diagnostic in Figure~\ref{fig:background-adc-precision-energy-cost} is narrower and includes only computation and peripheral energy within the IMC macro.


\paragraph{Energy Efficiency.}
We report the system-level energy efficiency for linear projections as
\begin{equation}
\eta_{\mathrm{proj}} =
\frac{N_{\mathrm{ops,proj}}}{E_{\mathrm{proj}}}
\times 10^{-12}
\quad \text{TOPS/W},
\end{equation}
where $N_{\mathrm{ops,proj}}$ counts two operations per multiply-accumulate across the seven linear projections, and $E_{\mathrm{proj}}$ includes their associated system-level energy overheads.

For the Pareto curves, we normalize this value by the digital INT8 counterfactual under the same workload and linear-projection boundary, $G_\eta=\eta_{\mathrm{proj}}/\eta_{\mathrm{INT8}}$. This reports the multiplicative system-level energy-efficiency gain for linear projections, whereas the task-level tables report absolute $\eta_{\mathrm{proj}}$ and $\eta_{\mathrm{INT8}}$ in TOPS/W.

\paragraph{Evaluation Methodology.}
All methods use the same hardware, workload mapping, and schedule. Budget matching counts physical ADC phase units and therefore accounts for both output grouping and input-row tiling. The resulting allocation determines $\eta_{\mathrm{proj}}$ and thus $G_\eta$.


\paragraph{Digital INT8 Counterfactual.}
The digital INT8 energy baseline covers the same seven linear projections and associated system-level overheads, replacing the analog IMC compute complex with digital MAC and accumulator units while preserving modeled memory, communication, and conversion traffic. Digital arithmetic energy is obtained from the 28-nm Aladdin models, and attributable leakage is included. This controlled, iso-traffic comparison isolates the energy difference between analog and digital projection computation; it is not intended to represent an independently optimized commercial accelerator.

\subsection{Energy Breakdown Across ADC Resolutions}
\label{appx:energy-breakdown-adc-resolutions}
\begin{figure*}[t]
    \centering
    \includegraphics[width=\textwidth]{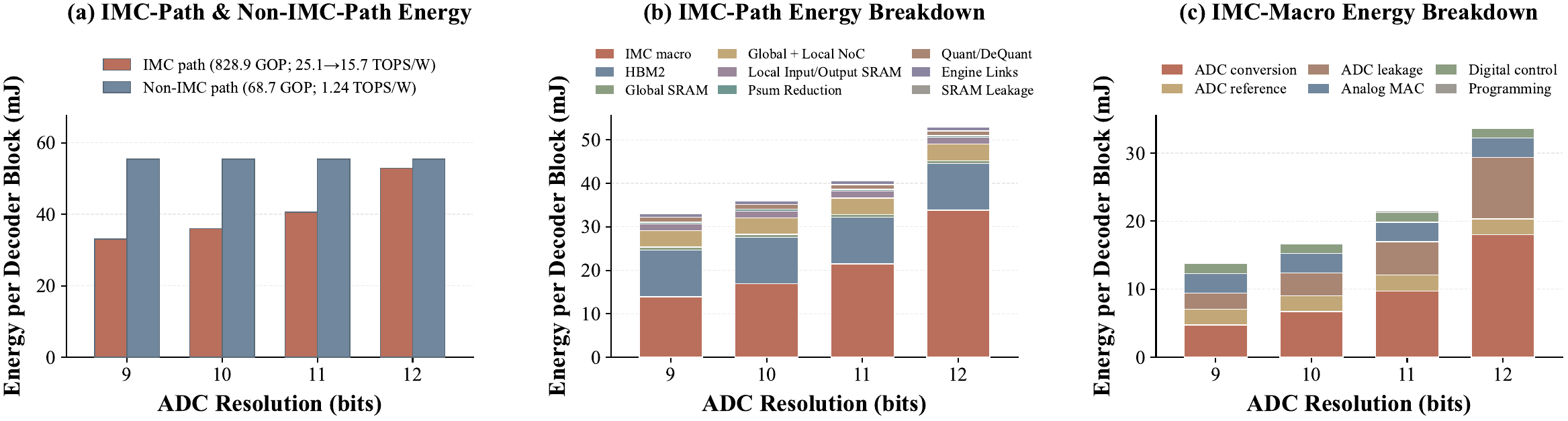}
    \caption{Hierarchical energy breakdown across ADC resolutions (9--12 bits). (a) Energy consumption of the IMC and non-IMC execution paths, with their respective operation counts. (b) Component-level breakdown of $E_{\mathrm{proj}}$. (c) IMC macro energy breakdown, showing that ADC-related energy is the primary resolution-dependent component. All energies are reported per decoder block.}
    \label{fig:energy-breakdown-adc-resolutions}
\end{figure*}

Figure~\ref{fig:energy-breakdown-adc-resolutions} presents a hierarchical breakdown of energy consumption across ADC resolutions. Figure~\ref{fig:energy-breakdown-adc-resolutions}(a) compares the IMC and non-IMC execution paths. Despite handling substantially more operations (828.9 GOP versus 68.7 GOP, approximately $12.1\times$), the IMC path consumes less energy across all evaluated ADC resolutions, highlighting the energy-efficiency benefits of mapping suitable workloads to IMC. However, $E_{\mathrm{proj}}$ increases superlinearly with ADC resolution, while higher resolution is also critical for preserving model accuracy. This underscores the importance of improving the accuracy--energy Pareto frontier through algorithmic mixed-ADC-resolution allocation. Figure~\ref{fig:energy-breakdown-adc-resolutions}(b) further decomposes $E_{\mathrm{proj}}$ into macro computation, memory, interconnect, and peripheral costs. Since ADC resolution does not affect workload mapping or data movement, only the IMC macro energy varies with ADC resolution. Finally, Figure~\ref{fig:energy-breakdown-adc-resolutions}(c) shows that this increase primarily originates from ADC conversion and leakage, while the remaining macro components exhibit nearly constant energy consumption.

%% file: appendices/app05_permutation_construction_for_channel_reordering.tex
\section{Permutation Construction for Channel Reordering}
\label{appx:permutation-construction}

\subsection{Permutations}
\paragraph{Reordering Residual Channels ($P_A$).}
To ensure zero runtime overhead, we apply a single, global permutation matrix across all decoder blocks. Specifically, channels are reordered at the outputs of the attention output projection ($W_O$) and the MLP down projection ($W_{\mathrm{down}}$). To maintain mathematical equivalence, the inverse permutation is identically applied to the input channels of the subsequent query, key, and value projections ($W_Q, W_K, W_V$) and the MLP gate and up projections ($W_{\mathrm{gate}}, W_{\mathrm{up}}$). Because this permutation is shared globally across the entire model, the sensitivity metric used to determine the optimal ordering is defined as the sum of the individual output-channel ACS values of the attention output and down projections across all blocks. As a design choice, a more granular, per-block permutation strategy could be achieved by fusing the permutations into the normalization layers \citep{RPTQ} with some runtime overhead.

\paragraph{Reordering Value Channels ($P_B$).}
For the value projection, the channel reordering strategy must account for the multi-head architecture. Specifically, channels are reordered within each individual attention head at the output of the value projection ($W_V$). Each head may use an independent within-head permutation; to preserve mathematical equivalence, its inverse is fused into the corresponding input columns of the attention output projection ($W_O$). This reordering is executed on a per-block basis. Consequently, the sensitivity metric used to derive the optimal permutation for each block is defined by the individual output-channel ACS values of its respective value projection.

\paragraph{Reordering Query/Key Channels ($P_C$).}
For the query and key projections, we first convert the split-half RoPE layout, which rotates channels $i$ and $i+d_{\mathrm{head}}/2$ together, into the interleaved layout, where each rotation pair consists of adjacent channels. This conversion allows the salient attention head channels, which often belong to the same Rotary Positional Embeddings (RoPE) \citep{RoPE} pair as observed in \citet{chen2026massive}, to be grouped in the same IMC unit under the granularity constraint. The conversion preserves mathematical equivalence through a static channel permutation and a corresponding adjustment to the RoPE layout.

We then permute the RoPE channel pairs together within each head at the outputs of the query and key projections ($W_Q$ and $W_K$), with the corresponding RoPE frequencies identically reordered to preserve mathematical equivalence. To ensure that one reordered RoPE frequency layout applies universally, every head within a block shares the same query/key pair permutation. This operation is executed on a per-block basis, guided by a sensitivity metric defined as the sum of the output-channel ACS values of the query and key projections. For architectures lacking RoPE, this constraint relaxes to the independent per-head permutation applied to the value projection.

\paragraph{Reordering Gate and Up Channels ($P_D$).}
For the MLP's internal hidden dimension, channel reordering is executed across the entire output channel space within each decoder block. Specifically, a shared permutation is applied to the outputs of the gate and up projections ($W_{\mathrm{gate}}, W_{\mathrm{up}}$), while the corresponding inverse permutation is identically applied to the input channels of the down projection ($W_{\mathrm{down}}$). The sensitivity metric governing this intra-block reordering is defined as the sum of the output-channel ACS values of the gate and up projections. For conventional architectures lacking SwiGLU \citep{SwiGLU}, this metric simplifies to consider only the ACS of the up projection.

Table \ref{tab:permutation_llama7b_config} summarizes the configurations for a RoPE-based SwiGLU LLM architecture.

\subsection{Compatibility with Rotations}
\label{appx:reordering-rotation-compatibility}
\begin{figure*}[t]
    \centering
    \includegraphics[width=\textwidth]{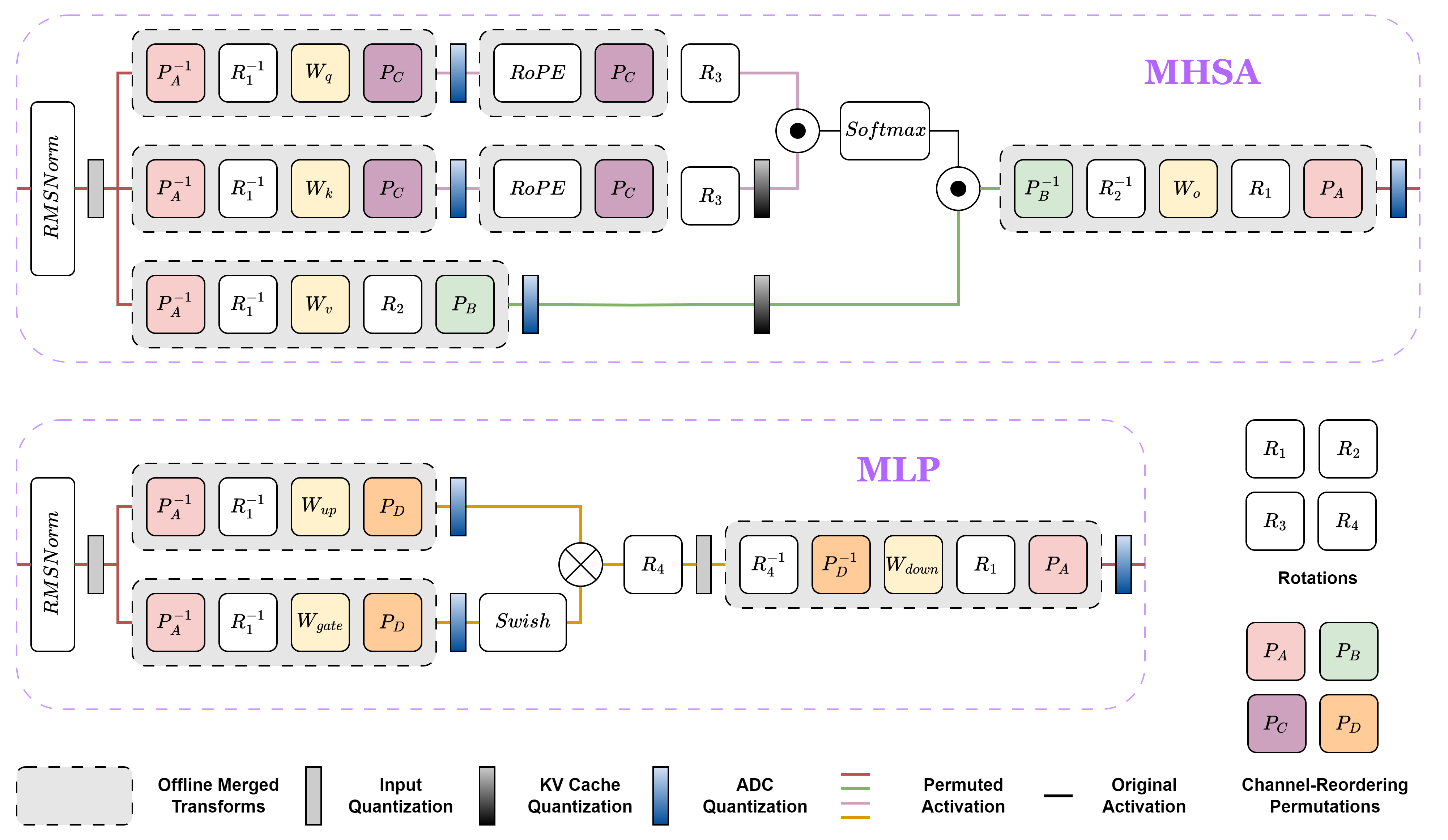}
    \caption{Channel reordering is compatible with rotation methods, allowing optimizations on both input quantization and ADC quantization.}
    \label{fig:reorder_with_rotations}
\end{figure*}

Rotation-based methods \citep{Quarot, Spinquant} are widely adopted to facilitate activation quantization by amortizing extreme outlier values across the channel dimension. In contrast, our channel reordering only permutes the spatial arrangement of these channels to optimize ADC quantization, leaving the numerical magnitudes entirely intact. This orthogonal mechanism guarantees that sensitivity-aware reordering is compatible with rotation techniques without inducing conflicting transformations. Figure \ref{fig:reorder_with_rotations} shows the permutations and rotations for a RoPE-based SwiGLU architecture.

In practice, we integrate both methods through a sequential pipeline. First, we apply the offline rotation matrices (e.g., $R_1, R_1^{-1}, R_2, R_2^{-1}, R_4^{-1}$) to the static weights and register the necessary online rotations (e.g., $R_3, R_4$). Next, we compute the Hessian-guided ACS values by executing forward passes on a calibration dataset using this fully rotated model. Once the optimal permutation matrices ($P_A, P_B, P_C, P_D$) are derived from these ACS values, they are statically fused into the weights. Notably, to properly integrate the inverse permutation $P_D^{-1}$ into the down projection ($W_{\mathrm{down}}$) without violating its previously applied offline rotation ($R_4^{-1}$), we strictly enforce the correct transformation sequence: we temporarily un-rotate the layer by applying $R_4$, insert the permutation $P_D^{-1}$, and then re-apply $R_4^{-1}$.

\begin{table}[htbp]
\centering
\caption{Permutation configurations for a RoPE-based SwiGLU LLM architecture.}
\label{tab:permutation_llama7b_config}
\resizebox{\columnwidth}{!}{%
\begin{tabular}{clcl}
\toprule
Permutations & Scope & Dimension & Sensitivity Metric \\
\midrule
$P_A$ & global & $d_{\text{out}}$ & O + Down \\
$P_B$ & per-block$^\dagger$ & $d_{\text{head}}$ & V \\
$P_C$ & per-block$^\dagger$ & $d_{\text{head}}/2$ & Q + K \\
$P_D$ & per-block$^\dagger$ & $d_{\text{out}}$ & Gate + Up \\
\bottomrule
\end{tabular}
}
\par\smallskip
{\footnotesize $^\dagger$\textit{block} denotes a decoder block in LLMs.}
\end{table}

%% file: appendices/app06_extended_experiments_and_results.tex
\section{Extended Experiments and Results}
\label{appx:extended-results}

\subsection{Extended Task-level Evaluation}
To complement the evaluations in the main text, we present an expanded set of zero-shot reasoning benchmarks to further validate the robustness of \ours{}.

\paragraph{Full LLM Accuracy Results}
Table \ref{tab:allocation_schemes_full_accuracy} reports zero-shot accuracy on all seven common-sense reasoning tasks used to compute the averages in Table \ref{tab:allocation_schemes}: ARC-Challenge and ARC-Easy \citep{ARC}, BoolQ \citep{BoolQ}, HellaSwag \citep{HellaSwag}, PIQA \citep{PIQA}, OpenBookQA (OBQA) \citep{openbookQA}, and Winogrande \citep{winogrande}. Under the matched O+V energy budget, \ours{} achieves the highest seven-task average accuracy for LLaMA-2-7B, LLaMA-3.2-3B, and Mistral-7B-v0.3. For Qwen2-1.5B-Instruct, its average accuracy is 0.20 percentage points below the heuristic baseline.


\begin{table}[htbp]
\centering
\scriptsize
\setlength{\tabcolsep}{2pt}
\caption{Full zero-shot accuracy (\%) on seven common-sense reasoning tasks under the matched O+V energy budget. Bold values denote the best reported mixed-ADC allocation result for each model and task.}
\label{tab:allocation_schemes_full_accuracy}
\resizebox{\columnwidth}{!}{%
\begin{tabular}{lcccccccc}
\toprule
Method & ARC-c & ARC-e & BoolQ & HellaSwag & PIQA & OBQA & Winogrande & Avg. \\
\midrule
\multicolumn{9}{l}{\textbf{LLaMA-2-7B}} \\
INT8 Baseline & 45.05 & 73.74 & 79.48 & 76.13 & 78.78 & 44.0 & 69.30 & 66.64 \\
Random Alloc. & 27.05 & 45.16 & 59.63 & 53.45 & 65.89 & 29.40 & 54.93 & 47.93 \\
Heuristic Alloc. & 38.48 & 64.44 & 68.96 & 69.46 & \textbf{73.99} & \textbf{38.60} & 60.77 & 59.24 \\
\ours{} & \textbf{39.76} & \textbf{65.82} & \textbf{70.12} & \textbf{69.64} & 73.72 & 38.20 & \textbf{62.12} & \textbf{59.91} \\
\midrule
\multicolumn{9}{l}{\textbf{LLaMA-3.2-3B}} \\
INT8 Baseline & 46.25 & 71.42 & 74.31 & 74.17 & 77.86 & 41.0 & 69.93 & 64.99 \\
Random Alloc. & 25.09 & 37.04 & 48.81 & 37.78 & 58.16 & 25.20 & 52.33 & 40.63 \\
Heuristic Alloc. & 33.70 & \textbf{55.22} & 52.81 & 58.45 & 67.52 & \textbf{32.20} & 55.80 & 50.81 \\
\ours{} & \textbf{34.13} & 54.92 & \textbf{55.23} & \textbf{60.67} & \textbf{70.02} & \textbf{32.20} & \textbf{59.43} & \textbf{52.37} \\
\midrule
\multicolumn{9}{l}{\textbf{Mistral-7B-v0.3}} \\
INT8 Baseline & 53.84 & 79.92 & 83.49 & 80.69 & 82.05 & 47.6 & 73.48 & 71.58 \\
Random Alloc. & 36.86 & 63.01 & 65.17 & 65.89 & 72.96 & 37.60 & 60.69 & 57.45 \\
Heuristic Alloc. & \textbf{45.90} & 72.14 & 72.08 & 74.49 & \textbf{78.07} & \textbf{42.20} & 64.64 & 64.22 \\
\ours{} & 45.82 & \textbf{72.73} & \textbf{73.91} & \textbf{75.01} & 77.97 & 41.80 & \textbf{66.69} & \textbf{64.85} \\
\midrule
\multicolumn{9}{l}{\textbf{Qwen2-1.5B-Instruct}} \\
INT8 Baseline & 39.59 & 66.75 & 76.42 & 65.97 & 75.57 & 36.4 & 64.64 & 60.76 \\
Random Alloc. & 25.94 & 39.52 & 52.20 & 44.91 & 58.38 & 27.20 & 51.46 & 42.80 \\
Heuristic Alloc. & 28.41 & \textbf{48.86} & 56.51 & \textbf{53.08} & \textbf{65.18} & 28.20 & \textbf{55.17} & \textbf{47.92} \\
\ours{} & \textbf{28.67} & 47.26 & \textbf{59.20} & 52.33 & 64.31 & \textbf{29.80} & 52.49 & 47.72 \\
\bottomrule
\end{tabular}
}
\end{table}


\paragraph{Performance Under an Alternative Energy Budget}
In the primary evaluations (Tables \ref{tab:allocation_schemes} and \ref{tab:allocation_schemes_full_accuracy}), the compute budget is aligned with the O+V heuristic configuration, which assigns high-resolution ADCs to the \textbf{attention output} and \textbf{value} projections. To evaluate \ours{} under a larger alternative budget, we consider the O+Down+V heuristic configuration, which additionally assigns the \textbf{down} projections to high resolution while all other projections use low resolution.

Table \ref{tab:allocation_schemes_alternative_budget} summarizes quality and system-level energy efficiency for linear projections under this alternative budget. Among the mixed-ADC schemes, \ours{} achieves the lowest KL divergence and perplexity across all four models. It also achieves the highest average accuracy for Mistral-7B-v0.3 and Qwen2-1.5B-Instruct; the heuristic baseline is higher by 0.03 and 0.61 percentage points for LLaMA-2-7B and LLaMA-3.2-3B, respectively. Unlike the discrete operating points provided by architecture-based heuristics, \ours{} can adapt its fine-grained allocation to different energy constraints.

\begin{table*}[t]
\centering
\footnotesize
\setlength{\tabcolsep}{1pt}
\caption{Summary of LLM quality and system-level energy efficiency for linear projections under the matched O+Down+V energy budget, defined by assigning high-resolution ADCs to the attention output, down, and value projections in the heuristic baseline. KL denotes KL divergence, PPL denotes WikiText-2 perplexity, and average accuracy is the unweighted mean over seven common-sense reasoning tasks. Lower KL and PPL are better; higher average accuracy and TOPS/W are better. Bold values denote the best reported quality among mixed-ADC allocation schemes at the matched energy budget.}
\label{tab:allocation_schemes_alternative_budget}
\begingroup
\renewcommand{\todo}[1]{\textcolor{cyan}{TODO}}
\renewcommand{\arraystretch}{1.08}
\resizebox{\textwidth}{!}{%
\begin{tabular}{l*{4}{cccc}}
\toprule
\multirow{2}{*}{Method}
& \multicolumn{4}{c}{LLaMA-2-7B}
& \multicolumn{4}{c}{LLaMA-3.2-3B}
& \multicolumn{4}{c}{Mistral-7B-v0.3}
& \multicolumn{4}{c}{Qwen2-1.5B-Instruct} \\
\cmidrule(lr){2-5}\cmidrule(lr){6-9}\cmidrule(lr){10-13}\cmidrule(lr){14-17}
& KL & PPL & Avg. Acc. & TOPS/W
& KL & PPL & Avg. Acc. & TOPS/W
& KL & PPL & Avg. Acc. & TOPS/W
& KL & PPL & Avg. Acc. & TOPS/W \\
\midrule
INT8 Baseline
& 0.0015 & 5.47 & 66.64 & 2.42
& 0.0024 & 7.82 & 64.99 & 2.41
& 0.0008 & 5.32 & 71.58 & 2.42
& 0.0123 & 10.12 & 60.76 & 2.40 \\
\midrule
Random Alloc.
& 0.5114 & 8.42 & 54.04 & 20.30
& 1.0914 & 21.69 & 48.40 & 19.90
& 0.3120 & 7.01 & 63.00 & 20.28
& 0.6594 & 18.54 & 49.56 & 18.92 \\
Heuristic Alloc.
& 0.1531 & 6.26 & \textbf{63.17} & 20.30
& 0.3389 & 10.71 & \textbf{57.78} & 19.90
& 0.1397 & 6.04 & 67.75 & 20.20
& 0.4578 & 14.97 & 52.13 & 18.92 \\
\ours{}
& \textbf{0.1374} & \textbf{6.16} & 63.14 & 20.30
& \textbf{0.3048} & \textbf{10.54} & 57.17 & 19.90
& \textbf{0.1178} & \textbf{5.91} & \textbf{67.79} & 20.28
& \textbf{0.3032} & \textbf{12.99} & \textbf{53.87} & 18.92 \\
\bottomrule
\end{tabular}
}
\endgroup
\end{table*}

\subsection{Quality of Sensitivity Estimation}
\label{appx:sensitivity-estimation-quality}

To test whether ACS better predicts how ADC perturbations affect task loss, Table~\ref{tab:sensitivity-estimation-quality} compares allocation and reordering driven by ACS, Hessian-only sensitivity~\citep{MixLLM, dong2020hawqv2}, and activation-magnitude-based sensitivity~\citep{ATOM,ashkboos2024quik}, all under the same physical ADC-energy budget. The first block compares complete allocation pipelines, while the second fixes ACS allocation to isolate the effect of reordering. Hessian-only sensitivity captures loss curvature but omits the ADC noise-propagation factors modeled by ACS, whereas activation magnitude does not account for task-loss curvature. The ACS pipeline achieves lower perplexity than the alternatives, and ACS-guided reordering provides a further improvement. These results show that ACS more accurately captures the task-level effect of ADC error, while sensitivity-aware reordering provides a complementary benefit.

\begin{table}[t]
\centering
\small
\setlength{\tabcolsep}{5pt}
\caption{Sensitivity metric and channel-reordering ablations on LLaMA-2-7B under the O+V budget with 9/12-bit ADCs.}
\label{tab:sensitivity-estimation-quality}
\begin{tabular}{llr}
\toprule
Allocation & Reordering & PPL ($\downarrow$) \\
\midrule
\multicolumn{3}{l}{\textit{Allocation pipelines}} \\
Heuristic & None & 7.3283 \\
Activation magnitude & Activation magnitude & 13.5037 \\
Hessian-only & Hessian-only & 7.5437 \\
\textbf{ACS} & \textbf{ACS} & \textbf{7.0130} \\
\midrule
\multicolumn{3}{l}{\textit{Reordering with ACS allocation}} \\
ACS & None & 7.0445 \\
ACS & Activation magnitude & 7.0377 \\
\textbf{ACS} & \textbf{ACS} & \textbf{7.0130} \\
\bottomrule
\end{tabular}
\end{table}

\subsection{Calibration Efficiency and Robustness}
\label{appx:calibration-robustness}

Table \ref{tab:calibration-robustness} evaluates the efficiency and robustness of the \ours{} calibration procedure on LLaMA-2-7B under the O+V budget with 9/12-bit ADCs. Each sweep varies one calibration parameter at a time, and all PPL values are measured on the full WikiText-2 evaluation set.

\begin{table}[htbp]
\centering
\footnotesize
\setlength{\tabcolsep}{5pt}
\caption{Calibration efficiency and robustness of \ours{} on LLaMA-2-7B. GPU denotes peak allocated GPU memory.}
\label{tab:calibration-robustness}
\begin{tabular}{lrrr}
\toprule
Configuration & Time (s) & GPU (GiB) & PPL ($\downarrow$) \\
\midrule
\multicolumn{4}{l}{\textbf{A. Calibration sample count}} \\
8 & 9.9 & 31.4 & 7.0270 \\
16 & 19.3 & 31.4 & 7.0113 \\
32 & 37.9 & 31.4 & 7.0065 \\
64 (default) & 74.9 & 31.4 & 7.0138 \\
128 & 150.0 & 31.4 & 7.0185 \\
\midrule
\multicolumn{4}{l}{\textbf{B. Calibration sequence length}} \\
128 & 49.2 & 28.7 & 7.0816 \\
256 & 57.0 & 29.6 & 7.0355 \\
512 (default) & 74.9 & 31.4 & 7.0138 \\
1,024 & 109.0 & 35.3 & 7.1062 \\
2,048 & 185.1 & 43.1 & 7.0884 \\
\midrule
\multicolumn{4}{l}{\textbf{C. Calibration data domain}} \\
WikiText-2 & 74.9 & 31.4 & 7.0138 \\
C4 & 74.9 & 31.4 & 7.0378 \\
Pile & 75.7 & 31.4 & 7.0153 \\
\bottomrule
\end{tabular}
\end{table}

The results show that only a few calibration samples are needed: using eight samples yields a PPL within 0.0205 of the best result in the sample-count sweep, while runtime scales approximately linearly. Longer sequences increase runtime and peak memory without consistently improving PPL, and changing the calibration domain varies PPL by only 0.0240. Overall, calibration is a low-cost, one-time offline procedure that is robust to reasonable changes in its configuration.

\subsection{Alternative ADC Resolution Configurations}
\label{appx:alternative-adc-resolution}

We evaluate \ours{} across binary and three-level ADC resolution configurations. The results show that \ours{} is not tied to a particular pair of ADC resolutions and can use additional resolution levels effectively.

\paragraph{Binary ADC Resolutions.}
We evaluate six low-/high-resolution ADC pairs under the O+V allocation budget. As shown in Table~\ref{tab:adc-resolution-sweep}, \ours{} achieves the lowest PPL among the compared allocation schemes for every pair. Configurations with an 8-bit low-resolution endpoint degrade sharply, indicating that this endpoint is insufficient for the high-row-parallelism IMC setting.

\begin{table}[t]
\centering
\footnotesize
\setlength{\tabcolsep}{3pt}
\caption{Binary ADC resolution configurations on LLaMA-2-7B under the O+V allocation budget. The ACS w/o reordering variant uses ACS allocation without channel reordering. Lower PPL is better.}
\label{tab:adc-resolution-sweep}
\resizebox{\columnwidth}{!}{%
\begin{tabular}{lrrrr}
\toprule
ADC Bits & \ours{} & ACS w/o Reord. & Heuristic & Random \\
\midrule
9/12 & \textbf{7.0130} & 7.0445 & 7.3283 & 11.9438 \\
9/11 & \textbf{7.1411} & 7.1773 & 7.4508 & 12.5052 \\
9/10 & \textbf{7.7380} & 7.7748 & 8.0997 & 13.2116 \\
8/12 & \textbf{309.8581} & 361.5514 & 634.2484 & 6{,}082.6392 \\
8/11 & \textbf{333.8069} & 392.6672 & 689.6928 & 6{,}197.6001 \\
8/10 & \textbf{476.3950} & 543.5512 & 892.2458 & 6{,}279.1348 \\
\bottomrule
\end{tabular}
}
\end{table}

\paragraph{Three-Level ADC Resolutions.}
We further extend our allocation strategy to three ADC resolutions. This extension can be incorporated into our existing framework because the expected ADC quantization error at each resolution can be characterized directly by its quantization step size. For IMC output group $l$, we rank an upgrade from $b$ to $b'$ bits by its predicted reduction in quantization-induced loss per additional ADC energy:
\begin{equation}
    \rho^{(l)}(b \!\rightarrow\! b') =
    \frac{
        \bar{\Omega}^{(l)}
        \left[\Delta_{\mathrm{ADC}}^2(b)-\Delta_{\mathrm{ADC}}^2(b')\right]
    }{
        K^{(l)}
        \left[e_{\mathrm{ADC}}(b')-e_{\mathrm{ADC}}(b)\right]
    }.
    \label{eq:three-level-benefit-cost}
\end{equation}
Here, $\bar{\Omega}^{(l)}$ is the aggregated ACS defined in Section~\ref{subsec:channel-imc-unit-sensitivity}, $\Delta_{\mathrm{ADC}}(b)$ is the quantization step size at resolution $b$, $e_{\mathrm{ADC}}(b)$ is the energy per ADC conversion, and $K^{(l)}$ is the number of physical input-row tiles for group $l$. All groups begin at the lowest resolution. We greedily select the feasible upgrade with the highest benefit-per-cost ratio, allowing an upgrade to the highest resolution only after selecting the corresponding intermediate-resolution upgrade.

\begin{table}[t]
\centering
\small
\caption{Three-level ADC resolution configurations on LLaMA-2-7B at the same physical ADC-energy budget as the binary 9/12-bit O+V setting. Lower PPL is better.}
\label{tab:three-level-adc}
\begin{tabular}{lr}
\toprule
ADC Bits & PPL ($\downarrow$) \\
\midrule
9/12 & 7.0130 \\
9/10/12 & \textbf{5.8338} \\
9/11/12 & 5.8462 \\
\bottomrule
\end{tabular}
\end{table}

Table~\ref{tab:three-level-adc} shows that both three-level configurations improve PPL over the binary 9/12-bit allocation at the same physical ADC-energy budget. This demonstrates that \ours{} can be easily extended to settings with even more ADC resolution choices.

\subsection{IMC Macro Dimension Sweep}
\label{appx:imc-macro-dimensions}

We evaluate \ours{} across different IMC macro dimensions by independently varying the row capacity $h$ and column capacity $w$ around the default $h=512$, $w=32$ configuration.

\begin{table}[t]
\centering
\footnotesize
\setlength{\tabcolsep}{4pt}
\caption{Full WikiText-2 PPL on LLaMA-2-7B under different IMC macro dimensions using 9/12-bit ADCs and the O+V physical ADC-energy budget. NaN indicates numerical collapse. Lower PPL is better.}
\label{tab:imc-macro-dimension-sweep}
\begin{tabular}{rrrr}
\toprule
Dimension & Random & Hessian-only & \ours{} \\
\midrule
\multicolumn{4}{l}{\textit{Column sweep ($h=512$)}} \\
$w=8$ & 11.8732 & 7.5672 & \textbf{6.9997} \\
$w=16$ & 11.9124 & 7.5616 & \textbf{7.0162} \\
$w=32$ & 11.9438 & 7.5672 & \textbf{7.0130} \\
$w=64$ & 12.2049 & 7.5709 & \textbf{7.0144} \\
$w=128$ & 12.0034 & 7.5642 & \textbf{7.0233} \\
$w=256$ & 12.2914 & 7.5596 & \textbf{7.0273} \\
$w=512$ & 12.7881 & 7.5729 & \textbf{7.0805} \\
$w=1{,}024$ & 11.7607 & 7.5752 & \textbf{7.0484} \\
\midrule
\multicolumn{4}{l}{\textit{Row sweep ($w=32$)}} \\
$h=128$ & 6.0211 & 5.8493 & \textbf{5.7314} \\
$h=256$ & 6.8692 & 6.3030 & \textbf{6.0675} \\
$h=512$ & 11.9438 & 7.5672 & \textbf{7.0130} \\
$h=1{,}024$ & 191.7455 & 13.3821 & \textbf{12.0586} \\
$h=2{,}048$ & 5{,}411.2236 & 129.0776 & 343.8246 \\
$h=4{,}096$ & NaN & NaN & 11{,}006.3721 \\
\bottomrule
\end{tabular}
\end{table}

\paragraph{Column Dimension.}
We vary $w$ from 8 to 1,024 while fixing $h=512$. A larger $w$ forces more output channels to share one ADC resolution, strengthening the allocation-granularity constraint. Nevertheless, \ours{} remains stable, spanning 6.9997--7.0805 PPL, and consistently outperforms both random and Hessian-only allocation.

\paragraph{Row Dimension.}
We vary $h$ from 128 to 4,096 while fixing $w=32$. Increasing row parallelism makes ADC quantization progressively more challenging, and \ours{} achieves the lowest PPL from $h=128$ through $h=1{,}024$. At $h=2{,}048$ and $h=4{,}096$, none of the methods provides reasonable perplexity, suggesting that this operating regime is unsuitable for 9/12-bit ADCs.

\subsection{Robustness Across ADC Energy Budgets}
\label{appx:adc-budget-robustness}

We further evaluate \ours{} under different physical ADC-energy budgets to examine whether its benefits generalize across operating points. Mixed-resolution allocation is most relevant at intermediate budgets, where high-resolution ADCs must be selectively assigned to sensitive computations.

\begin{table}[t]
\centering
\caption{Full WikiText-2 PPL on LLaMA-2-7B under different physical ADC-energy budgets using 9/12-bit ADCs. Lower PPL is better.}
\label{tab:adc-budget-robustness}
\small
\setlength{\tabcolsep}{5pt}
\begin{tabular}{lrrr}
\toprule
ADC Budget & Random & Heuristic & \ours{} \\
\midrule
O+V & 11.9438 & 7.3283 & \textbf{7.0130} \\
O+Down+V & 8.1960 & 6.2600 & \textbf{6.1584} \\
\bottomrule
\end{tabular}
\end{table}

As shown in Table~\ref{tab:adc-budget-robustness}, \ours{} consistently achieves lower perplexity than random and heuristic allocation under both budgets. The improvement over the heuristic is more pronounced under the tighter O+V budget, highlighting the benefit of sensitivity-aware resolution allocation when high-resolution ADC resources are limited.

\subsection{Effect of Analog Noise}
\label{appx:analog-noise}

Analog IMC circuits are subject to nonidealities such as thermal sampling noise, comparator noise, and reference-circuit noise. Although switched-capacitor IMC architectures substantially mitigate analog computation noise~\citep{valavi201964,lee2021fully}, residual noise remains present at the ADC input. Such random noise is commonly modeled as additive, zero-mean Gaussian noise~\citep{ASIM}. Following \citet{IMC-CLINIC}, we inject independent, zero-mean Gaussian noise into each slice-level analog partial sum before ADC quantization, with RMS magnitudes ranging from 0.1 to 0.4 ADC LSB.

Table~\ref{tab:analog-noise} evaluates the robustness of \ours{} on LLaMA-2-7B under the O+V ADC allocation budget. Although increasing analog noise degrades inference quality across all methods, \ours{} consistently achieves lower perplexity and higher average task accuracy than heuristic and random allocations. These results demonstrate that the benefits of sensitivity-guided ADC allocation extend beyond idealized ADC-quantization-only settings.

\begin{table}[t]
\centering
\caption{Robustness to analog noise on LLaMA-2-7B under the O+V ADC budget. Noise magnitude is measured in ADC LSB RMS.}
\label{tab:analog-noise}
\small
\setlength{\tabcolsep}{4pt}
\begin{tabular}{clcc}
\toprule
Noise & Allocation Scheme & PPL ($\downarrow$) & Avg. Acc. ($\uparrow$) \\
\midrule
\multirow{3}{*}{0.1} & Random Alloc. & 14.80 & 44.01 \\
 & Heuristic Alloc. & 7.72 & 57.20 \\
 & \ours{} & \textbf{7.31} & \textbf{58.26} \\
\midrule
\multirow{3}{*}{0.2} & Random Alloc. & 42.00 & 39.25 \\
 & Heuristic Alloc. & 9.36 & 54.13 \\
 & \ours{} & \textbf{8.47} & \textbf{54.26} \\
\midrule
\multirow{3}{*}{0.3} & Random Alloc. & 253.77 & 36.06 \\
 & Heuristic Alloc. & 16.08 & 46.99 \\
 & \ours{} & \textbf{12.50} & \textbf{47.45} \\
\midrule
\multirow{3}{*}{0.4} & Random Alloc. & 1424.73 & 35.05 \\
 & Heuristic Alloc. & 87.56 & 38.13 \\
 & \ours{} & \textbf{43.79} & \textbf{38.42} \\
\bottomrule
\end{tabular}
\end{table}

%% file: appendices/app07_experiment_setup_for_the_figures.tex
\section{Experiment Setup for the Figures}
\label{appx:figure_experiment_details}

\subsection{ADC Resolution Energy--Accuracy Trade-off (Figure \ref{fig:background-adc-precision-energy-cost})}
Figure \ref{fig:background-adc-precision-energy-cost} targets the reference accelerator in Appendix \ref{appx:energy-accelerator-architecture} and one interior LLaMA-2-7B decoder layer under batch-1, 2,048-token dense prefill. Query, key, value, attention output, gate, up, and down projections execute on IMC with the fixed mapping in Appendix \ref{appx:energy-workload-mapping-scheduling}. AccelForge validates that mapping at 10-bit ADC resolution and then reprices the identical actions, traffic, occupancy, and latency from 6 through 12 bits.

The IMC-system curve uses the complete projection-path boundary defined in Appendix \ref{appx:energy-accounting-evaluation-methodology}; the macro curve includes only analog MAC, ADC conversion/reference/leakage, macro control, and macro programming. Let $N_{\mathrm{MAC}}$ denote the projection MAC count for the modeled layer. With one multiply and one accumulate counted as two operations, the plotted value at resolution $b$ is
\begin{equation}
e(b)=\frac{E(b)}{2N_{\mathrm{MAC}}}\quad\mathrm{pJ/op}.
\end{equation}
Thus, both curves report absolute energy per INT8 operation rather than normalization to a selected ADC resolution.


\subsection{ADC Sensitivity of Decoder Blocks (Figure \ref{fig:layer-sensitivity-curve})}
To evaluate block-wise sensitivity to ADC resolution, we sweep the target decoder block's ADC resolution from 7 to 10 bits while fixing all other blocks at 12 bits. We evaluate the LLaMA-2-7B model on OpenBookQA—selected for rapid evaluation due to its small test set—and report the normalized accuracy.

\subsection{ADC Sensitivity of Channels (Figure \ref{fig:channel-sensitivity-heatmap})}
To analyze channel-wise ADC sensitivity, we perturb the channels of the attention output projection in decoder block 0 and measure the final-layer MSE against an 8-bit digital baseline (W8A8, without IMC) to observe channel-wise error propagation. While the layer comprises 4096 channels in total, we restrict our perturbation and visualization to the first 128 channels to ensure figure clarity. We evaluate LLaMA-2-7B using a calibration set of 8 randomly sampled 512-token sequences from the WikiText-2 training set.

\subsection{Channel Reordering for ADC Allocation (Figure \ref{fig:sensitivity-with-reordering})}
We analyze the ACS and mixed ADC resolution allocation for the up projection in decoder block 0 with and without channel reordering, assuming an allocation budget split evenly (50\%/50\%) between 9-bit and 12-bit ADCs. Notably, the ACS profile after channel reordering is not strictly decreasing; this occurs because the sorting permutation is determined by the combined sensitivities of both the gate and up projections as described in Appendix \ref{appx:reordering-rotation-compatibility}, rather than the up projection in isolation.

\subsection{KL Divergence and Perplexity vs. Energy-Efficiency Gain (Figure \ref{fig:pareto-curve})}
The fidelity sweep uses LLaMA-2-7B with 9-bit low-resolution and 12-bit high-resolution ADCs. All configurations are evaluated on the same fixed set of 512-token sequences drawn from a held-out 10\% portion of the WikiText-2 training split. KL divergence is evaluated according to Section~\ref{subsec:models-and-datasets}. The FP16 reference retains the same rotation and permutation transformations as the IMC model while disabling quantization and IMC emulation. The perplexity panel uses the same sampled validation data and is distinct from the full-corpus WikiText-2 test perplexity reported in the task-level tables.

For each sensitivity-based and Random curve, we sweep requested high-resolution budgets from 0\% to 100\% in 10-percentage-point increments. Allocation costs count physical ADC phase units, including both 32-channel output grouping and 512-row input tiling. The achieved high-resolution ratio determines the system-level energy efficiency for linear projections, $\eta_{\mathrm{proj}}$, and the horizontal axis reports its gain over the digital INT8 counterfactual, $G_\eta=\eta_{\mathrm{proj}}/\eta_{\mathrm{INT8}}$. The seven Heuristic markers are the nonempty subsets of attention output, value, and down projections; because different subsets can have equal energy but different fidelity, they are shown as individual operating points rather than a trajectory. System energy follows the LLaMA-2-7B projection workload and linear-projection accounting defined in Section~\ref{subsec:imc-accelerator}.

We compare the following five configurations against an INT8 digital baseline representing no ADC quantization. Its energy efficiency defines the denominator of $G_\eta$, and its fidelity is shown as a horizontal reference in each panel:
\begin{itemize}
    \item \texttt{cols=32}: Standard hardware-constrained Hessian-guided allocation
    \item \texttt{cols=32+reorder}: Hessian-guided allocation with sensitivity-aware channel reordering
    \item \texttt{cols=1}: Idealized per-channel allocation serving as the algorithmic upper bound
    \item \texttt{Heuristic}: Coarse-grained per-projection ADC resolution allocation
    \item \texttt{Random}: Random ADC resolution allocation under the same physical budget
\end{itemize}

Note: In the figure, several data points for the Random and Heuristic cases are omitted where their KL divergence or perplexity values exceed the corresponding panel's axis limits, indicating their resulting suboptimal model quality.

%% file: appendices/app08_discussions.tex
\section{Discussions}
\label{appx:discussions}

\subsection{Broader Impacts}
\label{appx:discussions-broader-implications}

Our work focuses on enabling highly energy-efficient inference of LLMs on analog IMC hardware.

\textbf{Positive Impacts:} This research promotes environmental sustainability by significantly reducing the carbon footprint of LLM inference. Furthermore, by lowering the energy and computational barriers for edge deployment, it establishes a foundational pathway toward democratizing AI access on future resource-constrained devices, ultimately reducing reliance on centralized cloud infrastructure.

\textbf{Potential Negative Impacts and Mitigation:} A potential fairness concern arises from \ours{}'s reliance on calibration datasets to compute resolution allocations. If the chosen calibration data underrepresents certain demographics, languages, or cultural contexts, the allocation may inadvertently assign lower resolution to the channels responsible for processing those specific features. Consequently, the quantized model could exhibit amplified biases or disproportionately degraded performance for underrepresented groups. To mitigate this risk, practitioners deploying \ours{} must ensure that calibration datasets are rigorously audited for diversity and representativeness.

\subsection{Compute Resources}
\label{appx:compute-resources}
A primary advantage of our IMC PTQ framework is its minimal computational overhead. We do not require massive distributed clusters necessary for QAT. All EFIM estimation, sensitivity computation, and downstream model evaluations can be fully executed on a single compute node equipped with one NVIDIA A100 GPU (80GB VRAM) and 32GB of system RAM. While the entire framework is reproducible on a single GPU, we distributed some parallel evaluation sweeps across up to four A100 GPUs strictly to accelerate experimental turnaround times; this multi-GPU setup is not a memory requirement.

Generating the Pareto frontiers and evaluating the quantized models on downstream tasks required approximately 1--2 GPU hours per model configuration. We estimate the total compute required to generate the final figures and tables reported in this manuscript to be roughly 20 GPU hours. Factoring in preliminary experiments, failed configurations, and exploratory hyperparameter sweeps during the development phase, the total compute footprint for this research project was approximately 300 GPU hours.

%% file: appendices/app09_documentation_of_assets_and_licenses.tex
\section{Documentation of Assets and Licenses}
\label{appendix:assets}

For reproducibility, we provide documentation for all external models, datasets, and software frameworks utilized in this work.

\subsection{Pretrained Models}
All models were obtained via the Hugging Face Model Hub. We utilized specific versions of these models as baselines for our quantization and IMC simulations.

\begin{table}[h]
\centering
\small
\resizebox{\columnwidth}{!}{%
\begin{tabular}{@{}lll@{}}
\toprule
\textbf{Model} & \textbf{License} & \textbf{Source} \\ \midrule
LLaMA-2-7B \cite{LLAMA2-7B} & Llama 2 Community & \href{https://huggingface.co/meta-llama/Llama-2-7b-hf}{[HuggingFace]} \\
LLaMA-3.2-3B \cite{LLaMA-3.2} & Llama 3.2 Community & \href{https://huggingface.co/meta-llama/Llama-3.2-3B}{[HuggingFace]} \\
Mistral-7B-v0.3 \cite{MISTRAL-7B-v0.3} & Apache 2.0 & \href{https://huggingface.co/mistralai/Mistral-7B-v0.3}{[HuggingFace]} \\
Qwen2-1.5B-Instruct \cite{QWEN2-1.5B-Instruct} & Apache 2.0 & \href{https://huggingface.co/Qwen/Qwen2-1.5B-Instruct}{[HuggingFace]} \\ \bottomrule
\end{tabular}
}
\caption{Documentation of pretrained language models used in this study.}
\end{table}

\subsection{Datasets}
We evaluate our methods across the following benchmarks:
\begin{itemize}[noitemsep,topsep=0pt,leftmargin=*]
    \item \textbf{WikiText-2}: CC BY-SA 3.0. \cite{WikiText-2}
    \item \textbf{Reasoning}: HellaSwag \cite{HellaSwag} (MIT), ARC (Easy/Challenge) \cite{ARC} (CC BY-SA 4.0), BoolQ \cite{BoolQ} (CC BY-SA 3.0), PIQA \cite{PIQA} (Apache 2.0), OpenBookQA \cite{openbookQA} (Apache 2.0), and Winogrande \cite{winogrande} (Apache 2.0).
    \item \textbf{Vision}: ImageNet-1K \cite{ImageNet-1K} (Custom; Non-Commercial).
\end{itemize}

\subsection{Software and Frameworks}
The following open-source frameworks were used for evaluation and hardware-aware simulation:
\begin{itemize}[noitemsep,topsep=0pt,leftmargin=*]
    \item \textbf{lm-evaluation-harness}: Zero-shot evaluation. MIT License. \href{https://github.com/EleutherAI/lm-evaluation-harness}{[GitHub]}
    \item \textbf{AccelForge}: System-level accelerator energy modeling and workload mapping. MIT License. \href{https://github.com/Accelergy-Project/accelforge}{[GitHub]} \cite{accelforge}
    \item \textbf{HF Transformers}: Model management and emulation. Apache 2.0. \href{https://github.com/huggingface/transformers}{[GitHub]} \cite{HuggingFace}
\end{itemize}

\subsection{License Compliance}
The authors certify that all assets were used in strict compliance with their respective licenses. We do not redistribute model weights with restrictive community licenses.

%% file: references.bib
@article{Murmann-mixed-signal,
  title={Mixed-signal computing for deep neural network inference},
  author={Murmann, Boris},
  journal={IEEE Transactions on Very Large Scale Integration (VLSI) Systems},
  volume={29},
  number={1},
  pages={3--13},
  year={2020},
  publisher={IEEE}
}

@article{programmableIMC,
  title={A programmable heterogeneous microprocessor based on bit-scalable in-memory computing},
  author={Jia, Hongyang and Valavi, Hossein and Tang, Yinqi and Zhang, Jintao and Verma, Naveen},
  journal={IEEE Journal of Solid-State Circuits},
  volume={55},
  number={9},
  pages={2609--2621},
  year={2020},
  publisher={IEEE}
}

@article{IMC-CLINIC,
  title={IMC-CLINIC: Coupled Loss-Informed Newton Iterations for Clipping in Analog In-Memory Computing},
  author={Chen, Yung-Chin and Chen, Chia-Yu and Verma, Naveen},
  journal={arXiv preprint arXiv:2609.35586},
  year={2026}
}

@article{valavi201964,
  title={A 64-tile 2.4-Mb in-memory-computing CNN accelerator employing charge-domain compute},
  author={Valavi, Hossein and Ramadge, Peter J and Nestler, Eric and Verma, Naveen},
  journal={IEEE Journal of Solid-State Circuits},
  volume={54},
  number={6},
  pages={1789--1799},
  year={2019},
  publisher={IEEE}
}

@inproceedings{lee2024switched,
  title={A switched-capacitor SRAM in-memory computing macro with high-precision, high-efficiency differential architecture},
  author={Lee, Jinseok and Zhang, Bonan and Verma, Naveen},
  booktitle={2024 IEEE European Solid-State Electronics Research Conference (ESSERC)},
  pages={357--360},
  year={2024},
  organization={IEEE}
}

@inproceedings{lee2021fully,
  title={Fully row/column-parallel in-memory computing SRAM macro employing capacitor-based mixed-signal computation with 5-b inputs},
  author={Lee, Jinseok and Valavi, Hossein and Tang, Yinqi and Verma, Naveen},
  booktitle={2021 Symposium on VLSI Circuits},
  pages={1--2},
  year={2021},
  organization={IEEE}
}

@inproceedings{MRAM-IMC,
  title={A 22nm 128-kb MRAM row/column-parallel in-memory computing macro with memory-resistance boosting and multi-column ADC readout},
  author={Deaville, Peter and Zhang, Bonan and Verma, Naveen},
  booktitle={2022 IEEE symposium on VLSI technology and circuits (VLSI technology and circuits)},
  pages={268--269},
  year={2022},
  organization={IEEE}
}

@article{RRAM-IMC,
  title={High-throughput in-memory computing for binary deep neural networks with monolithically integrated RRAM and 90-nm CMOS},
  author={Yin, Shihui and Sun, Xiaoyu and Yu, Shimeng and Seo, Jae-Sun},
  journal={IEEE Transactions on Electron Devices},
  volume={67},
  number={10},
  pages={4185--4192},
  year={2020},
  publisher={IEEE}
}

@INPROCEEDINGS{PCM-IMC,
  author={Khaddam-Aljameh, R. and Stanisavljevic, M. and Fornt Mas, J. and Karunaratne, G. and Braendli, M. and Liu, F. and Singh, A. and Müller, S. M. and Egger, U. and Petropoulos, A. and Antonakopoulos, T. and Brew, K. and Choi, S. and Ok, I. and Lie, F. L. and Saulnier, N. and Chan, V. and Ahsan, I. and Narayanan, V. and Nandakumar, S. R. and Le Gallo, M. and Francese, P. A. and Sebastian, A. and Eleftheriou, E.},
  booktitle={2021 Symposium on VLSI Circuits}, 
  title={HERMES Core – A 14nm CMOS and PCM-based In-Memory Compute Core using an array of 300ps/LSB Linearized CCO-based ADCs and local digital processing}, 
  year={2021},
  volume={},
  number={},
  pages={1-2},
  doi={10.23919/VLSICircuits52068.2021.9492362}
}

@article{edram-IMC,
  title={Scaling-CIM: EDRAM in-memory-computing accelerator with dynamic-scaling ADC and adaptive analog operation},
  author={Kim, Sangjin and Um, Soyeon and Jo, Wooyoung and Lee, Jingu and Ha, Sangwoo and Li, Zhiyong and Yoo, Hoi-Jun},
  journal={IEEE Journal of Solid-State Circuits},
  volume={59},
  number={8},
  pages={2694--2705},
  year={2024},
  publisher={IEEE}
}

@ARTICLE{current-domain-IMC,
  author={Zhang, Jintao and Wang, Zhuo and Verma, Naveen},
  journal={IEEE Journal of Solid-State Circuits}, 
  title={In-Memory Computation of a Machine-Learning Classifier in a Standard 6T SRAM Array}, 
  year={2017},
  volume={52},
  number={4},
  pages={915-924},
  doi={10.1109/JSSC.2016.2642198}
}

@INPROCEEDINGS{time-domain-IMC,
  author={Wu, Ping-Chun and Su, Jian-Wei and Chung, Yen-Lin and Hong, Li-Yang and Ren, Jin-Sheng and Chang, Fu-Chun and Wu, Yuan and Chen, Ho- Yu and Lin, Chen-Hsun and Hsiao, Hsu-Ming and Li, Sih-Han and Sheu, Shyh-Shyuan and Chang, Shih-Chieh and Lo, Wei-Chung and Lo, Chung-Chuan and Liu, Ren-Shuo and Hsieh, Chih-Cheng and Tang, Kea-Tiong and Wu, Chih-I and Chang, Meng-Fan},
  booktitle={2022 IEEE International Solid-State Circuits Conference (ISSCC)}, 
  title={A 28nm 1Mb Time-Domain Computing-in-Memory 6T-SRAM Macro with a 6.6ns Latency, 1241GOPS and 37.01TOPS/W for 8b-MAC Operations for Edge-AI Devices}, 
  year={2022},
  volume={65},
  number={},
  pages={1-3},
  doi={10.1109/ISSCC42614.2022.9731681}
}

@inproceedings{verma200625,
  title={A 25/spl mu/W 100kS/s 12b ADC for wireless micro-sensor applications},
  author={Verma, Naveen and Chandrakasan, Anantha P},
  booktitle={2006 IEEE International Solid State Circuits Conference-Digest of Technical Papers},
  pages={822--831},
  year={2006},
  organization={IEEE}
}

@article{yip2013resolution,
  title={A resolution-reconfigurable 5-to-10-bit 0.4-to-1 V power scalable SAR ADC for sensor applications},
  author={Yip, Marcus and Chandrakasan, Anantha P},
  journal={IEEE Journal of Solid-State Circuits},
  volume={48},
  number={6},
  pages={1453--1464},
  year={2013},
  publisher={IEEE}
}

@article{nuzzo2008noise,
  title={Noise analysis of regenerative comparators for reconfigurable ADC architectures},
  author={Nuzzo, Pierluigi and De Bernardinis, Fernando and Terreni, Pierangelo and Van der Plas, Geert},
  journal={IEEE Transactions on Circuits and Systems I: Regular Papers},
  volume={55},
  number={6},
  pages={1441--1454},
  year={2008},
  publisher={IEEE}
}

@article{Benchmarking-IMC,
  title={Benchmarking in-memory computing architectures},
  author={Shanbhag, Naresh R and Roy, Saion K},
  journal={IEEE Open Journal of the Solid-State Circuits Society},
  volume={2},
  pages={288--300},
  year={2022},
  publisher={IEEE}
}

@inproceedings{RAOQ,
  title={Reshape and adapt for output quantization (RAOQ): Quantization-aware training for in-memory computing systems},
  author={Zhang, Bonan and Chen, Chia-Yu and Verma, Naveen},
  booktitle={Forty-first International Conference on Machine Learning},
  year={2024}
}

@inproceedings{cambricon-cim,
  title={{Cambricon-CIM}: Enabling Energy-Efficient and Error-Resilient Analog {CIM} Acceleration via Reformation of Coding Bases},
  author={Guo, Hongrui and Ma, Tianrui and Du, Zidong and Zou, Mo and Hao, Yifan and Zhao, Yongwei and Zhang, Rui and Li, Wei and Hu, Xing and Xu, Zhiwei and Guo, Qi and Chen, Tianshi},
  booktitle={2026 IEEE International Symposium on High Performance Computer Architecture (HPCA)},
  pages={1--16},
  year={2026},
  publisher={IEEE},
  doi={10.1109/HPCA68181.2026.11408520},
  url={https://doi.org/10.1109/HPCA68181.2026.11408520}
}

@article{widrow-quantization,
  title={Statistical theory of quantization},
  author={Widrow, Bernard and Kollar, Istvan and Liu, Ming-Chang},
  journal={IEEE Transactions on instrumentation and measurement},
  volume={45},
  number={2},
  pages={353--361},
  year={1996},
  publisher={IEEE}
}

@article{grimm2022neural,
  title={Neural network training on in-memory-computing hardware with radix-4 gradients},
  author={Grimm, Christopher and Verma, Naveen},
  journal={IEEE Transactions on Circuits and Systems I: Regular Papers},
  volume={69},
  number={10},
  pages={4056--4068},
  year={2022},
  publisher={IEEE}
}

@inproceedings{kavishwar2026minimizing,
  title={Minimizing ADC Precision for Analog In-Memory Computing},
  author={Kavishwar, Mihir and Shanbhag, Naresh R},
  booktitle={ICASSP 2026-2026 IEEE International Conference on Acoustics, Speech and Signal Processing (ICASSP)},
  pages={20446--20450},
  year={2026},
  organization={IEEE}
}

@article{OCC,
  title={Signal processing methods to enhance the energy efficiency of in-memory computing architectures},
  author={Sakr, Charbel and Shanbhag, Naresh R},
  journal={IEEE Transactions on Signal Processing},
  volume={69},
  pages={6462--6472},
  year={2021},
  publisher={IEEE}
}

@inproceedings{lee2021charge,
  title={A charge-sharing based 8T SRAM in-memory computing for edge DNN acceleration},
  author={Lee, Kyeongho and Cheon, Sungsoo and Jo, Joongho and Choi, Woong and Park, Jongsun},
  booktitle={2021 58th ACM/IEEE Design Automation Conference (DAC)},
  pages={739--744},
  year={2021},
  organization={IEEE}
}

@inproceedings{ando20254541,
  title={A 4541 TOPS/W Saliency-Aware Analog Computing In-Memory Macro with Charge-Domain Saliency Detector},
  author={Ando, Shimpei and Miyagi, Satomi and Zhang, Wenlun and Chen, Yung-Chin and Yoshioka, Kentaro},
  booktitle={2025 IEEE European Solid-State Electronics Research Conference (ESSERC)},
  pages={465--468},
  year={2025},
  organization={IEEE}
}

@inproceedings{chen2024osa,
  title={OSA-HCIM: On-the-fly saliency-aware hybrid SRAM CIM with dynamic precision configuration},
  author={Chen, Yung-Chin and Ando, Shimpei and Fujiki, Daichi and Takamaeda-Yamazaki, Shinya and Yoshioka, Kentaro},
  booktitle={2024 29th Asia and South Pacific Design Automation Conference (ASP-DAC)},
  pages={539--544},
  year={2024},
  organization={IEEE}
}

@inproceedings{yoshioka202434,
  title={34.5 a 818-4094TOPS/W capacitor-reconfigured CIM macro for unified acceleration of CNNs and transformers},
  author={Yoshioka, Kentaro},
  booktitle={2024 IEEE International Solid-State Circuits Conference (ISSCC)},
  volume={67},
  pages={574--576},
  year={2024},
  organization={IEEE}
}

@inproceedings{DIANA,
  title={{DIANA}: An End-to-End Energy-Efficient {DIgital} and {ANAlog} Hybrid Neural Network {SoC}},
  author={Ueyoshi, Kodai and Papistas, Ioannis A. and Houshmand, Pouya and Sarda, Giuseppe M. and Jain, Vikram and Shi, Man and Zheng, Qilin and Giraldo, Sebastian and Vrancx, Peter and Doevenspeck, Jonas and Bhattacharjee, Debjyoti and Cosemans, Stefan and Mallik, Arindam and Debacker, Peter and Verkest, Diederik and Verhelst, Marian},
  booktitle={2022 IEEE International Solid-State Circuits Conference (ISSCC)},
  volume={65},
  pages={1--3},
  year={2022},
  publisher={IEEE},
  doi={10.1109/ISSCC42614.2022.9731716},
  url={https://doi.org/10.1109/ISSCC42614.2022.9731716}
}

@article{dash2021robust,
  title={Robust processing-in-memory with multibit ReRAM using Hessian-driven mixed-precision computation},
  author={Dash, Saurabh and Luo, Yandong and Lu, Anni and Yu, Shimeng and Mukhopadhyay, Saibal},
  journal={IEEE Transactions on Computer-Aided Design of Integrated Circuits and Systems},
  volume={41},
  number={4},
  pages={1006--1019},
  year={2021},
  publisher={IEEE}
}

@inproceedings{DIMC,
  title={DIMC: 2219TOPS/W 2569F2/b digital in-memory computing macro in 28nm based on approximate arithmetic hardware},
  author={Wang, Dewei and Lin, Chuan-Tung and Chen, Gregory K and Knag, Phil and Krishnamurthy, Ram K and Seok, Mingoo},
  booktitle={2022 IEEE International Solid-State Circuits Conference (ISSCC)},
  volume={65},
  pages={266--268},
  year={2022},
  organization={IEEE}
}

@inproceedings{pacim,
  title={Pacim: A sparsity-centric hybrid compute-in-memory architecture via probabilistic approximation},
  author={Zhang, Wenlun and Ando, Shimpei and Chen, Yung-Chin and Miyagi, Satomi and Takamaeda-Yamazaki, Shinya and Yoshioka, Kentaro},
  booktitle={Proceedings of the 43rd IEEE/ACM International Conference on Computer-Aided Design},
  pages={1--9},
  year={2024}
}

@article{jiang2020device,
  title={Device-circuit-architecture co-exploration for computing-in-memory neural accelerators},
  author={Jiang, Weiwen and Lou, Qiuwen and Yan, Zheyu and Yang, Lei and Hu, Jingtong and Hu, Xiaobo Sharon and Shi, Yiyu},
  journal={IEEE Transactions on Computers},
  volume={70},
  number={4},
  pages={595--605},
  year={2020},
  publisher={IEEE}
}

@article{cim2pq,
  title={Cim$^2$pq: An arraywise and hardware-friendly mixed precision quantization method for analog computing-in-memory},
  author={Sun, Sifan and Bai, Jinyu and Shi, Zhaoyu and Zhao, Weisheng and Kang, Wang},
  journal={IEEE Transactions on Computer-Aided Design of Integrated Circuits and Systems},
  volume={43},
  number={7},
  pages={2084--2097},
  year={2024},
  publisher={IEEE}
}

@article{ASIM,
  title={ASiM: Modeling and Analyzing Inference Accuracy of SRAM-Based Analog CiM Circuits},
  author={Zhang, Wenlun and Ando, Shimpei and Chen, Yung-Chin and Yoshioka, Kentaro},
  journal={IEEE Transactions on Very Large Scale Integration (VLSI) Systems},
  year={2025},
  publisher={IEEE}
}

@article{analog-foundation-models,
  title={Analog Foundation Models},
  author={B{\"u}chel, Julian and Chalas, Iason and Acampa, Giovanni and Chen, An and Fagbohungbe, Omobayode and Tsai, Sidney and Maghraoui, Kaoutar El and Gallo, Manuel Le and Rahimi, Abbas and Sebastian, Abu},
  journal={arXiv preprint arXiv:2505.09663},
  year={2025}
}

@article{ATOM,
  title={Atom: Low-bit quantization for efficient and accurate llm serving},
  author={Zhao, Yilong and Lin, Chien-Yu and Zhu, Kan and Ye, Zihao and Chen, Lequn and Zheng, Size and Ceze, Luis and Krishnamurthy, Arvind and Chen, Tianqi and Kasikci, Baris},
  journal={Proceedings of Machine Learning and Systems},
  volume={6},
  pages={196--209},
  year={2024}
}

@article{ResQ,
  title={Resq: Mixed-precision quantization of large language models with low-rank residuals},
  author={Saxena, Utkarsh and Sharify, Sayeh and Roy, Kaushik and Wang, Xin},
  journal={arXiv preprint arXiv:2412.14363},
  year={2024}
}

@article{MixLLM,
  title={Mixllm: Llm quantization with global mixed-precision between output-features and highly-efficient system design},
  author={Zheng, Zhen and Song, Xiaonan and Liu, Chuanjie},
  journal={arXiv preprint arXiv:2412.14590},
  year={2024}
}

@article{LLM-INT8,
  title={{GPT3.int8()}: 8-bit Matrix Multiplication for Transformers at Scale},
  author={Dettmers, Tim and Lewis, Mike and Belkada, Younes and Zettlemoyer, Luke},
  journal={Advances in Neural Information Processing Systems},
  volume={35},
  pages={30318--30332},
  year={2022},
  doi={10.52202/068431-2198},
  url={https://proceedings.neurips.cc/paper_files/paper/2022/hash/c3ba4962c05c49636d4c6206a97e9c8a-Abstract-Conference.html}
}

@article{slim-llm,
  title={SliM-LLM: Salience-driven mixed-precision quantization for large language models},
  author={Huang, Wei and Qin, Haotong and Liu, Yangdong and Li, Yawei and Liu, Qinshuo and Liu, Xianglong and Benini, Luca and Magno, Michele and Zhang, Shiming and Qi, Xiaojuan},
  journal={arXiv preprint arXiv:2405.14917},
  year={2024}
}

@article{FGMP,
  title={FGMP: fine-grained mixed-precision weight and activation quantization for hardware-accelerated LLM inference},
  author={Hooper, Coleman and Sakr, Charbel and Keller, Ben and Venkatesan, Rangharajan and Keutzer, Kurt and Shao, Sophia and Khailany, Brucek},
  journal={arXiv preprint arXiv:2504.14152},
  year={2025}
}

@inproceedings{dong2020hawqv2,
  title={{HAWQ-V2}: Hessian Aware Trace-Weighted Quantization of Neural Networks},
  author={Dong, Zhen and Yao, Zhewei and Arfeen, Daiyaan and Gholami, Amir and Mahoney, Michael W. and Keutzer, Kurt},
  booktitle={Advances in Neural Information Processing Systems},
  volume={33},
  pages={18518--18529},
  year={2020},
  publisher={Curran Associates, Inc.},
  url={https://proceedings.neurips.cc/paper/2020/hash/d77c703536718b95308130ff2e5cf9ee-Abstract.html}
}

@article{SpQR,
  title={Spqr: A sparse-quantized representation for near-lossless llm weight compression},
  author={Dettmers, Tim and Svirschevski, Ruslan and Egiazarian, Vage and Kuznedelev, Denis and Frantar, Elias and Ashkboos, Saleh and Borzunov, Alexander and Hoefler, Torsten and Alistarh, Dan},
  journal={arXiv preprint arXiv:2306.03078},
  year={2023}
}

@article{Squeezellm,
  title={Squeezellm: Dense-and-sparse quantization},
  author={Kim, Sehoon and Hooper, Coleman and Gholami, Amir and Dong, Zhen and Li, Xiuyu and Shen, Sheng and Mahoney, Michael W and Keutzer, Kurt},
  journal={arXiv preprint arXiv:2306.07629},
  year={2023}
}

@article{woodfisher,
  title={Woodfisher: Efficient second-order approximation for neural network compression},
  author={Singh, Sidak Pal and Alistarh, Dan},
  journal={Advances in Neural Information Processing Systems},
  volume={33},
  pages={18098--18109},
  year={2020}
}

@article{martens2020new,
  title={New insights and perspectives on the natural gradient method},
  author={Martens, James},
  journal={Journal of Machine Learning Research},
  volume={21},
  number={146},
  pages={1--76},
  year={2020}
}

@article{RPTQ,
  title={Rptq: Reorder-based post-training quantization for large language models},
  author={Yuan, Zhihang and Niu, Lin and Liu, Jiawei and Liu, Wenyu and Wang, Xinggang and Shang, Yuzhang and Sun, Guangyu and Wu, Qiang and Wu, Jiaxiang and Wu, Bingzhe},
  journal={arXiv preprint arXiv:2304.01089},
  year={2023}
}

@article{Quarot,
  title={Quarot: Outlier-free 4-bit inference in rotated llms},
  author={Ashkboos, Saleh and Mohtashami, Amirkeivan and Croci, Maximilian L and Li, Bo and Cameron, Pashmina and Jaggi, Martin and Alistarh, Dan and Hoefler, Torsten and Hensman, James},
  journal={Advances in Neural Information Processing Systems},
  volume={37},
  pages={100213--100240},
  year={2024}
}

@article{Spinquant,
  title={Spinquant: Llm quantization with learned rotations},
  author={Liu, Zechun and Zhao, Changsheng and Fedorov, Igor and Soran, Bilge and Choudhary, Dhruv and Krishnamoorthi, Raghuraman and Chandra, Vikas and Tian, Yuandong and Blankevoort, Tijmen},
  journal={arXiv preprint arXiv:2405.16406},
  year={2024}
}

@article{Prefixquant,
  title={Prefixquant: Eliminating outliers by prefixed tokens for large language models quantization},
  author={Chen, Mengzhao and Liu, Yi and Wang, Jiahao and Bin, Yi and Shao, Wenqi and Luo, Ping},
  journal={arXiv preprint arXiv:2410.05265},
  year={2024}
}

@misc{NVFP4,
  title={Introducing NVFP4 for Efficient and Accurate Low-Precision Inference},
  author={Alvarez, Eduardo and Almog, Omri and Chung, Eric and Layton, Simon and Stosic, Dusan and Krashinsky, Ronny and Aubrey, Kyle},
  year={2025},
  month={Jun},
  howpublished={NVIDIA Developer Blog},
  url={https://developer.nvidia.com/blog/introducing-nvfp4-for-efficient-and-accurate-low-precision-inference/}
}

@article{microscaling,
  title={Microscaling Data Formats for Deep Learning},
  author={Rouhani, Bita Darvish and Zhao, Ritchie and More, Ankit and Hall, Mathew and Khodamoradi, Alireza and Deng, Summer and Choudhary, Dhruv and Cornea, Marius and Dellinger, Eric and Denolf, Kristof and Stosic, Dusan and Elango, Venmugil and Golub, Maximilian and Heinecke, Alexander and James-Roxby, Phil and Jani, Dharmesh and Kolhe, Gaurav and Langhammer, Martin and Li, Ada and Melnick, Levi and Mesmakhosroshahi, Maral and Rodriguez, Andres and Schulte, Michael and Shafipour, Rasoul and Shao, Lei and Siu, Michael and Dubey, Pradeep and Micikevicius, Paulius and Naumov, Maxim and Verrilli, Colin and Wittig, Ralph and Burger, Doug and Chung, Eric},
  journal={arXiv preprint arXiv:2310.10537},
  year={2023},
  eprint={2310.10537},
  archivePrefix={arXiv},
  primaryClass={cs.LG},
  doi={10.48550/arXiv.2310.10537},
  url={https://arxiv.org/abs/2310.10537}
}

@inproceedings{ashkboos2024quik,
  title={Quik: Towards end-to-end 4-bit inference on generative large language models},
  author={Ashkboos, Saleh and Markov, Ilia and Frantar, Elias and Zhong, Tingxuan and Wang, Xincheng and Ren, Jie and Hoefler, Torsten and Alistarh, Dan},
  booktitle={Proceedings of the 2024 Conference on Empirical Methods in Natural Language Processing},
  pages={3355--3371},
  year={2024}
}

@article{chen2026massive,
  title={Massive Spikes in LLMs are Bias Vectors: Mechanistic Uncovering and Spike-Free Quantization},
  author={Chen, Yung-Chin and Lee, Chung Peng and Liou, Ze-Wei and Verma, Naveen},
  journal={arXiv preprint arXiv:2606.02288},
  year={2026}
}

@article{LLAMA2-7B,
  title={{Llama 2}: Open Foundation and Fine-Tuned Chat Models},
  author={Touvron, Hugo and Martin, Louis and Stone, Kevin and Albert, Peter and Almahairi, Amjad and Babaei, Yasmine and Bashlykov, Nikolay and Batra, Soumya and Bhargava, Prajjwal and Bhosale, Shruti and Bikel, Dan and Blecher, Lukas and Canton Ferrer, Cristian and Chen, Moya and Cucurull, Guillem and Esiobu, David and Fernandes, Jude and Fu, Jeremy and Fu, Wenyin and Fuller, Brian and Gao, Cynthia and Goswami, Vedanuj and Goyal, Naman and Hartshorn, Anthony and Hosseini, Saghar and Hou, Rui and Inan, Hakan and Kardas, Marcin and Kerkez, Viktor and Khabsa, Madian and Kloumann, Isabel and Korenev, Artem and Koura, Punit Singh and Lachaux, Marie-Anne and Lavril, Thibaut and Lee, Jenya and Liskovich, Diana and Lu, Yinghai and Mao, Yuning and Martinet, Xavier and Mihaylov, Todor and Mishra, Pushkar and Molybog, Igor and Nie, Yixin and Poulton, Andrew and Reizenstein, Jeremy and Rungta, Rashi and Saladi, Kalyan and Schelten, Alan and Silva, Ruan and Smith, Eric Michael and Subramanian, Ranjan and Tan, Xiaoqing Ellen and Tang, Binh and Taylor, Ross and Williams, Adina and Kuan, Jian Xiang and Xu, Puxin and Yan, Zheng and Zarov, Iliyan and Zhang, Yuchen and Fan, Angela and Kambadur, Melanie and Narang, Sharan and Rodriguez, Aurelien and Stojnic, Robert and Edunov, Sergey and Scialom, Thomas},
  journal={arXiv preprint arXiv:2307.09288},
  year={2023},
  eprint={2307.09288},
  archivePrefix={arXiv},
  primaryClass={cs.CL},
  doi={10.48550/arXiv.2307.09288},
  url={https://arxiv.org/abs/2307.09288}
}

@misc{LLaMA-3.2,
  author={{Meta AI}},
  title={{Llama 3.2}: Revolutionizing Edge {AI} and Vision with Open, Customizable Models},
  howpublished={Meta AI Blog},
  month=sep,
  year={2024},
  url={https://ai.meta.com/blog/llama-3-2-connect-2024-vision-edge-mobile-devices/},
  note={Accessed: 2026-10-18}
}

@misc{MISTRAL-7B-v0.3,
  author={{Mistral AI}},
  title={{Mistral-7B-v0.3} Model Card},
  howpublished={Hugging Face model repository},
  month=may,
  year={2024},
  url={https://huggingface.co/mistralai/Mistral-7B-v0.3},
  note={Accessed: 2026-10-18}
}

@article{QWEN2-1.5B-Instruct,
  title={{Qwen2} Technical Report},
  author={Yang, An and Yang, Baosong and Hui, Binyuan and Zheng, Bo and Yu, Bowen and Zhou, Chang and Li, Chengpeng and Li, Chengyuan and Liu, Dayiheng and Huang, Fei and Dong, Guanting and Wei, Haoran and Lin, Huan and Tang, Jialong and Wang, Jialin and Yang, Jian and Tu, Jianhong and Zhang, Jianwei and Ma, Jianxin and Yang, Jianxin and Xu, Jin and Zhou, Jingren and Bai, Jinze and He, Jinzheng and Lin, Junyang and Dang, Kai and Lu, Keming and Chen, Keqin and Yang, Kexin and Li, Mei and Xue, Mingfeng and Ni, Na and Zhang, Pei and Wang, Peng and Peng, Ru and Men, Rui and Gao, Ruize and Lin, Runji and Wang, Shijie and Bai, Shuai and Tan, Sinan and Zhu, Tianhang and Li, Tianhao and Liu, Tianyu and Ge, Wenbin and Deng, Xiaodong and Zhou, Xiaohuan and Ren, Xingzhang and Zhang, Xinyu and Wei, Xipin and Ren, Xuancheng and Liu, Xuejing and Fan, Yang and Yao, Yang and Zhang, Yichang and Wan, Yu and Chu, Yunfei and Liu, Yuqiong and Cui, Zeyu and Zhang, Zhenru and Guo, Zhifang and Fan, Zhihao},
  journal={arXiv preprint arXiv:2407.10671},
  year={2024},
  eprint={2407.10671},
  archivePrefix={arXiv},
  primaryClass={cs.CL},
  doi={10.48550/arXiv.2407.10671},
  url={https://arxiv.org/abs/2407.10671}
}

@article{WikiText-2,
  title={Pointer sentinel mixture models},
  author={Merity, Stephen and Xiong, Caiming and Bradbury, James and Socher, Richard},
  journal={arXiv preprint arXiv:1609.07843},
  year={2016}
}

@inproceedings{PIQA,
  title={{PIQA}: Reasoning about Physical Commonsense in Natural Language},
  author={Bisk, Yonatan and Zellers, Rowan and Le Bras, Ronan and Gao, Jianfeng and Choi, Yejin},
  booktitle={Proceedings of the AAAI Conference on Artificial Intelligence},
  volume={34},
  pages={7432--7439},
  year={2020},
  doi={10.1609/aaai.v34i05.6239},
  url={https://ojs.aaai.org/index.php/AAAI/article/view/6239}
}

@inproceedings{BoolQ,
  title={Boolq: Exploring the surprising difficulty of natural yes/no questions},
  author={Clark, Christopher and Lee, Kenton and Chang, Ming-Wei and Kwiatkowski, Tom and Collins, Michael and Toutanova, Kristina},
  booktitle={Proceedings of the 2019 conference of the north American chapter of the association for computational linguistics: Human language technologies, volume 1 (long and short papers)},
  pages={2924--2936},
  year={2019}
}

@article{ARC,
  title={Think you have solved question answering? try arc, the ai2 reasoning challenge},
  author={Clark, Peter and Cowhey, Isaac and Etzioni, Oren and Khot, Tushar and Sabharwal, Ashish and Schoenick, Carissa and Tafjord, Oyvind},
  journal={arXiv preprint arXiv:1803.05457},
  year={2018}
}

@inproceedings{openbookQA,
  title={Can a suit of armor conduct electricity? a new dataset for open book question answering},
  author={Mihaylov, Todor and Clark, Peter and Khot, Tushar and Sabharwal, Ashish},
  booktitle={Proceedings of the 2018 conference on empirical methods in natural language processing},
  pages={2381--2391},
  year={2018}
}

@article{winogrande,
  title={Winogrande: An adversarial winograd schema challenge at scale},
  author={Sakaguchi, Keisuke and Bras, Ronan Le and Bhagavatula, Chandra and Choi, Yejin},
  journal={Communications of the ACM},
  volume={64},
  number={9},
  pages={99--106},
  year={2021},
  publisher={ACM New York, NY, USA}
}

@inproceedings{HellaSwag,
  title={Hellaswag: Can a machine really finish your sentence?},
  author={Zellers, Rowan and Holtzman, Ari and Bisk, Yonatan and Farhadi, Ali and Choi, Yejin},
  booktitle={Proceedings of the 57th annual meeting of the association for computational linguistics},
  pages={4791--4800},
  year={2019}
}

@inproceedings{ImageNet-1K,
  title={Imagenet: A large-scale hierarchical image database},
  author={Deng, Jia and Dong, Wei and Socher, Richard and Li, Li-Jia and Li, Kai and Fei-Fei, Li},
  booktitle={2009 IEEE conference on computer vision and pattern recognition},
  pages={248--255},
  year={2009},
  organization={Ieee}
}

@inproceedings{cimloop,
  title={CiMLoop: A flexible, accurate, and fast compute-in-memory modeling tool},
  author={Andrulis, Tanner and Emer, Joel S and Sze, Vivienne},
  booktitle={2024 IEEE International Symposium on Performance Analysis of Systems and Software (ISPASS)},
  pages={10--23},
  year={2024},
  organization={IEEE}
}

@article{accelforge,
  title={AccelForge: Comprehensive Modeling and Co-Design Framework for AI Accelerators},
  author={Andrulis, Tanner and Gilbert, Michael and Sze, Vivienne and Emer, Joel S},
  journal={arXiv preprint arXiv:2609.11906},
  year={2026}
}

@misc{adc_survey,
   author = {Murmann, Boris},
   title = {{ADC Performance Survey 1997-2026}},
   year = {2026},
   note = {[Online]. Available: \url{https://github.com/bmurmann/ADC-survey}}
}

@inproceedings{muralimanohar2007cacti,
  title={{Optimizing NUCA Organizations and Wiring Alternatives for Large Caches with CACTI 6.0}},
  author={Muralimanohar, Naveen and Balasubramonian, Rajeev and Jouppi, Norman P.},
  booktitle={40th Annual IEEE/ACM International Symposium on Microarchitecture (MICRO 2007)},
  pages={3--14},
  year={2007},
  month=dec,
  publisher={IEEE},
  doi={10.1109/MICRO.2007.33},
  url={https://doi.org/10.1109/MICRO.2007.33}
}

@inproceedings{shafiee2016isaac,
  title={{ISAAC}: A Convolutional Neural Network Accelerator with In-Situ Analog Arithmetic in Crossbars},
  author={Shafiee, Ali and Nag, Anirban and Muralimanohar, Naveen and Balasubramonian, Rajeev and Strachan, John Paul and Hu, Miao and Williams, R. Stanley and Srikumar, Vivek},
  booktitle={2016 ACM/IEEE 43rd Annual International Symposium on Computer Architecture (ISCA)},
  pages={14--26},
  year={2016},
  month=jun,
  publisher={IEEE},
  doi={10.1109/ISCA.2016.12},
  url={https://doi.org/10.1109/ISCA.2016.12}
}

@inproceedings{oconnor2017fgdram,
  title={Fine-grained {DRAM}: Energy-efficient {DRAM} for Extreme Bandwidth Systems},
  author={O'Connor, Mike and Chatterjee, Niladrish and Lee, Donghyuk and Wilson, John and Agrawal, Aditya and Keckler, Stephen W. and Dally, William J.},
  booktitle={Proceedings of the 50th Annual IEEE/ACM International Symposium on Microarchitecture},
  series={MICRO-50},
  pages={41--54},
  year={2017},
  month=oct,
  publisher={Association for Computing Machinery},
  address={New York, NY, USA},
  doi={10.1145/3123939.3124545},
  url={https://doi.org/10.1145/3123939.3124545}
}

@article{shao2014aladdin,
  title={{Aladdin}: A Pre-{RTL}, Power-Performance Accelerator Simulator Enabling Large Design Space Exploration of Customized Architectures},
  author={Shao, Yakun Sophia and Reagen, Brandon and Wei, Gu-Yeon and Brooks, David},
  journal={ACM SIGARCH Computer Architecture News},
  volume={42},
  number={3},
  pages={97--108},
  year={2014},
  month=jun,
  publisher={Association for Computing Machinery},
  doi={10.1145/2678373.2665689},
  url={https://doi.org/10.1145/2678373.2665689}
}

@article{Slicegpt,
  title={Slicegpt: Compress large language models by deleting rows and columns},
  author={Ashkboos, Saleh and Croci, Maximilian L and Nascimento, Marcelo Gennari do and Hoefler, Torsten and Hensman, James},
  journal={arXiv preprint arXiv:2401.15024},
  year={2024}
}

@article{RoPE,
  title={Roformer: Enhanced transformer with rotary position embedding},
  author={Su, Jianlin and Ahmed, Murtadha and Lu, Yu and Pan, Shengfeng and Bo, Wen and Liu, Yunfeng},
  journal={Neurocomputing},
  volume={568},
  pages={127063},
  year={2024},
  publisher={Elsevier}
}

@article{SwiGLU,
  title={Glu variants improve transformer},
  author={Shazeer, Noam},
  journal={arXiv preprint arXiv:2002.05202},
  year={2020}
}

@inproceedings{HuggingFace,
  title={Transformers: State-of-the-Art Natural Language Processing},
  author={Wolf, Thomas and Debut, Lysandre and Sanh, Victor and Chaumond, Julien and Delangue, Clement and Moi, Anthony and Cistac, Pierric and Rault, Tim and Louf, Remi and Funtowicz, Morgan and Davison, Joe and Shleifer, Sam and von Platen, Patrick and Ma, Clara and Jernite, Yacine and Plu, Julien and Xu, Canwen and Le Scao, Teven and Gugger, Sylvain and Drame, Mariama and Lhoest, Quentin and Rush, Alexander},
  booktitle={Proceedings of the 2020 Conference on Empirical Methods in Natural Language Processing: System Demonstrations},
  month=oct,
  year={2020},
  address={Online},
  publisher={Association for Computational Linguistics},
  pages={38--45},
  doi={10.18653/v1/2020.emnlp-demos.6},
  url={https://aclanthology.org/2020.emnlp-demos.6/}
}
